\pdfoutput=1

\documentclass[smallcondensed]{svjour3}
\journalname{Empirical Software Engineering}
\usepackage[T1]{fontenc}
\usepackage[utf8]{inputenc}
\usepackage{lmodern}
\usepackage{microtype}
\usepackage{graphicx}
\usepackage{pdflscape}
\usepackage{xcolor}
\usepackage{booktabs}
\usepackage{array}
\usepackage{longtable}
\usepackage{tabularx}
\usepackage{multirow}
\usepackage{makecell}
\usepackage{amsmath,amssymb}
\usepackage{siunitx}
\usepackage{hyperref}
\usepackage[round]{natbib}
\makeatletter
\providecommand{\cl@chapter}{}
\let\cl@chapter\@empty
\makeatother
\usepackage{cleveref}
\usepackage{enumitem}
\usepackage{url}
\usepackage{orcidlink}
\usepackage{tikz}
\usetikzlibrary{shapes.geometric, positioning, arrows.meta}
\usepackage{pgfplots}
\pgfplotsset{compat=1.18}
\usepackage{tcolorbox}
\tcbuselibrary{breakable}
\usepackage{listings}
\usepackage{placeins}    
\hypersetup{
  colorlinks=true,
  linkcolor=black!70!blue,
  citecolor=black!70!green,
  urlcolor=black!70!purple,
  pdftitle={Do programming languages still matter to your AI coding agent teammate? Evidence at scale from chess engines}
}

\newcommand{\claudecode}{\textsc{Claude Code}}
\newcommand{\codex}{\textsc{Codex}}
\newcommand{\stockfish}{\textsc{Stockfish}}
\newcommand{\gpt}[1]{\textsc{gpt-#1}}
\newcommand{\opus}[1]{\textsc{Opus #1}}

\newcommand{\elo}[1]{#1~Elo}
\newcommand{\blitem}[1]{\textsc{bl-#1}}
\newcommand{\sbitem}[1]{\textsc{sb-#1}}
\newcommand{\RQ}[1]{\textbf{RQ#1}}
\newcommand{\agentbuilt}{agent-built}
\newcommand{\perft}{\texttt{perft}}
\newcommand{\uci}{\textsc{uci}}
\newcommand{\pgn}{\textsc{pgn}}

\newenvironment{rqanswer}[1]{%
  \par\medskip\noindent
  \colorbox{black!12}{\textbf{Answer to RQ#1.}}\hspace{0.5em}\ignorespaces
}{\par\medskip}

\usepackage[htt]{hyphenat}

\title{Do programming languages still matter to your AI coding
       agent teammate?\\Evidence at scale from chess engines}
\titlerunning{Do PLs still matter to AI coding agent teammates?}
\author{Mathieu Acher \and Jean-Marc J\'ez\'equel}
\authorrunning{M.\ Acher, J.-M.\ J\'ez\'equel}
\institute{%
  Mathieu Acher \at
    Univ Rennes, IRISA, CNRS, Inria, IUF, France\\
    \email{mathieu.acher@irisa.fr}
  \and
  Jean-Marc J\'ez\'equel \at
    Univ Rennes, IRISA, CNRS, Inria, IUF, France\\
    \email{jean-marc.jezequel@irisa.fr}
}
\date{Received: date / Accepted: date}

\begin{document}

\maketitle

\begin{abstract}
\noindent
Frontier coding agents promise to move software construction
from line-level autocompletion to end-to-end authorship of
complete software systems. Two empirical questions follow: do
AI coding-agent teammates have the ability to program in any
target programming language, including ones with no comparable
prior open-source artefact. And if so, does the choice of
language still shape the resulting artefact, and along which
dimensions? We investigate both questions through a polyglot
case study built around chess engines. Chess engines are non-trivial
multi-component software systems and admit a hierarchy of
language-agnostic oracles, ranging from exact
move-generation correctness to a strength
scale (Elo), that are observable whether the
implementation language is Rust or Brainfuck, an esoteric language with 8 commands.
We prompted two frontier coding agents (\textsc{Claude Code} and \textsc{Codex}, both run in
high-reasoning mode) at the capability level i) without
supplying chess knowledge or specific implementation
guidance ii) under a documented, systematic intervention and
stopping policy. Coding agents produced
\emph{34 agent-built chess engines spanning 17 primary
programming languages}, from mainstream to specialised,
domain-specific, legacy, and esoteric targets. We
combine a quantitative footprint (per-engine feature analysis, independent Elo
assessment, session trajectories) with a qualitative
analysis of source code and session transcripts, including a cross-language comparison of representative features (quiescence search, transposition
table, evaluation function).

Our results show that frontier coding agents are genuinely
\textbf{polyglot}: every language we tried produced at least
one feature-rich, working engine. For several of these
engines we could not identify any prior open-source chess
engine of comparable scope in the language, including
technically demanding targets like \LaTeX{}, while the COBOL
engines go well beyond the language's thin hobby precedent. Beyond raw
coverage, agents now take
on genuine software-engineering tasks end-to-end, and the
code they ship is synthesised from scratch rather than
copied. Even within this polyglot capability, we provide evidence that
\textbf{the choice of language still matters} in three
concrete ways. Strong playing strength is only
reachable in mainstream compiled languages. Cost and software engineering effort
grow sharply as the language becomes more exotic. The
chess features the agent picks, and how it codes them,
change from one language family to another. Agents also
check their own work without being asked: they write
correctness tests, play games against themselves, and
estimate their playing strength on their own. That
autonomy has limits, however. The self-estimates run too
fast, and a few engines cheated by calling a chess library
instead of implementing chess themselves.
Our case study suggests
that programming language is no longer about \emph{whether}
AI teammates can build a working system. It is now a choice
about performance (strength), cost, what features get built
and how, and how much humans still need to engineer the
agent's testing/benchmarking and check any cheating.

\end{abstract}

\section{Introduction}
\label{sec:intro}

\begin{tcolorbox}[
  colback=gray!4, colframe=gray!50!black, boxrule=0.4pt,
  sharp corners, left=8pt, right=8pt, top=4pt, bottom=4pt,
]
\small
\emph{``Code itself will go away in favor of just making the
binary directly.''} \\[1pt]
{\hfill-- Elon Musk, public remark, 2025--2026~\cite{musk2025codegoesaway}.}
\end{tcolorbox}

By integrating external tools, coding agents are redefining software development,
transitioning the role of \emph{large language models (LLMs)} from simple code completion to the end-to-end authorship of complete, functional, and performant systems.
The claim quoted above, made by Elon
Musk~\cite{musk2025codegoesaway} and widely circulated through
2025--2026, captures a recurring view in the public discourse
around this shift: that programming languages are an artefact
of the human-in-the-loop era and will become obsolete once a
sufficiently capable code-generating system arrives. With
agents now routinely completing multi-hour software tasks at
human-comparable reliability~\cite{kwa2026measuringaiabilitycomplete},
the claim is no longer abstract but becomes a falsifiable
empirical question: \emph{do programming languages (PLs) still matter to your
AI coding-agent teammate?} If yes, on which axes? If no, in
which cases? And what does ``mattering'' mean once the agent,
not a human, is the day-to-day implementer?

The Empirical Software Engineering special issue
\emph{``Agentic Software Engineering: The Rise of AI Teammates''},
rooted in the SE~3.0 vision of autonomous coding agents acting
as teammates rather than code-completion
aids~\cite{li2025riseaiteammatessoftware},
asks exactly this kind of question: what does an AI teammate
actually \emph{do} when given a non-trivial \emph{software engineering (SE)}
task, and how should the empirical-SE community measure it? The
existing polyglot-evaluation literature stops short of an
end-to-end answer. Function-synthesis benchmarks
(HumanEval~\cite{chen2021humaneval}, MBPP~\cite{austin2021mbpp},
MultiPL-E~\cite{cassano2023multipl}, HumanEval-XL~\cite{peng2024humanevalxl})
fix the language and limit the task to a single function or method.
Issue-repair benchmarks (SWE-bench~\cite{jimenez2024swebench},
Multi-SWE-bench~\cite{zan2025multiswebench},
Aider Polyglot~\cite{gauthier2024aiderpolyglot}) require an
existing repository to localise patches against, and span only
mainstream targets. A closest precedent for end-to-end
agent-driven \emph{system construction} is
Carlini--Anthropic's parallel-Claudes
compiler~\cite{carlini2026compiler} that targets a single mainstream
language (Rust), one large artefact, one agent team. None of these
combines \emph{from-scratch system construction} with a language
spectrum that crosses categories from mainstream general-purpose
to truly esoteric. We do.

Chess is unusually well-instrumented for this question. A working
chess engine is a feature-rich software system spanning board
representation, legal move generation, alpha--beta search, an
evaluation function, transposition tables, \emph{Universal Chess Interface} (\uci{}) protocol
compliance to play games, etc.
It is not a simple 8x8 matrix with some legal moves, but typically 500--5{,}000 lines that must compile,
run, and produce correct results to be usable at all. Chess also
admits a \emph{hierarchy of language-agnostic oracles}:
\perft{} node counts give exact move-generation correctness,
\uci{} gameplay records give legal-move compliance verifiable by
any compliant engine, and games against chess engines like \stockfish{} give a
comparable strength estimate on the Elo scale (a common chess rating that summarises ``how often does one player beat
another'' as a single number per player, see
\Cref{sec:background:oracles} for how to read it).
A chess engine written in Brainfuck (an esoteric language with 8 commands) and a Rust
engine can be measured on identical axes.
Beyond measurability, building a chess engine is a software engineering
task rather than a pure code-generation exercise. To reach a
working engine, an agent must comprehend and revise its own
code, keep the project building, test move generation against
exact oracles, debug the failures those oracles expose, and
benchmark the result through actual games. Our session data
confirms this breadth empirically: implementation and code
comprehension appear in every recovered session, and debugging,
testing, benchmarking, and build tooling each appear in a
majority of engines in every language category
(\Cref{sec:rq5:se-activities}).
This combination of a non-trivial feature-rich target,
language-agnostic external oracles, and broad
programming-language coverage is precisely what existing agent
benchmarks lack.

Working from chess as the case study, the first author prompted two frontier coding agents, \claudecode{} (Anthropic \opus{4.6}/\opus{4.7}) and
\codex{} (OpenAI \gpt{5-codex}, \gpt{5.3-codex}, \gpt{5.4}), both run in
high-reasoning mode, to construct chess
engines across 17 primary programming languages.\footnote{The
17 are counted over the 29 main-corpus from-scratch engines:
Python, Java, C, C\texttt{++}, Rust, Ruby, APL, Icon,
Lean~4, Why3, Rocq, LaTeX/TeX, CSS/HTML, SQL, COBOL, x86-64
assembly, and Brainfuck. LaTeX/TeX and CSS/HTML are each
counted once and the two OCaml-extracted formal targets
(Why3, Rocq) separately. The special-role ports, the GAMBIT
DSL experiment, and the underconverged Mojo session are
excluded from the count.} The corpus
contains \textbf{34 \agentbuilt{} chess engines}: 29 from-scratch
\emph{main-corpus} engines plus 5 \emph{special-role} experiments
(2 Java$\to$language ports, 1 DSL-design experiment, 1
unconverged attempt, 1 agent-heavy co-design case).
 Thanks to this unique data (including generated code, session transcripts, and analysis),
we address five research questions. Two are
\emph{qualitative} describing what the agents did. Three are
\emph{quantitative} measuring outcomes against external oracles or transcripts.

\begin{itemize}[leftmargin=1.4em,itemsep=2pt,topsep=2pt]
  \item \RQ{1} (quantitative, \emph{coverage}). \emph{Can AI
    coding agent teammates produce a working chess engine in
    every programming language we attempt?}
    Across which PL$\times$agent does a working chess engine
    emerge? 
  \item \RQ{2} (qualitative, \emph{synthesis \& features}). \emph{What
    do the agents build, and is it a copy?} Which feature sets are included per engine, how do agents adapt the same algorithm to different
    PLs, and is the code genuinely synthesised or memorised?
  \item \RQ{3} (qualitative+quantitative, \emph{validation}).
    \emph{How do agent teammates demonstrate that their software is
    correct?} Which oracles do they expose and implement (e.g., \perft{}, \uci{},
    self-Elo rating), when in the session are they introduced, and are there any attempts to \emph{cheat}, e.g., when the language
    constraint is relaxable through a library?
  \item \RQ{4} (quantitative, \emph{strength}). \emph{Does the PL
    matter for playing strength?} What playing strength (Elo)
    do the engines reach under a unified re-evaluation harness, and
    how do throughput, feature richness, and category track Elo?
  \item \RQ{5} (quantitative, \emph{cost \& effort}). \emph{Does the
    PL matter for cost and effort?} How many prompts, tokens, and
    dollars does a working engine cost? What fraction of the session
    is, e.g., debugging, and how does the breadth and intensity of
    software engineering activities the agent exercises differ across PL categories?
\end{itemize}

\RQ{1} establishes the \emph{existence} half of our answer: all
17 target languages \emph{can} host an \agentbuilt{} engine.
\RQ{4} and \RQ{5} establish the \emph{impact} half: PLs matter for
ceiling Elo, for cost, for the breadth and intensity of software engineering work the
agent has to do. \RQ{2} and \RQ{3} are the qualitative observations
that make the quantitative answers interpretable: we identify and characterise \emph{what} was built, \emph{how} it was synthesised across
languages, and \emph{how} the agents demonstrated correctness
to themselves and to us.

Before digging into the RQs, including the title question
itself, it is worth highlighting an important result: frontier coding agents now build complete chess engines in multiple, diverse PLs.
This
is a \emph{polyglot tour de force}. The agents did not stop at familiar
PLs such as C, Java, Python, or Rust (and even in such cases, agents did not generate pure copies, see hereafter).
They also produced working
engines in targets where chess is a challenging fit: pure TeX,
where data structures and control flow have to be encoded through
macros, or COBOL, a legacy PL rarely associated with advanced game-tree
search. For pure CSS, pure-TeX, APL, Icon, Lean~4,
Why3/Rocq, and Brainfuck we could not identify any previously
reported open-source chess engine of comparable scope, and
the corpus's x86-64 assembly engine has no comparable
open-source \uci{} predecessor we could find (we scope that
claim against older assembly prior art in \Cref{sec:rq1}).
 This is more than a
larger multilingual benchmark.
Prior polyglot evaluations mostly ask for functions, patches,
or edits~\cite{chen2021humaneval, cassano2023multipl, jimenez2024swebench, zan2025multiswebench}.
Here the agents build complete systems from scratch, without specific technical guidance
or interventions about architecture or chess knowledge.
 That is a capability shift: frontier coding agents can now
carry feature-rich system construction into PLs that were
previously unsupported by public examples, poorly suited to the
task, or both.

That shift refines the title question. Once agents can make even
LaTeX, COBOL, and Brainfuck work, the important question is no
longer only whether a PL can host the system. It is what the PL
changes once the system exists. Our evidence-based answer is as follows:
\textbf{Yes, programming languages still matter to your AI
coding agent teammate}.
Specifically, PLs set a \emph{ceiling} on playing strength (Elo rating): the
\elo{1900}--\elo{2100} band is mainstream-compiled only PLs.
Specialised, domain-specific, legacy, and esoteric engines
plateau hundreds to thousands of Elo below. PLs also multiply
\emph{cost and effort}. A typical mainstream first-pass engine
takes a handful of prompts and an estimated ${\sim}\$2$--\$30 of canonical
USD (\Cref{sec:corpus:evidence}). Engines in the harder PL
categories take ${\sim}$25--50 prompts at an estimated ${\sim}\$60$--\$480, with
debug-prompt fractions above 0.4 and broader SE activity from
the agent. PL choice therefore no longer answers
only ``can the agent make this language work?'' It answers
``how strong, how expensive, and with how much SE work?''.
%
%
Once every language we attempted is reachable and a candidate, choosing one becomes a first-class
engineering decision rather than a bottleneck check.
\emph{What matters when choosing a programming language for an
AI-teammate-built system?} The corpus suggests at least four
axes that move with PL category in this domain: the strength
ceiling the language allows, the cost multiplier on the
agent's session, the breadth of SE activities the agent has
to exercise, and the kind of features and verification the language affords. We do not turn this
into a fully predictive model, but we lay out the evidence
across \RQ{2}--\RQ{5}.

Another result, which we treat as one of the paper's
contributions (even though it is, in part, a negative one),
concerns Elo rating \emph{self-evaluation}. The agents wire up their own
strength procedures autonomously (\Cref{sec:rq3}), but the in-loop
self-Elo signal they optimise against is biased
\elo{200}--\elo{1100} above the external Elo assessment, with
extreme cases up to \elo{1078} of overestimation. To put those
numbers in scale: at a \elo{200} overestimate the favourite-to-win
prediction has already flipped, and the \elo{1000}+ cases describe
engines that lose essentially every game but report themselves
as credible opponents. This matters because the agent uses the
same signal to decide whether a code change helped and when to
stop iterating, so the bias pushes the agent to declare success
too early. The cause is the benchmarking setup the agents
converge to on their own: they run too few games per opponent
for the estimate to stabilise, and at time controls too short
to let their alpha--beta search reach reliable depth.
 The
implication for AI-teammate practice is direct: agents are
capable of validation, but precise and trustable Elo benchmarking still
requires more human guidance than current sessions get.
There is also the risk that agents cheat by implementing the software system in another PL
or reusing a library. It sits at the interface between \RQ{3}
(validation) and \RQ{4} (strength) and is easy to miss inside
either one alone.

Our work is an exploratory empirical field study. We observe and
characterise. We do not causally explain. We carefully separate
what we directly observe from what we describe but do not
causally attribute. Existence proofs and synthesis evidence rest
on direct observation. PL-vs-strength and PL-vs-cost
\emph{tendencies} are descriptive: six entangled confounders
(model mastery of the language, session stochasticity,
computational platform, budget, benchmarking strategy, and
feature--PL correlation) prevent any causal language
ranking.\footnote{The confounders are catalogued in
\Cref{sec:threats}. We treat them as a
contribution rather than an apology: they specify exactly what a
follow-up controlled experiment must hold fixed.}

\paragraph{Contributions.} Our contributions are as follows:

\begin{itemize}[leftmargin=1.4em,itemsep=3pt,topsep=2pt]
  \item \textbf{Polyglot, end-to-end system construction as a
    frontier-agent capability shift.} 34 \agentbuilt{}
    feature-rich chess engines (29 main-corpus + 5
    special-role) across 17 programming languages spanning
    five categories. Every PL we tried produced at least
    one working engine: a feasibility result not reported in
    the prior literature, plausibly because
    the previous generation of LLMs could not make it.
    No prior study combines from-scratch system
    construction with a language spectrum reaching strictly
    esoteric and previously non-implemented targets
    (\Cref{sec:related}).

  \item \textbf{First-of-kind \agentbuilt{} engines in PLs
    without prior chess engines of comparable scope.} The
    corpus contains, to our
    knowledge, the first chess engines of comparable scope ever
    produced in pure-TeX,
    APL, Icon, Lean~4, and Why3/Rocq, plus
    Brainfuck and the COBOL targets that have only thin hobby
    precedent, and an x86-64 assembly \uci{} engine without a
    comparable open-source predecessor we could identify.
    We treat these as a contribution in their own
    right: they show that AI coding-agent teammates can
    \emph{master} even unsuited or under-resourced PLs, and
    they reframe the empirical-SE question from ``which PLs
    are reachable?'' (in this corpus: all attempted) to \emph{``what
    matters when choosing a PL once every candidate is reachable?''}.
    It is the follow-up question that motivates the rest of the
    paper.

  \item \textbf{An empirical answer to ``do PLs still
    matter?''.} With existence (\RQ{1}), strength (\RQ{4}),
    and cost/effort (\RQ{5}) decomposed against language
    category, agent, feature set, and oracle rigor, we provide
    the first measurable statement of \emph{which} dimensions
    of ``mattering'' an AI teammate alters and which it does
    not, on the chess-engine domain.
    This answer runs counter to the code-obsolescence claim
    quoted at the start of the paper: even for AI teammates,
    the choice of programming language continues to shape
    outcomes.

  \item \textbf{A qualitative analysis of agent code synthesis
    across PLs.} A pattern feature catalog and a \emph{Chess Engine Feature Explorer}
    web tool (\Cref{app:method:explorer}) that browses variants
    sub-feature by sub-feature. The recurring pattern is that agents reproduce
    the same conceptual blueprint in every language but adapt
    its structure to the target PL's idiom and make markedly
    different sub-feature choices from one variant to the next.

  \item \textbf{A characterisation of agent validation behaviour
    in the wild.} Oracle-first iteration as a property of coding agents
    with evidence that self-Elo
    benchmarking can go wrong (and how it can be fixed) as well as cheating
    attempts under language constraints.
    It has implications for any developers, researchers, or agent harness intended to enforce a
    domain constraint.

  \item \textbf{A unified re-evaluation harness} (Docker images
    of every engine, a calibrated anchor ladder of five external
    references, and a round-robin runner) producing a single
    directly-comparable Elo per engine, plus the original in-tree
    PGN-derived numbers as historical context.

  \item \textbf{A reproducible analysis pipeline} that ingests
    \claudecode{} and \codex{} session transcripts, computes
    repository and interaction statistics, extracts chess-specific
    features over a catalog, executes games' tournaments,
    and produces every table and figure the paper reports.
    We also shared all source code and data of the 34 engines in dedicated repositories.

  \item \textbf{A research agenda} naming six confounders that
    jointly prevent causal language-effectiveness claims in any
    observational study of this form, and specifying exactly
    what a follow-up controlled experiment must hold fixed.
\end{itemize}

\paragraph{Remainder of the paper.}
\Cref{sec:background} introduces coding agents and chess-engine
anatomy. \Cref{sec:corpus} describes the corpus and method. The five RQs follow: \RQ{1} coverage
(\Cref{sec:rq1}), \RQ{2} synthesis and features
(\Cref{sec:rq2}), \RQ{3} validation
(\Cref{sec:rq3}), \RQ{4} strength (\Cref{sec:rq4}), and \RQ{5} cost
and effort (\Cref{sec:rq5}). \Cref{sec:threats} and
\Cref{sec:related} follow, and \Cref{sec:conclusion} discusses
the findings, concludes, and lays out future work.

\section{Background}
\label{sec:background}

\subsection{Frontier coding agents}
\label{sec:background:agents}

We use ``frontier coding agents'' to mean interactive, tool-using
LLM-powered systems that operate on a developer's local repository. The
two agents in our corpus are \claudecode{} (Anthropic) and \codex{}
(OpenAI). Both run a loop of \emph{plan} $\to$ \emph{tool use}
(shell, file edits, reads) $\to$ \emph{reflect}, persist a transcript of
every turn, and can invoke subagents for parallel exploration. In our
corpus we observe two model families: Anthropic's \opus{4.6} and
\opus{4.7}, and OpenAI's \gpt{5-codex}, \gpt{5.3-codex}, and
\gpt{5.4}. Both agents
were run in high-reasoning mode throughout (\claudecode{}'s
extended-thinking budget enabled and \codex{} with its
\texttt{reasoning=high} setting) so that the comparison is
between each frontier agent at its strongest configuration.

Both agents are fast-evolving commercial products, so we
record their configuration at the level the products expose.
\codex{} embeds the CLI version in every session header, and
the corpus sessions span versions 0.98.0 through 0.118.0.
The released derived snapshot does not record the
\claudecode{} CLI version, so we do not report a version
range for the \claudecode{} sessions.
Neither product exposes decoding parameters such as
temperature or random seeds, so all sessions ran at
vendor-default sampling settings, and an individual session
cannot be replayed exactly. Per-session timestamps are
preserved in the released data snapshot.

A \emph{session} is a sustained developer--agent conversation
that produces one engine (or fails trying). Both agents log
every session as a structured transcript. Each entry
carries a timestamp, the model used, token usage, and either
a developer message, the agent's reasoning and output, or a
tool invocation with its result.

\subsection{Chess-engine anatomy}
\label{sec:background:engine}

A standard chess engine is a pipeline of well-defined components~\cite{CPW}:
\begin{itemize}
  \item \textbf{Board representation.} Common choices: mailbox (8$\times$8
    or 0$\times$88), bitboards (64-bit masks per piece), or higher-level
    object models.
  \item \textbf{Move generation.} Pseudo-legal moves per piece, a
    \emph{legality filter} (king-in-check), and handling of special moves:
    castling, en passant, pawn promotion.
  \item \textbf{Evaluation.} Minimally material, typically extended with
    piece-square tables, tapered mid/endgame scoring, pawn-structure
    terms, king safety.
  \item \textbf{Search.} Negamax or minimax with alpha--beta pruning,
    iterative deepening, quiescence, move ordering (MVV-LVA, killer
    moves, history heuristic), and transposition tables keyed by a
    Zobrist hash. More advanced engines add null-move pruning and
    late-move reductions.
  \item \textbf{Protocols.} \uci{} is the de facto protocol to play games against other engines.
    \pgn{} is the game-record format used for measuring strength.
\end{itemize}

\subsection{An oracle hierarchy}
\label{sec:background:oracles}

Chess provides three layered oracles that are themselves encoded in code
or data, and that an agent can drive autonomously:
\begin{enumerate}
  \item \textbf{\perft{}.} Deterministic node counts of the
    move-generation tree at fixed depth from a fixed position. The
    reference values (up to depth~7) are published on the Chess
    Programming Wiki and admit exact comparison. Passing perft is strong
    evidence that move generation is correct.
  \item \textbf{Gameplay.} An engine that speaks \uci{} can play a
    reference engine (typically \stockfish{}) and produce \pgn{} records.
    Legal-move compliance is checkable mechanically: the record either
    is or is not a valid chess game.
  \item \textbf{Elo rating.} Against a calibrated opponent
    (e.g.\ \stockfish{} with \texttt{UCI\_LimitStrength}=true and
    \texttt{UCI\_Elo}=$R$), observing score $s$ over $N$ games yields
    an Elo estimate via the logistic
    $\Delta = -400 \log_{10}(\frac{1}{s} - 1)$ with standard error
    $\mathrm{SE}(\Delta) \approx
      \frac{400}{\ln 10 \cdot \sqrt{N \cdot s (1-s)}}$.
    Aggregating across opponents via inverse-variance weighting gives a
    confidence interval that shrinks with $N$.
\end{enumerate}

These three levels map onto three properties familiar from SE
testing: \textbf{functional correctness against a fixed
reference suite} (perft is a harsh, exhaustive
specification-conformance check), \textbf{functional
correctness across diverse operational scenarios} (every
gameplay record either parses as a valid chess game or does
not, across as many distinct game situations as the engine
plays), and \textbf{performance} (Elo grades how well the
engine plays on a cardinal, cross-engine scale). We use all
three to characterise the agents' validation behaviour
(\RQ{3}) and to assess playing strength (\RQ{4}).

\paragraph{How to read an Elo number.}
Elo is a way to put ``how good is this player?'' into a single
number. If player~A has 100 more Elo than player~B, A is
expected to win roughly two games out of three in the long run
(200 more is about three out of four, and 400 more is about ten
out of eleven). The numbers come from games actually played:
beating a stronger opponent moves your rating up, losing to a
weaker one moves it down, by an amount that depends on how
surprising the result was. Two practical points follow that
will matter when we read Elo numbers later in the paper. First,
an Elo number is learned from a particular community of
players, and only that community calibrates it: the
engine-vs-engine ladders (CCRL, CEGT), the FIDE human pool, and
the \stockfish{} gauntlets the agents run themselves are three
different communities that happen to use the same units, so an
\elo{1800} learned against one is not directly the same kind of
\elo{1800} as one learned against another. Second,
``\stockfish{}'' is not one opponent but a family: with
\texttt{UCI\_LimitStrength=true} and \texttt{UCI\_Elo=$R$} it
plays at approximately $R$ Elo, so two gauntlets that both say
``we played \stockfish{}'' can in fact be against very
different opponents depending on the target strength chosen.
Our re-evaluation harness (\Cref{sec:rq4}) handles both points
the same way: every corpus engine plays the same fixed panel of
opponents (Rustic at \elo{1820}, Asymptote at \elo{2150}, and
\stockfish{} at Skill 5/10/15) at the same time control, so the
Elo numbers we report in the \emph{External Elo Assessment}
column of \Cref{tab:rq4-elo} are on one consistent yardstick
rather than on whichever one each agent happened to pick.

\section{Corpus and method}
\label{sec:corpus}

\subsection{Corpus acquisition and interaction protocol}
\label{sec:corpus:acquisition}
\label{sec:corpus:protocol}
\label{sec:corpus:policy}

All projects in the corpus were created by the first author on a
single developer workstation over a period of about two and a
half months (2026-02 to 2026-04), each as a deliberate exercise
in ``can agent $A$ build a chess engine in language~$L$?''. No project was created to
pass a fixed benchmark. Each was a naturalistic attempt to
exercise the agents.

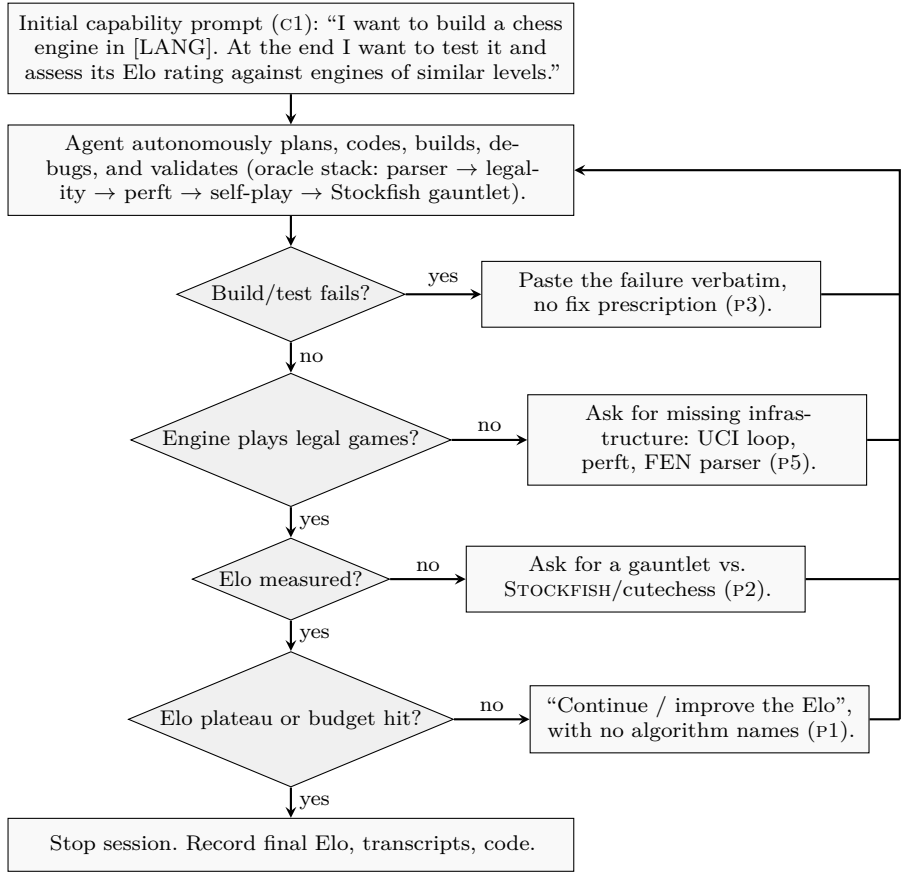
\begin{figure}[!htbp]
\centering
\begin{tikzpicture}[
  font=\small,
  node distance=4mm and 10mm,
  box/.style={draw, align=center,
              inner sep=4pt, minimum height=7mm, fill=gray!5},
  decision/.style={draw, diamond, aspect=2.4, align=center,
                   inner sep=1pt, fill=gray!12},
  flow/.style={->, >=stealth, thick},
  busline/.style={thick}
  ]
  \node[box, text width=72mm] (start) {%
    Initial capability prompt (\textsc{c1}):
    ``I want to build a chess engine in [LANG]. At the end I want to test it and assess its Elo
    rating against engines of similar levels.''};
  \node[box, below=of start, text width=72mm] (run) {%
    Agent autonomously plans, codes, builds, debugs,
    and validates (oracle stack: parser $\to$ legality
    $\to$ perft $\to$ self-play $\to$ Stockfish gauntlet).};
  \node[decision, below=of run]   (build)  {Build/test fails?};
  \node[box, right=of build, text width=42mm] (bug) {%
    Paste the failure verbatim, no fix prescription
    (\textsc{p3}).};
  \node[decision, below=of build] (play)   {Engine plays legal games?};
  \node[box, right=of play, text width=42mm]  (infra) {%
    Ask for missing infrastructure: UCI loop, perft,
    FEN parser (\textsc{p5}).};
  \node[decision, below=of play]  (assess) {Elo measured?};
  \node[box, right=of assess, text width=42mm] (eval) {%
    Ask for a gauntlet vs.\ \stockfish{}/cutechess
    (\textsc{p2}).};
  \node[decision, below=of assess](plat)   {Elo plateau or budget hit?};
  \node[box, right=of plat, text width=42mm]  (steer) {%
    ``Continue / improve the Elo'', with no algorithm
    names (\textsc{p1}).};
  \node[box, below=of plat, text width=72mm] (stop) {%
    Stop session. Record final Elo, transcripts, code.};

  \draw[flow] (start) -- (run);
  \draw[flow] (run)   -- (build);
  \draw[flow] (build) -- node[above]{\footnotesize yes} (bug);
  \draw[flow] (build) -- node[right]{\footnotesize no} (play);
  \draw[flow] (play)  -- node[above]{\footnotesize no} (infra);
  \draw[flow] (play)  -- node[right]{\footnotesize yes} (assess);
  \draw[flow] (assess)-- node[above]{\footnotesize no} (eval);
  \draw[flow] (assess)-- node[right]{\footnotesize yes} (plat);
  \draw[flow] (plat)  -- node[above]{\footnotesize no} (steer);
  \draw[flow] (plat)  -- node[right]{\footnotesize yes} (stop);
  \coordinate (bus) at ([xshift=4mm]steer.east);
  \draw[busline] (bug.east)   -- (bus |- bug.east);
  \draw[busline] (infra.east) -- (bus |- infra.east);
  \draw[busline] (eval.east)  -- (bus |- eval.east);
  \draw[busline] (steer.east) -- (bus |- steer.east);
  \draw[flow] (bus |- steer.east) -- (bus |- run.east) -- (run.east);
\end{tikzpicture}
\caption{Interaction protocol applied to every main-corpus
engine. The human's role is restricted to four follow-up
prompt classes: \textsc{p1}~capability-level steering,
\textsc{p2}~eval invocation, \textsc{p3}~bug/failure
report (verbatim, no fix prescription), and
\textsc{p5}~validation infrastructure. The audit taxonomy
also defines \textsc{p4}, the protocol-deviating class
(named-algorithm or named-feature prompts), which the
protocol forbids and the a posteriori audit found absent.
The human never names
a specific chess or SE algorithm or feature in their prompts.
Those choices are left entirely to the agent. Stop-rule: the session ends when the engine plays
legal games, has a measured Elo, and the agent's improvement
attempts have plateaued under a reasonable budget. We did
\emph{not} push for \elo{2800}-style targets
(\Cref{sec:discussion:ceiling-gap}).}
\label{fig:protocol}
\end{figure}

We followed a single interaction protocol across every engine.
The procedure has three phases
(initial capability prompt $\to$ minimal in-loop steering
$\to$ stop-rule). \Cref{fig:protocol} renders it as a flowchart.
The human role is intentionally restricted to capability-level
steering, error reporting, and evaluation invocation (if needs be).
Named-feature or named-algorithm prompts are excluded.
We verified \emph{a posteriori} that the prompting process of
\Cref{fig:protocol} was followed across the 24 engines with
recoverable prompt streams, over 374 follow-up prompts. The audit script and the per-prompt
classification are released with the replication artefact.
The handful of regex-flagged prompts are all the agent's own
roadmap echoed back by the user, port sessions naming
parent-engine features by design, or regex artefacts.

\paragraph{Three phases, expanded.}
Most sessions opened with one short, language-parametric
template (reproduced in \Cref{fig:protocol}'s top box). It is
deliberately \emph{open-ended on architecture} (no
instruction about search, eval, or representation) and
\emph{names an evaluation goal but not a method}. A small number of sessions used alternative entry points where
the canonical template was a poor fit, most notably the two
Java$\to$language ports, where the agent was given an existing
implementation to translate rather than asked to build from
scratch.

During the session itself, the human installed missing tools,
resolved environment issues, and explained agent errors, but
did \emph{not} propose algorithms, data structures, or
evaluation strategies unless the agent explicitly asked.
Capability-level nudges (``improve the Elo'') are in scope.
Code-level prescriptions (``use bitboards'') are out.

A session ended when Elo stabilised, when improvement stalled,
or when the engine could play end-to-end games against
\stockfish{}. No uniform budget cap per engine was imposed.

\subsection{Evidence sources and analysis}
\label{sec:corpus:evidence}

For each engine we combine four evidence sources: the
\emph{repository contents} (source files, tests, build
artefacts, PGN records, documentation), the \emph{git
history} where available, the \emph{\claudecode{} session
transcripts} (one JSONL per session, plus subagent
transcripts), and the \emph{\codex{} session transcripts}
(per-day JSONL rollouts, filtered to each project by
working-directory match).

These raw inputs are reduced by a Python extraction pipeline
into six per-engine derivatives, each chosen because it
underwrites a specific downstream research question rather
than for its own sake. The first is \emph{size and language
mix}, that is, LOC and file counts by language, so that
engines built in PLs with very different baseline verbosity
remain comparable. This is the precondition for any
cross-language claim in \RQ{1} and \RQ{5}. The second is a
\emph{feature fingerprint}: for each of 38 canonical
chess-engine features (board representation, move generation,
quiescence search, transposition table, evaluation, and so on)
curated from the established \emph{Chess Programming
Wiki} (CPW)~\cite{CPW} taxonomy, the pipeline records a yes/no
indicator of whether the engine contains that feature. This
compact representation is what lets us ask in \RQ{2} which
capabilities each agent reached for and which it skipped,
without privileging any one author's mental model of what a
chess engine should contain. The third is a \emph{session
ledger} of prompts, agent turns, tool uses, token counts, and
a per-engine cost estimate. This is the substrate that makes
\RQ{5}'s effort and cost analysis possible at all, since the
same engine can be cheap or expensive depending on how many
oracle-failure loops the agent had to traverse. The fourth is
a \emph{prompt-derived backlog}, split into a user-driven part
(\blitem{}) and an agent-centric step-wise part (\sbitem{})
extracted from session prompts. This split is what lets
\RQ{3} separate what the user asked for from what the agent
chose to do on its own, the central autonomy claim of the
paper. The fifth is the \emph{gameplay record}, a
\pgn{}/\perft{} index with Elo estimates against calibrated
opponents, which is the only direct evidence we have of how
well the engine actually plays and is used throughout
\RQ{4}. The sixth is a \emph{novelty audit} built from four
independent signals (manifest dependencies, source imports,
canonical-engine fingerprints, transcript authorship claims)
that flag whether an engine was written from scratch or
quietly leans on a chess library. The audit is the central
evasion check for \RQ{2} and \RQ{3}, since a capable agent
can satisfy the user's literal request while side-stepping
its spirit (\Cref{sec:rq3:evasion}).

Alongside the quantitative pipeline we also produced
in-depth qualitative case-studies of selected engines,
written up as standalone narratives that walk through a
session step by step (notably the TeX-based engine and the
Brainfuck engine~\cite{acher2026texccchess,acher2026bfchess}).
These case-studies give a textured, episode-by-episode
picture that complements the aggregate metrics, surface
anomalies the pipeline cannot label on its own, and serve as
a sanity check on the corpus-wide claims.

\paragraph{Cost reporting: canonical vs.\ vendor list-price.}
USD figures invoiced by the two agent vendors are not
directly comparable: Anthropic's \claudecode{} sessions
benefit from a sharp cache-read discount that dominates
long-context-cache-heavy runs, whereas \codex{} sessions are
billed primarily at standard input rates. Reporting either
vendor's invoice across the corpus would silently penalise
or favour one of them. We therefore report a single
\emph{canonical} cost per engine, computed uniformly from
the observed token volumes at a fixed reference rate
($\$1$ per million input tokens, $\$5$ per million output
tokens, ignoring vendor-specific cache discounts). The
column appears as \textbf{Norm.\ USD} in
\Cref{tab:rq1-engines}. The per-vendor list-price USD that
each agent vendor would have invoiced is preserved in the
replication artefact (\texttt{reports/usd-by-vendor.csv})
for transparency. Both forms are list-price (no account
discounts). Cross-vendor effort claims throughout the paper
are anchored on either canonical USD or token volume, never
on list-price USD.

\paragraph{LLM-assisted analysis, code first.}
The meta-analysis itself was conducted with LLM assistance,
mainly \claudecode{}, and we declare that usage in line with
the community guidelines for empirical studies involving
LLMs~\citep{baltes2025llmguidelines}. The division of labour
is deliberately code-first. Every quantitative figure in the
paper is computed by deterministic Python scripts over the
released transcripts, repositories, and game records, never
by an LLM reading those artefacts. LLM assistance went into
writing that code, drafting the per-engine reports and
qualitative narratives under an evidence-citation discipline,
and editing the manuscript, always under author review.
\Cref{app:llm-usage} details the LLM and author roles for
each analysis task, and \Cref{app:guidelines} self-assesses
the paper against the accompanying reporting checklist.

\subsection{Feature-level cross-engine analysis}
\label{sec:corpus:feature-analysis}

The file-level feature scan (\RQ{2}) records, for each of 38
canonical chess-engine features, whether the engine in question
contains that feature. The answer is a yes/no value obtained by
searching the project for the presence of patterns associated
with the feature (e.g.\ \texttt{quiescence}, \texttt{qsearch},
or related identifiers for the quiescence-search feature). The
catalog of 38 features was curated by the first author from the
Chess Programming Wiki (CPW)~\cite{CPW}, the de facto reference
taxonomy for chess-engine internals. The question we ask of
each engine is therefore not ``did the agent invent the right
concepts?'' but ``did it implement the concepts the
chess-engine community already agrees on?''.

A yes/no presence indicator is informative for coverage but
tells us little about \emph{depth}: which sub-options were
taken inside each feature, how the algorithm was adapted to
the target language, and whether two engines that both check
the box for ``transposition table'' are in fact implementing
similar things. To move from presence to depth we built a
second pipeline that, for each of three representative features
chosen to span the major component types of a chess engine
(quiescence search as a search algorithm, transposition table
as a data structure, and evaluation as an evaluation function),
extracts the per-engine \emph{hotspot}, that is, the function
body that implements the feature in each variant. Inside the
hotspot, around a dozen sub-feature checks per feature look for
specific, algorithmically meaningful design choices (delta
pruning inside quiescence, the transposition-table replacement
scheme, mid-/end-game tapering inside evaluation, and so on),
all drawn from the same Chess Programming Wiki taxonomy so the
per-feature analysis stays grounded in established chess-engine
vocabulary rather than in the analyst's choice of keywords.

Because regex-driven extraction can be inaccurate on unusual
languages or hand-rolled idioms, we triangulate the
algorithmic pipeline three ways. First, the catalog itself is
\emph{expert-curated}: the 38-feature list and the sub-feature
checks were authored and revised by the first author against
the chess-programming reference, not auto-generated. Second,
for each of the three case-study features we
\emph{hand-audited every per-variant extraction}: where the
algorithmic pipeline returned a suspicious result, we read the
source ourselves and corrected the verdict, so the final
per-variant verdicts reflect qualitative judgement informed by
the algorithmic baseline rather than the algorithmic baseline
alone. Third, we built a \emph{Chess Engine Feature Explorer},
a small tool that lets a reader browse one feature at a
time across all engines, view the extracted hotspot side by
side, and inspect pairwise distances between variants (a
lexical similarity over normalised tokens and an algorithmic
similarity over the sub-feature presence vector).

\subsection{SE-activity coding frame}
\label{sec:corpus:se-activities}

Beyond ``what did the agent build?'', the session transcripts
let us ask \emph{what did the agent do to build it?}. We
therefore developed a second coding frame, orthogonal to the
feature scan, that tags every step of the session with the
software-engineering activities it exhibits. Throughout the
paper, a \emph{step} is one user prompt together with all the
agent work that follows it, up to the next user prompt.
Step counts are computed over the full recoverable trajectory
record, which is a different accounting from the prompt
counts of \Cref{tab:rq1-engines}: the cost columns exclude
post-mortem analysis sessions, while engines whose main
transcripts were compacted retain only subagent records, so
per-engine step and prompt counts can differ in either
direction. The 10 activity
categories were \emph{induced} bottom-up from the trajectory
data rather than imposed a priori from a generic SE textbook.
The resulting frame is itself one of the paper's reusable
contributions, naming the SE work that is empirically
observable in frontier-agent chess-engine sessions. Each
engine is summarised as a 10-dimensional intensity profile
using a 4-level Likert scale (absent, light, moderate, heavy)
on the share of the session tagged with each capability.
Both the 10-category frame and the per-step tagging rules
that map a session step to a category were developed
iteratively. A first 13-category draft was tested against a
stratified sample of sessions and then pruned: pairs of
categories that proved indistinguishable in practice (for
example refactoring vs.\ implementation, since most agent
``refactors'' were file-edit sequences inside ongoing
implementation) were merged, until the 10 categories the paper
uses remained. The tagging rule for each category is a small
deterministic predicate over the structured per-step record
(filename touched, tool used, bash command issued, vocabulary
of the user's prompt), so a session can be re-coded by anyone
with access to the same trajectory data.

\subsection{Qualitative supervision against agent cheating}
\label{sec:corpus:supervision}

A capable agent can satisfy a user's literal request while
side-stepping its spirit, i.e., it is cheating. The canonical case in the corpus is
\texttt{chess-css-codex}, a CSS-styled chess engine whose
early drafts silently imported \texttt{python-chess} to do
the actual chess work (more details in \Cref{sec:rq3:evasion}). To guard
against this we apply a four-step qualitative supervision
protocol on every engine: read the engine-core diff, grep for
chess-library imports, verify that the oracle the engine is
tested against is independent of the engine under evaluation,
and redirect the session when evasion is detected.

\subsection{Canonical re-evaluation: a cross-PL chess-engine evaluation platform}
\label{sec:corpus:canonical-eval}

A central methodological concern underlies this paper: when
an agent self-reports that ``my engine reaches Elo $X$ against
\stockfish{}'', how should we read that number? Each session
ran its own self-administered evaluation under its own
protocol (different time controls, different opponent ladders,
different game counts). The Elo
numbers are not directly comparable across engines and are not
directly comparable to public engine ratings either. Worse,
those numbers are the same ones the agent itself optimised
against, so they inherit whatever bias the agent's chosen
setup carries. There is also the risk that agents overestimate their own performance
(e.g., on purpose and thus it can be seen as a form of cheating).


We therefore engineered an agent-independent evaluation
platform: a canonical re-evaluation harness in which every
engine in the corpus plays against the same fixed opponents,
at the same time control, with the same opening book and the
same adjudication rules. Concretely, every engine is matched
against \stockfish{} at a calibrated ladder of strengths
(\texttt{UCI\_Elo} from 800 to 2600 in steps of 200) at
TC=120+1, against a fixed 20-position opening book, and is
additionally compared to a five-engine cross-vendor reference
panel (Rustic, Asymptote, and \stockfish{} at three Skill
levels) whose own ratings are independently calibrated.
\Cref{sec:rq4} reports the resulting external Elo assessment
for every engine.

Two engineering points are worth naming. First, this evaluation infrastructure
is the only thing in the paper that makes cross-PL strength
claims defensible: without it, comparing a CSS engine's
self-claimed \elo{1800} against a Rust engine's self-claimed
\elo{1800} would be a comparison of two different rulers, with
no way to tell which one is rescaled, which is biased, or
which is honest. Second, when
the canonical number diverges substantially from an engine's
session-internal self-Elo, the gap is a finding, not noise
(\Cref{sec:rq3,sec:rq4}).
\FloatBarrier
\section{RQ1 (Coverage): every attempted programming language can be mastered}
\label{sec:rq1}

\paragraph{Question.} Across which programming languages and
coding agents can frontier agents produce a working chess
engine? Is there a PL category that resists \agentbuilt{}
construction at this scale?

The Musk-style claim that ``code goes away''
collapses, in software-engineering terms, to a single empirical
test: \emph{is there a programming language a sufficiently
capable agent cannot master?} If any PL still resists, the
agent loop is not a polyglot system-construction tool, only a
polyglot synthesiser of small artefacts~\cite{cassano2023multipl,peng2024humanevalxl,jimenez2024swebench,zan2025multiswebench,gauthier2024aiderpolyglot}.
 The chess corpus does test it, and the answer is consequential: with
frontier coding agents, no language category in our taxonomy
is empty. Several experiments in fact contain the first
open-source chess engine ever reported in that target.
It should be noted that each engine is its own data point: we do not group sessions by
language, because two sessions in the same language can
produce substantively different engines.

\paragraph{Language taxonomy.} We organise the corpus into five
categories:
\textbf{mainstream general-purpose} (Python, Java, C, C\texttt{++},
Rust, Ruby, Mojo), \textbf{specialised / academic} (APL,
Icon, Lean, OCaml, Why3, Rocq: real general-purpose or
formal-methods languages), \textbf{domain-specific / markup} (LaTeX/TeX, CSS/HTML,
SQL), \textbf{legacy} (COBOL, Assembly), and \textbf{esoteric / constrained-execution-model} (Brainfuck).

\begin{table*}[!htbp]\centering\scriptsize
\caption{The 34 \agentbuilt{} chess engines in scope, with
metadata (\RQ{1}: role, language, primary-developer agent,
files, LOC) and effort columns
(\RQ{5}: prompts, input/output tokens, both forms of USD).
\textbf{Role}: \emph{main} = from-scratch construction (29
engines, the population for \RQ{1}--\RQ{5}), \emph{port} =
Codex translation of the \texttt{chess-java-cc} reference
engine into another target language, \emph{DSL} = the agent
designs a new chess-engine DSL and its C\texttt{++}
transpiler, \emph{co-design} = pure-CSS engine built under
sustained expert steering, and \emph{underconverged} = a working
but very weak engine that did not reach competitive playing
strength within the session budget. \textbf{Primary agent} is the agent whose session(s)
authored the engine source. Rows marked $\flat$ also have
secondary-agent sessions in the repository that we classify
as \emph{post-mortem analysis} (no engine-core writes,
analysis vocabulary in the first prompt) and exclude from the
prompt, token, and cost aggregates per
\Cref{sec:corpus:evidence}. The recomputed columns reported
here strip those contributions.
\textbf{In/Out (kTok)} = thousands of input (incl.\ cache-read
and cache-creation) and output tokens.
\textbf{Norm.\ USD} is a canonical reference cost computed
uniformly across the corpus at $\$1$/MTok input $+$ $\$5$/MTok
output, ignoring vendor-specific cache discounts. It is
\emph{intra-corpus} comparable by construction and is the form
the body uses for cross-engine claims.
\textbf{List USD} is the vendor's own list-price estimate
(Anthropic for \claudecode{}, OpenAI for \codex{}). We report
it for situational interpretability (e.g., what a developer
would actually be invoiced) but it is \emph{not} cross-vendor
comparable, since Anthropic cache-read rates and OpenAI
standard-input rates do not align. \textbf{n/a} = main session
transcripts compacted by \claudecode{} before capture. The
source on disk plus subagent sidecars confirm the engine
\emph{was} agent-built. Only the main-loop counts are lost
(\Cref{sec:threats}).}
\label{tab:rq1-engines}
\resizebox{\textwidth}{!}{%
\begin{tabular}{l l l l r r r r r r r}
\toprule
Engine & Role & Lang. & Primary agent & Files & LOC & Prompts &
  In (kTok) & Out (kTok) & Norm.\ USD & List USD \\
\midrule
COBOL-chess$^\flat$ & main & COBOL & Codex & 76 & 8{,}890 & 24 & 463{,}436 & 2{,}013 & \$473.50 & \$332.02 \\
chess-revisit-java-toRust-codex & port & Rust & Codex & 45 & 11{,}141 & 33 & 467{,}345 & 1{,}019 & \$472.44 & \$330.57 \\
chess-revisit-java-toCOBOL-codex & port & COBOL & Codex & 6 & 6{,}770 & 28 & 339{,}290 & 889 & \$343.73 & \$241.16 \\
chess-brainfuck$^\flat$ & main & Brainfuck & Codex & 56 & 6{,}920 & 30 & 231{,}397 & 641 & \$234.60 & \$165.43 \\
chess-cobol-cc & main & COBOL & CC & 4 & 11{,}578 & 46 & 178{,}659 & 592 & \$181.62 & \$518.44 \\
lean-chess & main & Lean 4 & Codex & 15 & 2{,}333 & 21 & 172{,}738 & 487 & \$175.17 & \$123.23 \\
chess-latex-codex-replication & main & LaTeX & Codex & 46 & 5{,}222 & 15 & 167{,}596 & 532 & \$170.26 & \$119.73 \\
chess-newlang-codex & DSL & C++ (host) & Codex & 11 & 4{,}658 & 21 & 159{,}132 & 351 & \$160.89 & \$113.04 \\
chess-apl-codex54$^\flat$ & main & APL & Codex & 10 & 3{,}652 & 26 & 152{,}765 & 550 & \$155.52 & \$110.68 \\
cplusplus-chess & main & C++ & Codex & 34 & 4{,}420 & 16 & 109{,}611 & 800 & \$113.61 & \$81.01 \\
chess-rust-cc-redo & main & Rust & CC & 18 & 3{,}329 & 55 & 95{,}901 & 1{,}458 & \$103.19 & \$63.94 \\
chess-ruby-codex & main & Ruby & Codex & 43 & 124{,}443 & 7 & 85{,}463 & 176 & \$86.34 & \$60.76 \\
chess-assembly-codex & main & Asm & Codex & 9 & 12{,}310 & 12 & 64{,}717 & 269 & \$66.06 & \$47.01 \\
chess-brainfuck-cc$^\flat$ & main & Brainfuck & CC & 18 & 7{,}879 & 51 & 57{,}169 & 86 & \$57.60 & \$154.24 \\
chess-ruby-cc & main & Ruby & CC & 22 & 2{,}964 & 46 & 47{,}320 & 224 & \$48.44 & \$148.17 \\
chess-py & main & Python & Codex & 22 & 3{,}890 & 10 & 38{,}045 & 480 & \$40.45 & \$29.29 \\
chess-css-codex & main & CSS/HTML & Codex & 22 & 38{,}684 & 17 & 35{,}635 & 315 & \$37.21 & \$27.14 \\
chess-mojo$^\ddagger$ & underconv. & Mojo & Codex & 28 & 3{,}472 & 7 & 32{,}720 & 187 & \$33.65 & \$24.06 \\
chess-why3 & main & Why3$\to$OCaml & Codex & 46 & 4{,}530 & 9 & 32{,}832 & 127 & \$33.47 & \$23.92 \\
chess-java & main & Java & Codex & 22 & 2{,}835 & 7 & 29{,}848 & 136 & \$30.53 & \$21.85 \\
chess-icon-codex & main & Icon & Codex & 22 & 3{,}217 & 6 & 26{,}565 & 133 & \$27.23 & \$19.51 \\
chess-purec-codex & main & C & Codex & 16 & 3{,}095 & 4 & 21{,}553 & 140 & \$22.25 & \$15.99 \\
chess-rust-codex & main & Rust & Codex & 8 & 1{,}819 & 3 & 20{,}787 & 114 & \$21.36 & \$15.32 \\
chess-css-codex-guided & main & CSS/HTML & Codex & 4 & 2{,}077 & 9 & 14{,}597 & 130 & \$15.25 & \$11.10 \\
latex-chess-engine & main & TeX & Codex & 23 & 41{,}185 & 14 & 7{,}493 & 50 & \$7.74 & \$5.71 \\
chess-java-cc & main & Java & CC & 35 & 4{,}884 & 3 & 1{,}890 & 14 & \$1.96 & \$7.74 \\
chess-Rocq & main & Rocq$\to$OCaml & CC & 51 & 6{,}335 & 4 & 1{,}410 & 16 & \$1.49 & \$4.75 \\
\midrule
chess-cplusplus-claude & main & C++ & CC & 28 & 4{,}248 & n/a$^\dagger$ & n/a$^\dagger$ & n/a$^\dagger$ & n/a$^\dagger$ & n/a$^\dagger$ \\
chess-purec & main & C & CC & 28 & 3{,}158 & n/a$^\dagger$ & n/a$^\dagger$ & n/a$^\dagger$ & n/a$^\dagger$ & n/a$^\dagger$ \\
chess-py-cc & main & Python & CC & 22 & 2{,}995 & n/a$^\dagger$ & n/a$^\dagger$ & n/a$^\dagger$ & n/a$^\dagger$ & n/a$^\dagger$ \\
chess-rust-cc & main & Rust & CC & 27 & 5{,}872 & n/a$^\dagger$ & n/a$^\dagger$ & n/a$^\dagger$ & n/a$^\dagger$ & n/a$^\dagger$ \\
chess-sql & main & SQL & CC & 15 & 2{,}016 & n/a$^\dagger$ & n/a$^\dagger$ & n/a$^\dagger$ & n/a$^\dagger$ & n/a$^\dagger$ \\
chess-why3-cc & main & Why3$\to$OCaml & CC & 234 & 91{,}558 & n/a$^\dagger$ & n/a$^\dagger$ & n/a$^\dagger$ & n/a$^\dagger$ & n/a$^\dagger$ \\
chess-css-cc (\textit{a.k.a.} test-superset) & co-design & CSS+JS & CC & 46 & 224{,}772 & n/a$^\dagger$ & n/a$^\dagger$ & n/a$^\dagger$ & n/a$^\dagger$ & n/a$^\dagger$ \\
\bottomrule
\end{tabular}
}
\end{table*}

\Cref{tab:rq1-engines} lists 34 agent-built engines spanning
five language categories, with both frontier agents
(\claudecode{} and \codex{}) present across all five (multiple
engines in mainstream, specialised, domain-specific, and
legacy, and one shared language, Brainfuck, in the esoteric
category, where the \codex{} engine only partially meets the
pure-Brainfuck constraint, see below).

\paragraph{Special-role experiments.} Beyond the main corpus,
five engines follow protocols different enough to warrant
separate treatment. In two Java$\to$language ports
(\texttt{chess-revisit-java-toRust-codex},
\texttt{chess-revisit-java-toCOBOL-codex}), the agent
was given an existing Java reference engine to translate
rather than asked to build from scratch. In a DSL-design
experiment (\texttt{chess-newlang-codex}), the agent
designs the GAMBIT DSL and its C\texttt{++} transpiler before
writing the engine. An underconverged case
(\texttt{chess-mojo}) produced a working but very weak engine that we
keep separate from main-corpus aggregates. Finally, an agent-heavy
co-design case (\texttt{test-superset}, ChessCSS), built
under sustained expert steering, is discussed as the
positive counterexample to the CSS evasion in
\Cref{sec:rq3:evasion}. The bottom block of
\Cref{tab:rq1-engines} gives their metadata. We cite each
where it bears on a specific RQ rather than dedicate a
standalone section to them.

\paragraph{Size patterns by language category.}
Repository size (LOC) is informative because it tracks how
much code a category needs to express the same chess-engine
capabilities. Mainstream general-purpose engines cluster
tightly in the 2--12\,kLOC band. The legacy row (COBOL,
Assembly) reaches 10--120\,kLOC because each primitive does
less work per line and because derived C and build artefacts
inflate the count. Brainfuck takes a different shape again:
roughly 5--10\,kLOC of Python code-generator that compiles to
$\approx$5.6\,MB of BF. Four repositories exceed 40\,kLOC.
Two of them are not substantively larger engine \emph{cores} but
artefact-inflated repositories: \texttt{chess-ruby-codex}
ships a large PGN game database alongside the engine, and
\texttt{chess-why3-cc} vendors the C source trees of several
third-party opponent engines for in-repo Elo benchmarking
(while its own engine extracts from Why3 to OCaml). The
other two are large by construction: the co-design ChessCSS
engine compiles its move logic into $\approx$55k generated
CSS rules, and TeXCCChess spreads engine code and tables
across $\approx$40\,kLOC of TeX. The point
to take away is not the absolute LOC numbers but how LOC
tracks the category: legacy and esoteric targets need more
code per equivalent capability, and very large totals are
usually data or vendored infrastructure, not algorithm.

\paragraph{Claim supported by the table.} Agents produce a
working chess engine in at least two variants of most
mainstream languages and in at least one variant of every
category in our taxonomy. Coverage is \emph{per session}, not
\emph{per language} since two engines in the same language can lead to very different outcomes
and are thus two data points, not one.

\paragraph{Languages where the agents pushed past prior art.}

The first-of-kind and no-comparable-reference claims below
rest on a four-step search: (1) general-purpose web search
(Google), (2) GitHub code- and repository-search restricted to
the relevant language, (3) reading of relevant community
discussions (chess-engine forums, language-community fora,
public Reddit threads), and (4) cross-checking with an
LLM-powered deep-research tool (OpenAI Deep Research) to
surface references the first three steps may have missed.
Throughout, ``first-of-kind'' and ``no comparable public
reference'' denote the absence of a prior open-source
artefact \emph{of comparable scope} (a playable engine with
at least move generation, legal-move checking, and a search
routine), not the absence of any mention.

A subset of the corpus is consequential beyond a coverage
checkmark. In \texttt{latex-chess-engine} (TeXCCChess) the
agent produced what is, to our knowledge, the first chess
engine of comparable scope written in \texttt{expl3}/TeX
macros, with a self-reported $\approx$\elo{1300}.
\texttt{chess-css-cc} and the strict-CSS companion of
\texttt{chess-css-codex} are the first chess engines we are
aware of where rules and move generation are expressed in
pure CSS (with a minimal JavaScript harness for I/O and the
game loop). The APL, Icon, Lean~4, Why3, and Rocq
engines are, to our knowledge, likewise first-of-kind: in
none of these languages does an open-source chess engine of
comparable scope appear to have been previously published.
For APL, the closest prior art our search surfaced is a 2008
Dyalog chess GUI demonstration without an engine, board-puzzle
solvers (knight's tour, $n$-queens), and a three-player
variant ruleset, none of comparable scope.
Brainfuck is a slightly different case: discussions and
prototype gestures toward a Brainfuck chess engine exist in
the public record, and private exchanges with experts
confirm that the topic has been raised more than once, but
no surviving open-source artefact of comparable scope has
come to light through our four-step search. Of the two corpus
engines, the \claudecode{} one is the first we are aware of
that ships a complete \uci{}-compliant pure-Brainfuck engine
playing \stockfish{} gauntlets. The \codex{} engine plays as
well but kept part of its engine logic in the Python
toolchain, a shortfall of the pure-language constraint that
its repository documents (\Cref{app:github-repos}). The COBOL novelty is similar: small hobby
experiments exist, but each of our three COBOL
engines ships its own move generator, search, \uci{} loop,
and Elo harness. The Mojo case is the weakest in the
corpus: the special-role \texttt{chess-mojo} session
produced a working but weak engine, near \elo{900} on its
in-session gauntlets (the \elo{1100} of \Cref{tab:rq4-elo} is
the agent's pre-measurement planning estimate), and
stopped under the user's budget. It plays legal chess and
satisfies the ``working engine'' criterion, but it sits
below every main-corpus engine in playing strength, so we
classify it as an \emph{underconverged} case
(\Cref{app:special-role}).
\emph{Beyond demonstrating polyglot coverage, the agents
produced engines in programming languages where, to the best of our
knowledge, no chess engine had previously been developed.}

\paragraph{Boundary cases: where the language is the ceiling.}
Four engines in the main corpus illustrate the engineering
challenges of building under restrictive PL characteristics and constraints.

\textbf{Brainfuck} has eight instructions over a flat memory
tape with a single data pointer. There are no variables, no
functions, and no random memory access (no \texttt{board[i]}
with runtime \texttt{i})~\cite{acher2026bfchess}. The agent's
strategy was correspondingly indirect: a Python code generator
($\approx$7{,}400 LOC) tracks a virtual data pointer at
generation time and emits the corresponding \texttt{>} and
\texttt{<} runs plus pattern-encoded higher-level operations.
Each board read or write expands to $\approx$20\,kB of BF
because indexed access has to be reconstructed as a 64-way
switch over fixed-cell positions. The resulting
\texttt{chess.bf} weighs 5.6\,MB and reads as long sequences
of pointer movement and cell zeroing:
\begin{small}
\begin{verbatim}
>>>>>>>>>>>>>>[-][-]+[>...]>[-]>[-]>[-]>[-]> ...
\end{verbatim}
\end{small}
\noindent Search looks only three half-moves ahead, where
competitive engines routinely look 15 or more, and the
session's own gauntlet placed the engine at \elo{578}
(\Cref{tab:rq4-elo}).

\textbf{LaTeX} engines (an \texttt{expl3} replication at
$\approx$\elo{550} and TeXCCChess at $\approx$\elo{1300}) look
only two half-moves ahead, the shallowest depth in the corpus.
\TeX{} is a macro-expansion language designed for typesetting:
no native arrays, no recursion in the conventional sense, no
boolean expressions, and no return statement. A procedure
``returns'' by mutating a globally-scoped register. The
quiescence-search routine in
TeXCCChess~\cite{acher2026texccchess} illustrates the cost:
\begin{small}
\begin{verbatim}
\def\mm@quiesce{%
  \evaluate
  \mm@standpat=\evalscore\relax
  \ifnum\mm@standpat<\mm@qbeta\relax
    \ifnum\mm@standpat>\mm@qalpha\relax
      \mm@qalpha=\mm@standpat\relax
    \fi
    \generatelegal \mm@copycapturesonly  ...
  \else
    \global\mm@retval=\mm@qbeta\relax  % stand-pat cutoff
  \fi
}
\end{verbatim}
\end{small}
\noindent What in a mainstream language is a one-line
\texttt{if standpat >= qbeta: return qbeta} unfolds into
nested \verb|\ifnum| groups around scarce \verb|\count|
registers, with the return value plumbed through a
\verb|\global| mutation of \verb|\mm@retval| (the only way
to pass a value out of a recursive expansion).

\textbf{COBOL} (more specifically GnuCOBOL) imposes constraints the agent must handle: tables require static upper bounds via \texttt{OCCURS}, even with \texttt{OCCURS DEPENDING ON}, and recursive subprograms must be explicitly marked \texttt{RECURSIVE} at the \texttt{PROGRAM-ID} level. The three corpus engines therefore use conventional COBOL layouts: nested character tables for the $8{\times}8$ board, fixed-size move lists, and \texttt{REDEFINES} for alternate storage views. This contrasts with mainstream chess engines, where bitboards and native bitwise operators express many board operations as word-level computations.
\textbf{Assembly}
(\texttt{chess-assembly-codex}) ships a full \uci{} engine
that is predominantly x86-64 assembly ($\approx$9.5\,kLOC,
including the performance-critical move-generation and search
core), with a smaller $\approx$1.9\,kLOC C support layer
hosting the search and evaluation data structures. Both
layers were produced by the agent. Prior art in assembly
exists: Sargon (1978) was published as Z80 assembly source in
book form, and boot-sector engines such as BootChess (2015)
fit in 512 bytes. We therefore scope the claim: we could not
identify a prior open-source x86-64 assembly engine of
comparable scope with full \uci{} support.

What unifies the four cases is that the agent produces
working move generation, legality filtering, and a shallow
search in each. What varies is how deep the search can
reasonably go given the language's execution model. These are
the cases where the \emph{programming language, not the
agent}, is the binding constraint (\Cref{sec:rq4}).

Besides, APL, Icon, Lean, and
OCaml (via Rocq/Why3 extraction) are \emph{specialised}
languages with established communities and production use in
other domains. They do not target branch-heavy numerical
workloads well, but they are real general-purpose or
formal-methods languages. Our corpus shows them producing
working engines that sit in the middle of \Cref{tab:rq4-elo},
not at the bottom.

\begin{rqanswer}{1}
Frontier coding agents produce working chess engines
across all five language categories, with multiple independent
sessions per mainstream language and at least one per
specialised, domain-specific, legacy, and esoteric
category. Neither \claudecode{} nor \codex{} monopolises the
space, and several engines
are first-of-kind (pure TeX, pure CSS, APL,
Icon, Lean~4, Why3/Rocq, and, scoped to open-source x86-64
\uci{} engines, Assembly). On
the title question, this RQ delivers the existence half:
programming language no longer acts as a hard
\emph{coverage} barrier, and
even unsuited or under-resourced PLs are reachable at modest
spend. How PL still matters (ceiling, cost,
SE-workload shape, verification affordance) is what
\RQ{2}--\RQ{5} unpack.
\end{rqanswer}
\FloatBarrier
\section{RQ2 (Synthesis): code synthesis and feature adaptation}
\label{sec:rq2}

\paragraph{Question.} Which chess-engine features does each
\agentbuilt{} engine implement, how do those choices vary across
languages and agents, and is the resulting code a recognisable
copy of an existing engine or a genuine PL-adapted synthesis?

\paragraph{One question, three views.}
\Cref{sec:rq2:features} reports the per-engine feature
footprint over a 38-pattern catalog, that is, what each agent
built. \Cref{sec:rq2:novelty} runs a four-signal novelty
audit at the \emph{file} level: is the engine a copy?
\Cref{sec:rq2:qsearch}~zooms in to the \emph{algorithm} level
through the quiescence/TT/eval triad: when the agents
re-implement the same algorithm in 11 different PLs, do they
copy a canonical source, derive from a shared specification, or
adapt to the language's idiom?

\subsection{What was built: the feature footprint}
\label{sec:rq2:features}

We expand the feature vocabulary to 38 regex patterns across
six groups (rules \& protocol, board representation, search
core, search extensions, evaluation, strong-engine features),
detecting each per engine on source and documentation.
Unusual syntax (Brainfuck, compiled-out macros, binary
artefacts) gets a manual audit pass.
\Cref{tab:rq2-core} reports the rules, protocol, and
board-representation matrix, the layer where every engine is
dense. \Cref{tab:rq2-searcheval} reports the search-extension
and evaluation/strong-engine matrix, the parts where engines
diverge and where the \elo{1500}-vs-\elo{2000} variance lives.

\begin{table}[!htbp]\centering\scriptsize
\caption{\RQ{2} Core features: rules, protocol, and board representation. A \checkmark indicates the feature was detected by regex in the repository's source or documentation. For constrained-execution-model engines (Brainfuck, LaTeX), missing \checkmark may reflect detection limits rather than absence. We cross-reference the per-engine report before drawing claims. Notice the diversity of board representations (0$\times$88 mailbox, bitboards, 8$\times$8 arrays): agents pick per engine, not per language.}
\label{tab:rq2-core}
\resizebox{\textwidth}{!}{%
\begin{tabular}{l l cccccccccccc}
\toprule
Engine & Language & FEN & UCI & PGN & Cas & EP & Pro & Chk & 0x88 & BB & MBB & MB8 & MB10 \\
\midrule
chess-apl-codex54 & APL & \checkmark & \checkmark & \checkmark & \checkmark & \checkmark & \checkmark & \checkmark &  &  &  &  &  \\
chess-assembly-codex & Assembly & \checkmark & \checkmark &  & \checkmark &  & \checkmark & \checkmark &  &  &  & \checkmark &  \\
chess-brainfuck & Brainfuck & \checkmark & \checkmark & \checkmark & \checkmark & \checkmark & \checkmark & \checkmark &  &  &  &  &  \\
chess-brainfuck-cc & Brainfuck & \checkmark & \checkmark & \checkmark & \checkmark & \checkmark & \checkmark & \checkmark &  &  &  & \checkmark &  \\
chess-purec & C & \checkmark & \checkmark & \checkmark & \checkmark & \checkmark & \checkmark & \checkmark &  & \checkmark & \checkmark & \checkmark &  \\
chess-purec-codex & C & \checkmark & \checkmark & \checkmark & \checkmark &  & \checkmark & \checkmark &  &  &  &  &  \\
chess-cplusplus-claude & C++ & \checkmark & \checkmark & \checkmark & \checkmark & \checkmark & \checkmark & \checkmark &  & \checkmark & \checkmark &  &  \\
cplusplus-chess & C++ & \checkmark & \checkmark & \checkmark & \checkmark & \checkmark & \checkmark & \checkmark &  & \checkmark &  &  &  \\
COBOL-chess & COBOL & \checkmark & \checkmark & \checkmark & \checkmark & \checkmark & \checkmark & \checkmark & \checkmark & \checkmark &  & \checkmark &  \\
chess-cobol-cc & COBOL & \checkmark & \checkmark &  & \checkmark & \checkmark & \checkmark & \checkmark &  &  &  &  &  \\
chess-css-codex & CSS/HTML & \checkmark & \checkmark &  &  & \checkmark & \checkmark & \checkmark &  &  &  &  &  \\
chess-css-codex-guided & CSS/HTML & \checkmark &  &  & \checkmark & \checkmark & \checkmark & \checkmark &  &  &  &  &  \\
chess-icon-codex & Icon & \checkmark & \checkmark & \checkmark & \checkmark & \checkmark & \checkmark & \checkmark & \checkmark &  &  &  &  \\
chess-java & Java & \checkmark & \checkmark & \checkmark & \checkmark & \checkmark & \checkmark & \checkmark &  &  &  &  &  \\
chess-java-cc & Java & \checkmark & \checkmark & \checkmark & \checkmark & \checkmark & \checkmark & \checkmark &  & \checkmark & \checkmark & \checkmark &  \\
chess-latex-codex-replication & LaTeX & \checkmark & \checkmark & \checkmark & \checkmark & \checkmark & \checkmark & \checkmark &  &  &  &  &  \\
latex-chess-engine & LaTeX (TeX) & \checkmark & \checkmark & \checkmark & \checkmark & \checkmark & \checkmark & \checkmark &  &  &  & \checkmark &  \\
lean-chess & Lean 4 & \checkmark & \checkmark & \checkmark & \checkmark & \checkmark & \checkmark & \checkmark &  &  &  &  &  \\
chess-py & Python & \checkmark & \checkmark & \checkmark & \checkmark & \checkmark & \checkmark & \checkmark & \checkmark &  &  &  &  \\
chess-py-cc & Python & \checkmark & \checkmark & \checkmark & \checkmark & \checkmark & \checkmark & \checkmark &  & \checkmark &  & \checkmark &  \\
chess-Rocq & Rocq$\to$OCaml & \checkmark & \checkmark & \checkmark & \checkmark & \checkmark & \checkmark & \checkmark &  & \checkmark & \checkmark &  &  \\
chess-ruby-cc & Ruby & \checkmark & \checkmark & \checkmark & \checkmark & \checkmark & \checkmark & \checkmark & \checkmark &  &  &  &  \\
chess-ruby-codex & Ruby & \checkmark & \checkmark &  & \checkmark & \checkmark & \checkmark & \checkmark &  &  &  &  &  \\
chess-rust-cc & Rust & \checkmark & \checkmark & \checkmark & \checkmark & \checkmark & \checkmark & \checkmark &  & \checkmark & \checkmark & \checkmark &  \\
chess-rust-cc-redo & Rust & \checkmark & \checkmark & \checkmark & \checkmark & \checkmark & \checkmark & \checkmark &  & \checkmark &  & \checkmark &  \\
chess-rust-codex & Rust & \checkmark & \checkmark & \checkmark & \checkmark & \checkmark & \checkmark & \checkmark &  & \checkmark &  &  &  \\
chess-sql & SQL & \checkmark & \checkmark & \checkmark & \checkmark & \checkmark & \checkmark & \checkmark &  &  &  &  &  \\
chess-why3-cc & Why3$\to$OCaml & \checkmark & \checkmark & \checkmark & \checkmark & \checkmark & \checkmark & \checkmark & \checkmark & \checkmark & \checkmark & \checkmark &  \\
chess-why3 & Why3$\to$OCaml & \checkmark & \checkmark & \checkmark & \checkmark &  & \checkmark & \checkmark &  &  &  &  &  \\
\bottomrule
\end{tabular}
}
\end{table}

\begin{table}[!htbp]\centering\tiny
\caption{\RQ{2} Per-engine search and evaluation feature matrix. \textbf{Search core}: Negamax,
$\alpha\beta$, iterative deepening, quiescence, transposition
table, Zobrist hashing, perft. \textbf{Search extensions}:
MVV-LVA ordering, killer moves, history heuristic, PV move,
null-move pruning, LMR, LMP, aspiration windows, futility,
razoring. \textbf{Evaluation \& strong-engine features}:
material, PST, tapered mid/endgame scoring, king safety,
pawn-structure terms, mobility, opening book, endgame tables,
time management, NNUE. The right-block extensions are what
typically separate an~\elo{1500} from an~\elo{2000}
\agentbuilt{} engine in the same language. Tapered evaluation,
king safety, and pawn structure appear in roughly half the
mainstream cells, NNUE in one engine, and endgame tables only where
C/C\texttt{++} tooling makes them practical.}
\label{tab:rq2-searcheval}
\resizebox{\textwidth}{!}{%
\begin{tabular}{l l ccccccc ccccccccccc cccccccccc}
\toprule
& & \multicolumn{7}{c}{Search core} &
\multicolumn{10}{c}{Search extensions} &
\multicolumn{10}{c}{Evaluation \& strong-engine}\\
\cmidrule(lr){3-9}\cmidrule(lr){10-19}\cmidrule(lr){20-29}
Engine & Lang & Neg & $\alpha\beta$ & ID & Qs & TT & Zob & Pft &
MVV & Kil & His & PV & NMP & LMR & LMP & Asp & Fut & Raz &
Mat & PST & Tap & KS & PS & Mob & Bk & EGT & TM & NNUE\\
\midrule
chess-apl-codex54 & APL & \checkmark & \checkmark & \checkmark & \checkmark & \checkmark & \checkmark & \checkmark & \checkmark &  & \checkmark &  & \checkmark &  &  &  &  &  & \checkmark & \checkmark & \checkmark & \checkmark & \checkmark & \checkmark & \checkmark &  & \checkmark &  \\
chess-assembly-codex & Asm &  & \checkmark & \checkmark & \checkmark & \checkmark & \checkmark &  &  &  &  & \checkmark & \checkmark &  &  & \checkmark & \checkmark &  & \checkmark &  & \checkmark &  & \checkmark &  & \checkmark &  & \checkmark &  \\
chess-brainfuck & BF & \checkmark & \checkmark & \checkmark & \checkmark & \checkmark &  & \checkmark &  &  &  & \checkmark &  & \checkmark &  &  &  &  & \checkmark & \checkmark &  & \checkmark &  &  & \checkmark &  & \checkmark &  \\
chess-brainfuck-cc & BF & \checkmark & \checkmark & \checkmark &  &  &  & \checkmark & \checkmark &  &  &  &  &  &  &  &  &  & \checkmark &  &  & \checkmark & \checkmark &  & \checkmark & \checkmark & \checkmark &  \\
chess-purec & C & \checkmark &  & \checkmark & \checkmark & \checkmark & \checkmark & \checkmark & \checkmark & \checkmark & \checkmark & \checkmark & \checkmark & \checkmark & \checkmark & \checkmark & \checkmark & \checkmark & \checkmark &  & \checkmark & \checkmark & \checkmark & \checkmark & \checkmark &  & \checkmark &  \\
chess-purec-codex & C & \checkmark & \checkmark & \checkmark & \checkmark & \checkmark & \checkmark & \checkmark &  & \checkmark & \checkmark & \checkmark & \checkmark & \checkmark &  & \checkmark &  &  & \checkmark &  & \checkmark &  & \checkmark & \checkmark & \checkmark & \checkmark & \checkmark &  \\
chess-cplusplus-claude & C++ &  & \checkmark & \checkmark & \checkmark & \checkmark & \checkmark & \checkmark & \checkmark & \checkmark & \checkmark & \checkmark & \checkmark & \checkmark & \checkmark & \checkmark & \checkmark & \checkmark & \checkmark & \checkmark & \checkmark & \checkmark & \checkmark & \checkmark &  &  & \checkmark &  \\
cplusplus-chess & C++ & \checkmark &  &  & \checkmark & \checkmark & \checkmark & \checkmark & \checkmark &  &  & \checkmark & \checkmark &  &  & \checkmark & \checkmark &  & \checkmark & \checkmark &  & \checkmark & \checkmark & \checkmark & \checkmark &  & \checkmark &  \\
COBOL-chess & COBOL & \checkmark & \checkmark & \checkmark & \checkmark & \checkmark & \checkmark & \checkmark & \checkmark & \checkmark & \checkmark & \checkmark & \checkmark & \checkmark & \checkmark & \checkmark &  &  & \checkmark & \checkmark & \checkmark & \checkmark & \checkmark & \checkmark & \checkmark & \checkmark & \checkmark &  \\
chess-cobol-cc & COBOL & \checkmark & \checkmark & \checkmark & \checkmark & \checkmark &  &  & \checkmark & \checkmark & \checkmark & \checkmark &  & \checkmark &  & \checkmark & \checkmark &  & \checkmark & \checkmark & \checkmark &  & \checkmark &  &  &  & \checkmark &  \\
chess-css-codex & CSS/HTML & \checkmark & \checkmark & \checkmark & \checkmark & \checkmark &  &  &  &  & \checkmark & \checkmark &  &  &  &  &  &  & \checkmark & \checkmark &  &  & \checkmark & \checkmark &  &  & \checkmark &  \\
chess-css-codex-guided & CSS/HTML &  &  &  &  &  &  & \checkmark &  &  &  &  &  &  &  &  &  &  & \checkmark &  &  &  &  & \checkmark & \checkmark &  &  &  \\
chess-icon-codex & Icon &  & \checkmark & \checkmark & \checkmark & \checkmark &  & \checkmark & \checkmark & \checkmark &  &  & \checkmark & \checkmark &  & \checkmark &  &  & \checkmark &  &  &  & \checkmark &  &  &  & \checkmark &  \\
chess-java & Java & \checkmark & \checkmark & \checkmark & \checkmark & \checkmark & \checkmark & \checkmark &  &  &  &  &  &  &  &  &  &  & \checkmark & \checkmark & \checkmark & \checkmark &  & \checkmark & \checkmark & \checkmark & \checkmark &  \\
chess-java-cc & Java & \checkmark & \checkmark & \checkmark & \checkmark & \checkmark & \checkmark & \checkmark & \checkmark & \checkmark & \checkmark & \checkmark & \checkmark & \checkmark &  & \checkmark & \checkmark &  & \checkmark & \checkmark & \checkmark & \checkmark & \checkmark & \checkmark & \checkmark &  & \checkmark &  \\
chess-latex-codex-replication & LaTeX & \checkmark &  &  &  &  &  & \checkmark &  &  &  &  &  &  &  &  &  &  & \checkmark &  &  & \checkmark &  &  & \checkmark &  & \checkmark &  \\
latex-chess-engine & TeX & \checkmark & \checkmark & \checkmark & \checkmark &  &  &  & \checkmark &  &  &  &  &  &  &  &  &  & \checkmark & \checkmark &  &  &  &  &  &  & \checkmark &  \\
lean-chess & Lean 4 & \checkmark & \checkmark & \checkmark & \checkmark & \checkmark &  & \checkmark &  &  &  &  & \checkmark &  &  &  &  &  & \checkmark & \checkmark &  &  & \checkmark &  &  &  & \checkmark &  \\
chess-py & Python & \checkmark & \checkmark & \checkmark & \checkmark & \checkmark & \checkmark & \checkmark &  &  &  &  & \checkmark &  &  &  & \checkmark &  & \checkmark & \checkmark & \checkmark &  & \checkmark & \checkmark &  &  & \checkmark &  \\
chess-py-cc & Python &  & \checkmark & \checkmark & \checkmark & \checkmark & \checkmark & \checkmark & \checkmark & \checkmark & \checkmark & \checkmark & \checkmark & \checkmark &  & \checkmark & \checkmark & \checkmark & \checkmark & \checkmark & \checkmark & \checkmark & \checkmark & \checkmark &  &  & \checkmark &  \\
chess-Rocq & Rocq$\to$OCaml & \checkmark & \checkmark & \checkmark & \checkmark & \checkmark &  & \checkmark & \checkmark & \checkmark & \checkmark &  &  &  &  &  &  &  & \checkmark & \checkmark &  & \checkmark & \checkmark & \checkmark & \checkmark &  & \checkmark &  \\
chess-ruby-cc & Ruby & \checkmark &  & \checkmark & \checkmark & \checkmark & \checkmark & \checkmark & \checkmark & \checkmark & \checkmark & \checkmark & \checkmark & \checkmark & \checkmark & \checkmark & \checkmark & \checkmark & \checkmark & \checkmark & \checkmark & \checkmark & \checkmark & \checkmark &  &  & \checkmark &  \\
chess-ruby-codex & Ruby & \checkmark & \checkmark &  & \checkmark & \checkmark & \checkmark & \checkmark &  & \checkmark &  &  & \checkmark & \checkmark &  &  &  &  & \checkmark & \checkmark &  & \checkmark & \checkmark &  & \checkmark &  & \checkmark &  \\
chess-rust-cc & Rust &  & \checkmark & \checkmark & \checkmark & \checkmark & \checkmark & \checkmark & \checkmark & \checkmark & \checkmark & \checkmark & \checkmark & \checkmark & \checkmark & \checkmark & \checkmark & \checkmark & \checkmark & \checkmark & \checkmark & \checkmark & \checkmark & \checkmark & \checkmark &  & \checkmark &  \\
chess-rust-cc-redo & Rust & \checkmark &  & \checkmark & \checkmark & \checkmark & \checkmark & \checkmark &  & \checkmark & \checkmark & \checkmark & \checkmark & \checkmark & \checkmark & \checkmark & \checkmark &  & \checkmark &  & \checkmark & \checkmark & \checkmark & \checkmark & \checkmark &  & \checkmark &  \\
chess-rust-codex & Rust & \checkmark & \checkmark & \checkmark & \checkmark & \checkmark &  & \checkmark &  & \checkmark & \checkmark & \checkmark & \checkmark & \checkmark &  &  &  &  & \checkmark &  &  & \checkmark & \checkmark & \checkmark & \checkmark &  & \checkmark &  \\
chess-sql & SQL & \checkmark &  &  & \checkmark &  &  & \checkmark & \checkmark &  &  &  &  &  &  &  &  &  & \checkmark & \checkmark &  & \checkmark & \checkmark &  &  &  & \checkmark &  \\
chess-why3-cc & Why3$\to$OCaml & \checkmark & \checkmark & \checkmark & \checkmark & \checkmark & \checkmark & \checkmark & \checkmark & \checkmark & \checkmark & \checkmark & \checkmark & \checkmark & \checkmark & \checkmark & \checkmark & \checkmark & \checkmark & \checkmark & \checkmark & \checkmark & \checkmark & \checkmark & \checkmark & \checkmark & \checkmark & \checkmark \\
chess-why3 & Why3$\to$OCaml & \checkmark & \checkmark & \checkmark & \checkmark & \checkmark &  & \checkmark &  &  &  &  & \checkmark &  &  & \checkmark &  &  & \checkmark & \checkmark & \checkmark &  &  & \checkmark &  &  & \checkmark &  \\
\bottomrule
\end{tabular}
}
\end{table}

\paragraph{Board representation varies \emph{within} a language.}
This was invisible in the per-language aggregate but clear per
engine:
\begin{itemize}
  \item In C, \texttt{chess-purec} uses bitboards while
    \texttt{chess-purec-codex} uses an 8$\times$8 mailbox. Same
    language, different agent, different board representation.
  \item In Rust, \texttt{chess-rust-cc} mixes mailbox with
    bitboard helpers and \texttt{chess-rust-codex} is
    bitboard-first.
  \item Ruby engines converge on 0$\times$88 mailbox across both
    agent sessions, and Java engines do likewise.
  \item The COBOL engines split: \texttt{chess-cobol-cc} uses an
    8$\times$8 \texttt{OCCURS}-indexed array,
    \texttt{COBOL-chess} uses a flat mailbox,
    \texttt{chess-revisit-java-toCOBOL-codex} mirrors the Java
    reference.
\end{itemize}
The paper's language-is-a-ceiling-not-a-point claim
(\Cref{sec:discussion}) sits here: the agent picks a board
representation per session, not per language.

\paragraph{Search is where the strength variance lives.}
Every engine in the mainstream-GP rows implements alpha--beta
$+$ iterative deepening $+$ at least one of \{quiescence,
transposition table\}. The interesting variance is in the
\emph{extension} layer: aspiration windows, LMR, LMP, history
heuristic, null-move. These are agent-initiated improvements
that surface after the user says ``improve the Elo'' without
specifying what to improve. The correlation with the external
Elo assessment (\Cref{tab:rq4-elo}) is positive, and we discuss
the confound in \Cref{sec:rq4}. The right block of \Cref{tab:rq2-searcheval}
shows this directly: mainstream engines where every extension
fires (the \texttt{chess-rust-cc}, \texttt{chess-purec},
\texttt{chess-py-cc} rows are 9--10 of 10 extensions present)
sit at the top of \Cref{tab:rq4-elo}, whereas engines where
extensions are sparse plateau lower.

\paragraph{Evaluation and advanced features.} Material and PST
are near-universal. Tapered evaluation, king safety, and pawn
structure appear in about half the mainstream engines and are
absent from the constrained-execution-model rows. NNUE and
endgame tables appear in at most one or two engines each: the
effort budget of a typical frontier-agent session does not
reach these. The right-most block of
\Cref{tab:rq2-searcheval} confirms: NNUE fires once
(\texttt{chess-why3-cc}), and endgame tables appear only in
C/C\texttt{++}-tooled engines.

\subsubsection{Feature-introduction trajectory}
\label{sec:rq2:trajectory}

\begin{figure}[!htbp]
\centering
\begin{tikzpicture}
\begin{axis}[
  width=0.95\linewidth, height=0.5\linewidth,
  xlabel={Session progress (step / total steps)},
  ylabel={Cumulative distinct features introduced},
  xmin=0, xmax=1.02,
  ymin=0, ymax=36,
  grid=both, grid style={gray!20},
  legend style={font=\footnotesize, at={(0.98,0.02)}, anchor=south east, draw=none, fill=white, fill opacity=0.8, text opacity=1},
  tick label style={font=\footnotesize},
  label style={font=\footnotesize},
]
  \addplot[color=blue!70!black, mark=*, mark options={fill=blue!70!black}, thick] coordinates { (0.062,0) (0.125,0) (0.188,23) (0.250,23) (0.312,23) (0.375,23) (0.438,29) (0.500,29) (0.562,32) (0.625,32) (0.688,32) (0.750,32) (0.812,32) (0.875,32) (0.938,32) (1.000,32) };
  \addlegendentry{Ruby (mainstream)}
  \addplot[color=red!70!black, mark=square*, mark options={fill=red!70!black}, thick] coordinates { (0.027,0) (0.054,0) (0.081,0) (0.108,6) (0.135,6) (0.162,8) (0.189,8) (0.216,10) (0.243,10) (0.270,10) (0.297,10) (0.324,10) (0.351,10) (0.378,10) (0.405,10) (0.432,10) (0.459,13) (0.486,13) (0.514,13) (0.541,13) (0.568,15) (0.595,15) (0.622,15) (0.649,15) (0.676,15) (0.703,19) (0.730,19) (0.757,19) (0.784,19) (0.811,19) (0.838,19) (0.865,21) (0.892,22) (0.919,22) (0.946,22) (0.973,24) (1.000,24) };
  \addlegendentry{COBOL (legacy)}
  \addplot[color=purple!70!black, mark=triangle*, mark options={fill=purple!70!black}, thick] coordinates { (0.067,0) (0.133,0) (0.200,6) (0.267,6) (0.333,6) (0.400,6) (0.467,6) (0.533,6) (0.600,6) (0.667,6) (0.733,6) (0.800,6) (0.867,6) (0.933,6) (1.000,6) };
  \addlegendentry{LaTeX (Codex, scratch)}
  \addplot[color=green!45!black, mark=diamond*, mark options={fill=green!45!black}, thick] coordinates { (0.005,0) (0.010,0) (0.015,0) (0.020,0) (0.025,0) (0.030,0) (0.035,0) (0.040,0) (0.045,0) (0.051,0) (0.056,0) (0.061,0) (0.066,0) (0.071,0) (0.076,0) (0.081,0) (0.086,0) (0.091,0) (0.096,0) (0.101,0) (0.106,0) (0.111,0) (0.116,0) (0.121,0) (0.126,0) (0.131,0) (0.136,0) (0.141,0) (0.146,0) (0.152,0) (0.157,0) (0.162,0) (0.167,0) (0.172,0) (0.177,0) (0.182,0) (0.187,0) (0.192,0) (0.197,0) (0.202,0) (0.207,0) (0.212,0) (0.217,0) (0.222,0) (0.227,0) (0.232,0) (0.237,0) (0.242,0) (0.247,0) (0.253,0) (0.258,0) (0.263,0) (0.268,0) (0.273,0) (0.278,0) (0.283,0) (0.288,0) (0.293,0) (0.298,0) (0.303,0) (0.308,0) (0.313,0) (0.318,0) (0.323,0) (0.328,0) (0.333,0) (0.338,0) (0.343,0) (0.348,0) (0.354,0) (0.359,0) (0.364,0) (0.369,0) (0.374,0) (0.379,0) (0.384,0) (0.389,0) (0.394,0) (0.399,0) (0.404,0) (0.409,0) (0.414,0) (0.419,0) (0.424,0) (0.429,0) (0.434,0) (0.439,0) (0.444,0) (0.449,0) (0.455,0) (0.460,0) (0.465,0) (0.470,0) (0.475,0) (0.480,0) (0.485,0) (0.490,0) (0.495,0) (0.500,0) (0.505,0) (0.510,0) (0.515,0) (0.520,0) (0.525,0) (0.530,0) (0.535,0) (0.540,0) (0.545,0) (0.551,0) (0.556,0) (0.561,0) (0.566,0) (0.571,0) (0.576,0) (0.581,0) (0.586,0) (0.591,0) (0.596,0) (0.601,0) (0.606,0) (0.611,0) (0.616,0) (0.621,0) (0.626,0) (0.631,0) (0.636,0) (0.641,0) (0.646,0) (0.652,0) (0.657,0) (0.662,0) (0.667,0) (0.672,0) (0.677,0) (0.682,0) (0.687,0) (0.692,0) (0.697,0) (0.702,0) (0.707,0) (0.712,0) (0.717,0) (0.722,0) (0.727,0) (0.732,0) (0.737,0) (0.742,0) (0.747,5) (0.753,5) (0.758,5) (0.763,5) (0.768,5) (0.773,5) (0.778,5) (0.783,5) (0.788,5) (0.793,6) (0.798,6) (0.803,6) (0.808,6) (0.813,6) (0.818,6) (0.823,6) (0.828,7) (0.833,7) (0.838,7) (0.843,18) (0.848,18) (0.854,18) (0.859,18) (0.864,18) (0.869,18) (0.874,18) (0.879,18) (0.884,18) (0.889,18) (0.894,18) (0.899,18) (0.904,18) (0.909,19) (0.914,19) (0.919,19) (0.924,19) (0.929,19) (0.934,19) (0.939,19) (0.944,19) (0.949,19) (0.955,19) (0.960,19) (0.965,19) (0.970,19) (0.975,19) (0.980,19) (0.985,19) (0.990,19) (0.995,19) (1.000,20) };
  \addlegendentry{Brainfuck (esoteric)}
\end{axis}
\end{tikzpicture}
\caption{Cumulative distinct chess-engine features introduced into the repository over normalised session progress, for four representative engines, one per language category (mainstream general-purpose, legacy, domain-specific markup, esoteric). Feature counts are first-appearances under the 38-pattern catalog of \Cref{sec:rq2}, matched against the \emph{new} content of every Write / Edit / apply\_patch event. The four trajectories take qualitatively different shapes, discussed in \Cref{sec:rq2:trajectory}.}
\label{fig:feature-growth}
\end{figure}
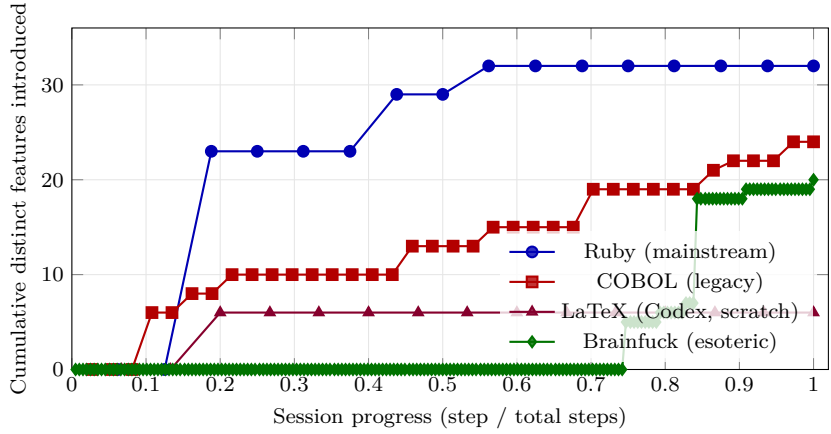

Which features are implemented is one question. \emph{When}
during the session they are introduced is a different one and,
for the language-is-not-destiny argument, the more revealing.
We scanned every authoring event in the transcripts of four
representative engines spanning the paper's main language
categories (Ruby, COBOL, LaTeX, Brainfuck), matched new content
against the 38-pattern catalogue, and recorded the first step
at which a pattern landed. The cumulative-count curve over
normalised progress is depicted in \Cref{fig:feature-growth}.
Ruby and COBOL converge on a similar
final feature set along different paths (a single burst vs.\ a
37-step accretion), whereas LaTeX and Brainfuck reach
different final sets because their languages prevent the agent
from spending effort on the search-extension tier without
crushing runtime. Languages affect both what the agent can
reach and how long it takes to get there, but they do not
affect which \emph{core} (FEN, castling, alpha--beta, material
eval) the agent reaches for first.

\subsubsection{Feature implementation depth: sub-feature analysis}
\label{sec:rq2:subfeature}

The 38-pattern catalogue measures presence but not depth.
A complementary pipeline (\Cref{sec:corpus:feature-analysis})
extracts a per-engine \emph{hotspot} for three features
(quiescence search, transposition table, evaluation/PST) and
applies 13 sub-feature regexes. Per-feature tier breakdowns
and per-engine matrices are in the replication artefact
(\texttt{reports/features/},
\hyperref[sec:data-availability]{Data and code availability}).
Two findings carry forward to
the rest of the paper. First, the rare sub-features in
quiescence (delta pruning, SEE pruning, check evasion in
qsearch) concentrate in engines that received ``improve the
Elo'' prompt iteration, confirming that prompt pressure
drives exploration of the extension layer. Second, across
all three features the same second-tier layer is missing
(check evasion in quiescence, age-based and depth-preferred
replacement in TT, king safety and tapered evaluation in
PST): LLMs implement the core specification reliably and
miss the addenda that require additional bookkeeping. The
five-tier evaluation taxonomy shows that the most structural
diversity sits in evaluation, not in search:
Tier~0 (material only),
Tier~1 (basic PST lookup),
Tier~2 (incremental MG/EG tables),
Tier~3 (full tapered evaluation, reached only by
\texttt{chess-ruby-cc}; the tapered-scoring column of
\Cref{tab:rq2-searcheval}
marks regex-level \emph{presence} of tapering vocabulary,
whereas the tier records the hand-audited \emph{depth} of the
implementation), and
Tier~4 (computed/structural, e.g.\ \texttt{lean-chess} via
geometric functions and \texttt{chess-sql} as a SQL
\texttt{WITH} query).

\subsection{Is it a copy? File-level novelty audit}
\label{sec:rq2:novelty}
\label{sec:rq2:audit}

A fair concern about any LLM-authored artefact is that the
model has emitted memorised code from its training corpus
rather than a new program. Chess is a hard test of this worry
because canonical open-source engines (\stockfish{},
\texttt{python-chess}, \texttt{chess.js}, Sunfish, TSCP, Crafty,
GNU Chess) dominate the domain and are almost certainly in
pre-training data. For the languages where no pre-existing
open-source chess engine of comparable scope appears to exist
(named in \Cref{sec:rq1}: Brainfuck, pure-TeX LaTeX, pure CSS,
APL, Icon, Lean~4, Why3, and Rocq), the no-copy claim is
trivial since there is no public reference for the agent to
have copied from. For the mainstream languages where prior
engines do exist, the claim is non-trivial, and we audit it
directly. We audit each such engine on four evidence streams:
\emph{manifest
dependencies}, \emph{engine-core chess-library imports},
\emph{canonical-engine fingerprints} (a curated catalogue of
distinctive constants and API surfaces from \stockfish{},
Sunfish, \texttt{python-chess}, \texttt{chess.js}, TSCP, and a
handful of others), and \emph{transcript authorship claims}
(strong patterns such as ``ported from'' or ``adapted from''
and weak patterns such as ``inspired by'' or ``based on''). The
per-stream procedure, the engine-core vs.\ tooling-file split
that removes false positives from PGN converters and test
harnesses, the regex caveats for compiled-out unusual
syntaxes, and the per-engine audit table itself are in
\Cref{app:novelty-audit} (\Cref{tab:rq3-novelty}).

The main finding is that the corpus contains no
recognisable copies. Of the 29 main-corpus engines, 27 are
classified \emph{scratch}, with no engine-core chess-library
import, no fingerprint match, and no self-reported copy claim.
One is \emph{library-assisted}: \texttt{chess-rust-codex} uses
the Rust \texttt{chess} crate for board state and move
generation but writes search and evaluation itself, a
legitimate dependency rather than a copy. And one is the
\emph{evasion} case (\texttt{chess-css-codex}, discussed in
\Cref{sec:rq3:evasion}). No engine triggers a \emph{strong}
transcript copy claim. Weak references like
``write something like Sunfish'' surface only as design
targets, not as port instructions. Read alongside the
within-language feature diversity documented earlier in this
section (different board representations, different
search-extension layers, different evaluation tiers from the
same agent in the same language) and the agent-trajectory
evidence in \RQ{3}, the audit supports a no-copy reading of
the corpus rather than a no-copy proof: copies would have to
evade four independent signals plus the qualitative
supervision protocol, and only one engine, the evasion case,
visibly does so.

\subsection{Algorithm-level synthesis: the quiescence triad}
\label{sec:rq2:qsearch}

The four-signal audit above operates at the \emph{file} level:
it asks whether a canonical engine's source appears in the
project tree. A complementary and finer-grained question is
whether agents synthesise a specific \emph{algorithm} from
scratch or essentially reproduce a known specification, a
distinction invisible to import-level auditing.

We probe this at the level of \textbf{quiescence search},
chosen for three reasons: (i) it is a non-trivial but
well-bounded algorithm with a canonical description on the
Chess Programming Wiki, (ii) it is present in every
main-corpus engine that reaches meaningful strength, and
(iii) its implementation admits a natural set of optional
optimisations (delta pruning, SEE pruning, check evasion, TT
probe, \ldots), making it possible to distinguish a verbatim
template instantiation from an agent that genuinely selects
which optimisations to apply.

\paragraph{Methodology.}
For each engine we extract the source-code region that
implements quiescence search (the \emph{hotspot}), score it
against a dozen sub-feature checks drawn from the Chess
Programming Wiki taxonomy (delta pruning, SEE pruning, check
evasion in qsearch, TT probe, depth limit, MVV-LVA ordering,
\ldots), and compare every pair of hotspots on two
similarity measures. \emph{Token-Jaccard} treats each hotspot
as a multiset of normalised source tokens (language keywords
filtered) and reports the size of the intersection over the
size of the union: it measures \emph{textual} proximity, that
is, how much vocabulary two implementations share. The
\emph{feature-Jaccard} treats each hotspot as a binary vector
over the sub-feature checks and applies the same
intersection-over-union ratio: it measures \emph{algorithmic}
proximity, independently of surface syntax. Two engines with
high feature-Jaccard but low token-Jaccard implement the same
algorithmic ideas in different languages or styles. Of the
main-corpus engines, 28 have at least one quiescence match,
and the analysis rests on 26 valid extractions after manual
audit.

\paragraph{Three representative variants.}
\Cref{fig:qsearch-variants} shows quiescence search in three
contrasting languages drawn from the corpus: Java
(\texttt{chess-java}), Coq/Gallina (\texttt{chess-Rocq}), and
COBOL (\texttt{COBOL-chess}). The sub-feature comparison is
summarised in \Cref{tab:qsearch-trio}.

\begin{figure*}[!htbp]
\centering
\begin{minipage}[t]{0.30\textwidth}
\lstset{language=Java,morekeywords={var}}
\begin{lstlisting}
// [1] stand-pat
int standPat =
  evaluator.evaluate(board);
boolean inCheck =
  board.isInCheck(board.sideToMove());
if (!inCheck) {
  // [2] beta-cutoff
  if (standPat >= beta) return beta;
  // [3] alpha-update
  if (standPat > alpha) alpha = standPat;
}
// [4] capture-only (or all if in check)
List<Move> moves =
  board.generateLegalMoves(!inCheck);
if (moves.isEmpty())
  return inCheck ? -MATE_SCORE + ply : alpha;
for (Move move : moves) {
  if (!inCheck && !isTactical(move)) continue;
  board.makeMove(move, undo);
  // [5] negamax sign-flip
  int score = -quiescence(-beta, -alpha, ply+1);
  board.unmakeMove(move, undo);
  if (score >= beta) return beta;
  if (score > alpha) alpha = score;
}
return alpha;
\end{lstlisting}
\centerline{\small\textbf{(a) Java}: \texttt{chess-java}}
\end{minipage}%
\hfill
\begin{minipage}[t]{0.34\textwidth}
\lstset{language={},morekeywords={Fixpoint,match,with,end,let,in,if,then,else,nat,list}}
\begin{lstlisting}
(* qdepth:nat -> termination proved *)
Fixpoint quiesce (pos : Position)
    (alpha beta : int) (qdepth : nat) : int :=
  match qdepth with
  | O => evaluate pos  (* [1] base case *)
  | S qdepth' =>
    let stand_pat := evaluate pos in  (* [1] *)
    if leb beta stand_pat then stand_pat  (* [2] *)
    else if ltb (add stand_pat delta_margin) alpha
         then stand_pat  (* delta pruning *)
    else
      let a := if ltb alpha stand_pat  (* [3] *)
               then stand_pat else alpha in
      let captures := gen_capture_moves pos in  (* [4] *)
      let fix go (mvs : list Move)
                 (cur_a : int) (fuel : nat) : int :=
        match fuel with
        | O => cur_a
        | S fuel' =>
          match mvs with
          | nil => cur_a
          | m :: rest =>
            let score := negate_score  (* [5] *)
              (quiesce (make_move pos m)
                (negate_score beta)
                (negate_score cur_a) qdepth') in
            if leb beta score then score
            else go rest
                   (if ltb cur_a score
                    then score else cur_a) fuel'
          end
        end
      in go captures a 256%nat
  end.
\end{lstlisting}
\centerline{\small\textbf{(b) Coq/Gallina}: \texttt{chess-Rocq}}
\end{minipage}%
\hfill
\begin{minipage}[t]{0.30\textwidth}
\lstset{language={},morekeywords={CALL,IF,MOVE,COMPUTE,PERFORM,VARYING,UNTIL,END-IF,END-PERFORM,GOBACK}}
\begin{lstlisting}
*> [1] stand-pat
CALL "EVAL" USING GAME-STATE STANDPAT
IF STANDPAT >= L-BETA *> [2] beta-cutoff
    MOVE L-BETA TO L-OUT
    GOBACK
END-IF
IF STANDPAT > L-ALPHA *> [3] alpha-update
    MOVE STANDPAT TO L-ALPHA
END-IF
*> delta pruning: 950 = queen value (cp)
IF (STANDPAT + 950) < L-ALPHA
    MOVE L-ALPHA TO L-OUT
    GOBACK
END-IF
*> [4] capture-only move generation
MOVE 1 TO CAP-ONLY
CALL "MOVEGEN" USING GAME-STATE LS-LIST CAP-ONLY
PERFORM VARYING I FROM 1 BY 1
    UNTIL I > ML-COUNT OF LS-LIST
    PERFORM SCORE-Q-MOVE
END-PERFORM
PERFORM VARYING I FROM 1 BY 1
    UNTIL I > ML-COUNT OF LS-LIST
    PERFORM PICK-BEST-Q
    CALL "MAKEMOVE" USING GAME-STATE LS-MOVE OK
    IF OK = 1
        COMPUTE CHILD-ALPHA = 0 - L-BETA
        COMPUTE CHILD-BETA  = 0 - L-ALPHA
        CALL "QUIESCE" USING GAME-STATE *> [5]
            CHILD-ALPHA CHILD-BETA L-SS CHILD-SCORE
        COMPUTE CHILD-SCORE = 0 - CHILD-SCORE
        CALL "UNMAKEMOVE" USING GAME-STATE LS-MOVE
        IF CHILD-SCORE > L-ALPHA
            MOVE CHILD-SCORE TO L-ALPHA
            IF L-ALPHA >= L-BETA
                MOVE L-BETA TO L-OUT
                GOBACK
            END-IF
        END-IF
    END-IF
END-PERFORM
MOVE L-ALPHA TO L-OUT
GOBACK.
\end{lstlisting}
\centerline{\small\textbf{(c) COBOL}: \texttt{COBOL-chess}}
\end{minipage}
\caption{Quiescence search in three languages. Labels [1]--[5]
mark the five invariant algorithmic steps shared by all 26
valid corpus extractions: \textbf{[1]}~stand-pat evaluation,
\textbf{[2]}~$\beta$-cutoff, \textbf{[3]}~$\alpha$-update,
\textbf{[4]}~capture-only move generation, \textbf{[5]}~negamax
sign-flip. PL-specific adaptations are discussed in the text.
Sub-feature scores and pairwise distances in
\Cref{tab:qsearch-trio}.}
\label{fig:qsearch-variants}
\end{figure*}

\begin{table}[!htbp]
\centering\small
\begin{tabular}{l ccc cc}
\toprule
& \multicolumn{3}{c}{Sub-features present} & \multicolumn{2}{c}{Pairwise sim.}\\
\cmidrule(lr){2-4}\cmidrule(lr){5-6}
Engine & count & features & missing key & feat-J & tok-J\\
\midrule
\texttt{chess-java} & 7/13 & sp, bc, au, co, dl, ce, nr &
  delta, see & --- & ---\\
\texttt{chess-Rocq} & 4/13 & sp, co, dp, dl &
  bc$^*$, au$^*$, nr, ce & 38\% & 10\%\\
\texttt{COBOL-chess} & 4/13 & sp, co, dp, mv &
  bc$^*$, au$^*$, nr, ce & 22\% & 9\%\\
\midrule
\multicolumn{4}{l}{\textit{Rocq $\leftrightarrow$ COBOL}} & 60\% & 8\%\\
\bottomrule
\end{tabular}
\caption{Sub-feature profiles and pairwise distances for the
three variants in \Cref{fig:qsearch-variants}. Abbreviations:
sp=stand\_pat, bc=beta\_cutoff, au=alpha\_update,
co=capture\_only, dp=delta\_pruning, dl=depth\_limit,
ce=check\_evasion, nr=negamax\_recurse, mv=mvv\_lva,
see=see\_pruning. $^*$Coq uses \texttt{leb}/\texttt{ltb}
predicates that detect the concepts but use different tokens,
explaining the low token-Jaccard. feat-J = feature-Jaccard,
tok-J = token-Jaccard, both relative to \texttt{chess-java}
for the first two rows and between Rocq and COBOL for the
last.}
\label{tab:qsearch-trio}
\end{table}

\noindent\textbf{What the three variants share, and how they
adapt.}
Every implementation traces the same five steps labelled in
\Cref{fig:qsearch-variants}: static stand-pat evaluation,
$\beta$-cutoff, $\alpha$-update, capture-only move
generation, and a negamax sign-flipped recursive call. The
PL-specific adaptations go beyond mechanical syntax
translation. Java adds a check-evasion branch unavailable in
the functional and procedural variants. Coq/Gallina declares
\texttt{Fixpoint} with a \texttt{qdepth~:~nat} argument
(termination is \emph{proved}, not assumed) and uses
\texttt{let~\ldots~in} bindings throughout because Coq has no
mutable variables, with an inner \texttt{fix~go} carrying its
own \texttt{fuel~:~nat} for the capture loop's termination
proof. COBOL replaces the outer capture loop with a
\texttt{PERFORM~VARYING} iterator (the recursive call to
\texttt{QUIESCE} still exists at line~[5], but the loop over
captures is unrolled into an explicit index-driven iteration)
and computes child bounds into named working-storage fields,
reproducing the queen-value delta-pruning constant
\textbf{950}~centipawns from the CPW article exactly. The Rocq--COBOL pair scores 60\%
feature-Jaccard against 8\% token-Jaccard, two
implementations that converge on the same sub-feature subset
of the CPW specification while sharing almost no tokens.

\paragraph{Canonical skeleton, no textual copy.}
Every valid extraction follows the same four-line skeleton, and
the terminology confirms a common source: the wiki community
term \texttt{stand\_pat}/\texttt{standpat}/\texttt{STANDPAT}
appears in 19 of 26 extractions across 11 PLs. Yet at the character level the implementations are not copies
of each other. Token-Jaccard peaks at 47\% (a Java engine and
its direct Rust port) and is typically 15--30\% across
language boundaries. Algorithmic similarity is substantially
higher:
\texttt{chess-java-cc}~$\leftrightarrow$~\texttt{chess-purec}
(Java and~C, different models) reaches 88\% feature-Jaccard,
and four pairs across C/C++, Java, Rust, and Ruby reach
$\geq$80\% on cross-language, cross-model pairs that share no
token-level similarity. The clearest smoking gun is the delta-pruning
threshold: where present (11 of 26 engines), the constant is
\emph{always} 900--950 centipawns, the exact value in the
CPW delta-pruning article, which independent derivation would
not reproduce so precisely. The pattern is a shared community
specification \emph{plus} substantial per-variant divergence
in sub-feature selection and PL idiom rather than textual
inheritance from any canonical engine. We make this divergence
inspectable through the \emph{Chess Engine Feature Explorer}
web tool (\Cref{app:method:explorer}), which lets a reviewer
browse the 28 corpus variants and compare any pair side by
side at the source-code level with the 13 sub-feature chips
rendered as a header band.

\paragraph{Extending the triad: TT and evaluation.}
The same hotspot pipeline applied to transposition tables
(21 valid extractions) finds a different divergence pattern.
Zobrist hashing is universal, but architectural choices
diverge: C/C++/Java engines allocate flat power-of-two
arrays, \texttt{chess-ruby-cc} uses Ruby's native hash,
\texttt{lean-chess} stores entries as \texttt{Option Move} in
a \texttt{HashMap}, and \texttt{COBOL-chess} uses
\texttt{OCCURS 500000 TIMES}. For evaluation/PST (10 valid
extractions out of 19 pattern matches), the most direct novelty test is whether any
engine reproduces a canonical engine's material constants:
Sunfish (\texttt{R=479, N=280, Q=929}) and the CPW simplified
values.
\emph{No} generated chess engines reproduce either, across all extractions.
\texttt{lean-chess} replaces lookup tables with
geometric functions, and \texttt{chess-sql} expresses
evaluation as a SQL \texttt{WITH} query.

\begin{rqanswer}{2}
Every \agentbuilt{} engine implements a core of rules
and a minimal search, and above the core, feature choice
varies substantially within a language. On the novelty
question, 27 of 29 main-corpus engines are scratch-built on
the four-signal audit, 1 uses a language-native chess crate,
1 evaded the language constraint (\Cref{sec:rq3:evasion}),
while no agent ever self-reported porting or adapting a
canonical engine.
Overall, \textbf{coding agents have absorbed an abstract chess programming specification deeply enough to
regenerate it in any language, but variants differ markedly
in which sub-features they actually implement.}
The generated code is not a textual copy of any canonical
source: numerical constants are generated fresh, and
PL-specific structural adaptations (e.g., Coq termination proofs,
COBOL working-storage, SQL WITH-query evaluation) go well
beyond syntax translation.
\end{rqanswer}

\FloatBarrier
\section{RQ3 (Validation): how agent teammates demonstrate correctness}
\label{sec:rq3}

\paragraph{Question.} How do AI coding-agent teammates
\emph{demonstrate} that the software they produce is correct?
Which oracles do they wire up, and when in the session do those
oracles fire? Do they even ``cheat'', and what supervision is
needed to keep the experiment honest?

Most polyglot evaluation benchmarks
(MultiPL-E~\cite{cassano2023multipl},
SWE-bench~\cite{jimenez2024swebench},
Aider Polyglot~\cite{gauthier2024aiderpolyglot}) take the
test suite as given, making the validation layer invisible.
In our corpus the agents construct their own oracles, which
makes validation behaviour observable. This section provides
an empirical baseline.

\subsection{The validation stack the agents wire up}
\label{sec:rq3:stack}

The background (\Cref{sec:background:oracles}) introduced the
three oracle levels chess offers: \perft{} batteries that pin
down move-generation correctness against published reference
counts, gameplay against a calibrated reference engine
through the \uci{} protocol, and Elo aggregation across those
games. The question here is which of those levels each
\emph{agent} actually wired up on its own initiative, and what
shape that wiring takes across the corpus. We fold \uci{}
conformance into the gameplay layer rather than tracking it
separately, because an engine that fails the handshake never
enters a gauntlet in the first place. The handshake itself
is a non-trivial achievement in constrained-execution targets
such as Brainfuck and \LaTeX{}, where a UCI shim has to be
hand-built around the engine core, but we track it implicitly
through gauntlet entry rather than as a dedicated column.

\Cref{tab:eval-infra} reports the per-engine result. The
table tracks five observable validation artefacts: a
\perft{} harness for the move-generation oracle, three faces
of the gameplay layer (a \stockfish{} gauntlet, a
cutechess-cli runner, and a random-mover baseline), and a
bespoke Elo script for the strength layer. The few engines
without a gauntlet column either never reached \uci{}
conformance (\texttt{chess-css-codex},
\texttt{chess-css-codex-guided}) or routed validation through
a non-UCI bespoke runner (\texttt{chess-brainfuck},
\texttt{chess-brainfuck-cc}, the two Java$\to$language ports
that scored against their parent engine rather than
\stockfish{}).

\begin{table*}[!htbp]\centering\scriptsize
\caption{Validation infrastructure each agent set up
\emph{without explicit user instruction}, detected from
repository contents at corpus snapshot. Each column flags a
specific oracle or harness the agent built or wired up,
ordered roughly from \emph{exact} to \emph{ordinal}.
\textbf{Perft harness}: the engine ships its own
\perft{} (move-generation count) test code, comparing against
published reference values to verify move generation is
correct (independent of search and evaluation).
\textbf{Stockfish gauntlet}: a tournament against
\stockfish{} as the reference opponent (with or without
\texttt{UCI\_LimitStrength}). It requires the engine to honour
the \uci{} protocol and play legal moves under arbitrary
positions for a full game.
\textbf{cutechess-cli runner}: the gauntlet (or any
multi-engine match) is driven through cutechess-cli, the
de-facto chess-engine tournament harness. Running cleanly
through cutechess-cli is itself a non-trivial conformance test
because the engine must answer \texttt{uci}, \texttt{isready},
\texttt{position}, and \texttt{go} correctly across hundreds
of positions and emit \texttt{bestmove} on time.
\textbf{Random-mover baseline}: a random-legal-move
opponent invoked when \stockfish{} trivially wins (used as a
floor reference for the weakest engines).
\textbf{Bespoke Elo script}: a per-repo Python or shell
script that runs the engine, computes a logistic Elo from the
score, and prints a number. Together with the gauntlet, this
is the agent's own self-evaluation pipeline.
\textbf{Self-play harness}: engine versus older versions of
itself, present only in engines strong enough to learn from it.
\textbf{Test files}: count of files matching
\texttt{tests?/} or \texttt{*\_test*} at any depth in the
repository (a coarse build-system--integration signal, not
specific to chess oracles).}
\label{tab:eval-infra}
\resizebox{\textwidth}{!}{%
\begin{tabular}{l l c c c c c c r}
\toprule
Engine & Language & \makecell{Perft\\harness} &
\makecell{Stockfish\\gauntlet} &
\makecell{cutechess\\runner} &
\makecell{Random\\baseline} &
\makecell{Bespoke\\Elo script} &
\makecell{Self-play\\harness} &
\makecell{\#Test\\files} \\
\midrule
\texttt{chess-apl-codex54} & APL & $\checkmark$ & $\checkmark$ & $\checkmark$ & $\checkmark$ & $\checkmark$ &   & 3 \\
\texttt{chess-assembly-codex} & Assembly &   & $\checkmark$ & $\checkmark$ &   & $\checkmark$ &   & 0 \\
\texttt{chess-brainfuck} & Brainfuck & $\checkmark$ & $\checkmark$ &   &   & $\checkmark$ &   & 6 \\
\texttt{chess-brainfuck-cc} & Brainfuck & $\checkmark$ & $\checkmark$ &   & $\checkmark$ &   &   & 1 \\
\texttt{chess-cobol-cc} & COBOL &   & $\checkmark$ & $\checkmark$ &   &   &   & 1 \\
\texttt{chess-cplusplus-claude} & C++ & $\checkmark$ & $\checkmark$ & $\checkmark$ &   &   & $\checkmark$ & 4 \\
\texttt{chess-css-cc} & CSS/HTML &   & $\checkmark$ & $\checkmark$ &   &   &   & 0 \\
\texttt{chess-css-codex} & CSS/HTML &   &   &   &   & $\checkmark$ &   & 1 \\
\texttt{chess-css-codex-guided} & CSS/HTML & $\checkmark$ &   &   &   &   &   & 0 \\
\texttt{chess-icon-codex} & Icon & $\checkmark$ & $\checkmark$ & $\checkmark$ &   & $\checkmark$ &   & 0 \\
\texttt{chess-java} & Java & $\checkmark$ & $\checkmark$ & $\checkmark$ &   & $\checkmark$ &   & 0 \\
\texttt{chess-java-cc} & Java & $\checkmark$ & $\checkmark$ & $\checkmark$ &   &   &   & 0 \\
\texttt{chess-latex-codex-replication} & LaTeX &   & $\checkmark$ & $\checkmark$ &   & $\checkmark$ &   & 0 \\
\texttt{chess-mojo}$^\ddagger$ & Mojo & $\checkmark$ & $\checkmark$ & $\checkmark$ &   & $\checkmark$ & $\checkmark$ & 2 \\
\texttt{chess-newlang-codex} & DSL (newlang) & $\checkmark$ & $\checkmark$ & $\checkmark$ &   &   &   & 0 \\
\texttt{chess-purec} & C & $\checkmark$ & $\checkmark$ & $\checkmark$ &   &   &   & 4 \\
\texttt{chess-purec-codex} & C & $\checkmark$ & $\checkmark$ & $\checkmark$ &   &   &   & 0 \\
\texttt{chess-py} & Python & $\checkmark$ & $\checkmark$ & $\checkmark$ &   &   &   & 5 \\
\texttt{chess-py-cc} & Python & $\checkmark$ & $\checkmark$ & $\checkmark$ &   &   &   & 5 \\
\texttt{chess-revisit-java-toCOBOL-codex} & Java$\to$COBOL port &   &   & $\checkmark$ &   &   &   & 0 \\
\texttt{chess-revisit-java-toRust-codex} & Java$\to$Rust port & $\checkmark$ &   & $\checkmark$ &   &   &   & 6 \\
\texttt{chess-Rocq} & Rocq$\to$OCaml & $\checkmark$ & $\checkmark$ & $\checkmark$ &   &   &   & 3 \\
\texttt{chess-ruby-cc} & Ruby & $\checkmark$ & $\checkmark$ & $\checkmark$ &   & $\checkmark$ &   & 7 \\
\texttt{chess-ruby-codex} & Ruby & $\checkmark$ & $\checkmark$ &   &   & $\checkmark$ &   & 5 \\
\texttt{chess-rust-cc} & Rust & $\checkmark$ & $\checkmark$ & $\checkmark$ &   & $\checkmark$ &   & 1 \\
\texttt{chess-rust-cc-redo} & Rust & $\checkmark$ & $\checkmark$ & $\checkmark$ &   & $\checkmark$ &   & 1 \\
\texttt{chess-rust-codex} & Rust & $\checkmark$ & $\checkmark$ & $\checkmark$ &   & $\checkmark$ &   & 0 \\
\texttt{chess-sql} & SQL & $\checkmark$ & $\checkmark$ & $\checkmark$ &   &   &   & 7 \\
\texttt{chess-why3} & Why3$\to$OCaml &   & $\checkmark$ & $\checkmark$ &   & $\checkmark$ &   & 0 \\
\texttt{chess-why3-cc} & Why3$\to$OCaml & $\checkmark$ & $\checkmark$ & $\checkmark$ &   & $\checkmark$ &   & 308 \\
\texttt{COBOL-chess} & COBOL & $\checkmark$ & $\checkmark$ & $\checkmark$ &   &   &   & 15 \\
\texttt{cplusplus-chess} & C++ & $\checkmark$ & $\checkmark$ & $\checkmark$ &   &   &   & 1 \\
\texttt{latex-chess-engine} & LaTeX (TeX) &   & $\checkmark$ & $\checkmark$ &   &   &   & 0 \\
\texttt{lean-chess} & Lean 4 & $\checkmark$ & $\checkmark$ & $\checkmark$ &   &   &   & 0 \\
\bottomrule
\end{tabular}}
\end{table*}

\paragraph{What the audit shows.}
The modal validation footprint pairs the \perft{} oracle
with the gameplay-gauntlet layer. 26 of the 34 engines ship
a \perft{} harness and 30 wire up a \stockfish{} gauntlet,
and the combination is present across every PL category from
mainstream to esoteric. The gauntlet itself almost always
runs through cutechess-cli, with 29 of the 34 engines using
it as their tournament harness. The five exceptions either
run a bespoke runner or skip the gameplay layer altogether
(\texttt{chess-brainfuck}, \texttt{chess-brainfuck-cc},
\texttt{chess-css-codex},
\texttt{chess-css-codex-guided}, \texttt{chess-ruby-codex}).
Running cleanly under cutechess-cli is itself a non-trivial
conformance test, because the harness throws arbitrary
middlegame and endgame positions at the engine for hundreds
of games and expects legal, on-time moves throughout.

Self-strength estimation is where agents take the most
freedom, and where the most variance shows up. 15 of the 34
engines ship a custom Elo-estimation script of their own
design, but only four come close to the truth.
\texttt{chess-Rocq} ($\Delta=-1$), \texttt{chess-ruby-cc}
($\Delta=-87$), \texttt{chess-java-cc} ($\Delta=-116$), and
\texttt{chess-rust-cc-redo} ($\Delta=+110$) all land within
\elo{150} of the external Elo assessment of
\Cref{tab:rq4-elo}. The \texttt{chess-Rocq} agreement is
ruler-dependent: its $\Delta=-1$ holds against the Phase~B
anchor measurement (\elo{1499~$\pm$~171}), while the unified
Bradley-Terry refinement places the engine lower
(\elo{1315~$\pm$~83}, \Cref{app:anchor-eval:bt}). Two of those four,
\texttt{chess-Rocq} and \texttt{chess-rust-cc-redo},
triangulated across multiple \texttt{UCI\_Elo} levels with
$\geq$100 games per level, which is the single methodological
choice that most reliably keeps the self-claim near the
truth. The other eleven of the 15 overestimate by \elo{200}
to \elo{1100}, a bias we dissect below
(\Cref{sec:rq3:first-call}) and revisit from the cost angle in
\Cref{sec:rq5:reward}.

Two further patterns round out the picture. Self-play
harnesses are rare in the corpus, with only
\texttt{chess-cplusplus-claude} and \texttt{chess-mojo} wiring
one up. And two engines (\texttt{chess-apl-codex54},
\texttt{chess-brainfuck-cc}) fall back to a random-legal-move
opponent in addition to or instead of \stockfish{}, so that
engines too weak to ever beat \stockfish{} still receive a
measurable signal rather than a floor of 0/$N$.

\subsection{Oracle-first as a population property}
\label{sec:rq3:first-call}

The validation finding with the sharpest population-level
signature is \emph{when} the oracles fire. Looking at the \emph{first} bash call of each kind in
each session gives a sharp statistic: \emph{the median engine
runs its first \texttt{cargo test} / \texttt{make} at step~2,
its first \perft{} test at step~2, and its first
cutechess/custom-Elo gauntlet at step~1}. Agents reach for the
strong-oracle stack essentially from the first substantive
response, and they do not wait to be asked to evaluate. This is
\emph{oracle-first behaviour}, and the corpus-wide number is
the strongest single behavioural signal of the agentic-SE
shift~\cite{li2025riseaiteammatessoftware}: pre-agent
code-generation tools produced code and waited
for a human evaluator. Current agents produce code and then
\emph{check their own work against external oracles}, where the
domain admits them.
 A second pattern, harder to spot in aggregate but visible
per session, is that when the gameplay plateaus,
the strongest agents switch to self-Elo triangulation.
\texttt{chess-Rocq}'s agent ran three calibrated
\texttt{UCI\_Elo} levels (1320, 1500, 1700), each with 100
games at TC$=$10s$+$0.1s, explicitly triangulated around the
level scoring closest to 50\%, and explicitly acknowledged
uncertainty in the published number.
\emph{The fixable form of self-Elo
bias is methodology choice, not the existence of
\stockfish{} as an opponent.}

\Cref{sec:rq4} reports that most agents' self-reported Elo is
biased \elo{200} to \elo{1100} above the external assessment.
Through the validation lens, the same numbers are an
empirical statement about \emph{which oracle the agent
optimises against}, rather than about reporting accuracy alone. Three
sources compound. The first is \stockfish{}'s
\texttt{UCI\_LimitStrength} drift at the high end of the
ladder. The second is time-control choice, since most agent
gauntlets used 50--200~ms per move, where shallow fixed-depth
search is structurally favoured. The third is small
per-opponent sample size, with a median of 10--30 games whose
$\approx$\elo{300} standard error is reported as a point
estimate. The arithmetic is direct.
On this hardware, nominal \texttt{UCI\_Elo$=$2400} plays at
$\approx$\elo{2050} effective and \texttt{UCI\_Elo$=$2600}
at $\approx$\elo{2150}. An agent that self-evaluates only
at the two highest rungs measures real \elo{20} eval
improvements as noise (false negatives), while small
``style'' improvements at the compressed configuration
register as \elo{200} jumps (false positives). The reward
signal the agent climbs is not the gradient of true Elo, and
the extreme case is \texttt{chess-assembly-codex}, which
claims \elo{2481} against an external \elo{1403}
($\Delta=-1078$, reproducible across configurations). An
11-engine self-reproduction sweep, which re-ran each engine's
original eval methodology unchanged (953 games,
\Cref{app:anchor-eval:q2}), confirms that the bias is embedded
in the methodology rather than in any one run: 9 of 11 engines
reproduce within $\pm$\elo{150} of their original self-claim,
yet the reproduced numbers sit on average \elo{348} above the
external assessment. The self-Elo oracle \emph{layer} an agent
constructs thus determines whether its own loop converges to a
generalising or non-generalising strength claim. What the
biased signal did to the spend is taken up in
\Cref{sec:rq5:reward}.

\subsection{Do agents try to cheat?}
\label{sec:rq3:evasion}

A qualitatively different validation question arises when the
experiment is about \emph{language}: can the agent satisfy
the user's literal request while side-stepping its spirit by
importing a permissive library that does the chess work? In
our corpus the answer is \emph{yes, occasionally, in
ways the supervision protocol catches}.

\paragraph{CSS evasion (\texttt{chess-css-codex}).}
Asked for a CSS chess engine, Codex's early drafts silently
imported \texttt{python-chess} in
\texttt{src/chess\_css/engine.py}, \texttt{search.py}, and
\texttt{uci.py} as the rules$+$move-gen backbone while CSS
only rendered the board, leaving a Python engine with CSS
decorations. The novelty audit (\Cref{sec:rq2:audit}) flags
the engine as \emph{fingerprint-match}, the only such case in
the corpus.
\label{sec:rq3:evasion-rust}\paragraph{Rust crate evasion
(\texttt{chess-rust-codex}).} A subtler case in a mainstream
language. Codex's first Rust attempt did not paste any
canonical engine's source, but engine-core files
(\texttt{src/uci.rs} and \texttt{src/engine.rs}) import the
Rust \texttt{chess} crate and delegate board representation,
move generation, and legal-move checking to that third-party
library. The agent authored only search and evaluation. The
audit classifies the engine as \emph{library-assisted} and the
crate dependency is genuinely idiomatic Rust, but by the
study's intent half the engine had been outsourced. The
manifest-and-import grep caught this, and we re-ran Rust
from a blank slate as \texttt{chess-rust-cc-redo} (deep-dive in
\Cref{app:deepdive-rust-redo}), the strongest from-scratch
Rust engine in the corpus.
\paragraph{Counterexample (ChessCSS) and a near-evasion the
agent refused (Brainfuck).}
The same target language can hold the constraint under a
different protocol. \texttt{test-superset}
(``ChessCSS''~\cite{acher2026chesscss-readme}) is a
CSS-rule-based engine ($\approx$55k CSS rules, with a minimal
JavaScript harness for I/O and the game loop) built over 31
commits of progressive JS$\to$CSS migration under sustained
expert steering, the README explicitly noting that the human had to
drive the agent with technical expertise to keep the
constraint. Conversely, the Brainfuck engine
(\texttt{chess-brainfuck-cc}, deep-dive report in the
replication artefact) is authored as a Python code-generator
that emits Brainfuck.
When the user asked whether the chess logic should ``just
live in Python and the engine call out to BF for the I/O
loop'', the agent argued in the session log for keeping all
chess decisions inside the BF tape and accepted the
additional cost. The episode shows agents \emph{can} hold a
language constraint when it is named.

\paragraph{The supervision protocol.}
The lesson is not that agents are adversarial (they are
not), but that ``write a chess engine in $L$'' is
under-specified once a library $M$ exists in training data
that trivialises the language-specific work. The asymmetry
between the CSS and Rust cases is informative. In CSS the
library is structurally inappropriate and the evasion stands
out at a glance. In Rust the library is idiomatic and the
evasion blends in with how a Rust engineer would normally
start a project. We therefore apply a four-step qualitative
supervision protocol on every engine (read the engine-core
diff, grep for chess-library imports, verify the oracle is
independent, redirect when evasion is detected). The full
protocol is in \Cref{app:method:supervision}. Under that
protocol we found exactly one engine-core fingerprint match
(\texttt{chess-css-codex}) and one library-assisted engine
(\texttt{chess-rust-codex}). The no-further-evasion result
is \emph{conditional on the protocol having been applied}.
ProgramBench~\cite{yang2026programbench} corroborates the
structural risk at scale: under their adversarial sandbox,
frontier models still attempt source lookup in 1--36\% of runs
across four model families.




\begin{rqanswer}{3}
AI coding-agent teammates demonstrate correctness in
this domain by autonomously constructing a stacked oracle
pipeline: \perft{} for move-generation
correctness, \stockfish{} matches over the \uci{} protocol
for gameplay-level legality and result aggregation, self-Elo
estimates for performance strength, and occasionally
head-to-head round-robins between sibling engines. Agents
build it typically in that order, and typically without
being asked. The median engine runs its first \perft{} test
at session step~2 and its first gauntlet at step~1.
Validation behaviour is, however, uneven in quality since most agents'
self-Elo is methodologically biased (\Cref{sec:rq3:first-call}).
 Validation behaviour also has a
language-constraint failure mode: cheating via
permitted-library imports, with one cleanly cheating engine
(\texttt{chess-css-codex}), one positive
counterexample under expert co-design (ChessCSS), and one
near-evasion the agent refused when explicitly named
(Brainfuck). Overall, \textbf{frontier
agents \emph{are} capable of validation, their validation
methodology is uneven, and language constraints must be controlled by the developers.}
\end{rqanswer}
\FloatBarrier
\section{RQ4 (Strength): does the PL matter for Elo?}
\label{sec:rq4}

\paragraph{Question.} What playing strength does each engine
reach under a unified re-evaluation harness, and how does the
resulting Elo distribution relate to programming language,
implemented feature set, and the validation rigor we
characterised in \RQ{3}?

After \RQ{1}--\RQ{3} settled \emph{whether} and \emph{how} agents
build engines across PLs, the title question's \emph{impact} half
asks whether the choice of PL still moves the engine's reachable
strength (Elo). The answer is \emph{yes}: PL category sets a loose
upper bound on reachable Elo, and within-language variance lives
strictly below it. The tendency is descriptive, not causal
(\Cref{sec:threats}), but it is robust.

\paragraph{Two numbers per engine.}
The external number comes from the canonical re-evaluation platform
of \Cref{sec:corpus:canonical-eval}: every UCI-conformant engine
plays the same five externally-rated references (Rustic~\elo{1820},
Asymptote~\elo{2150}, and calibrated \stockfish{} Skill 5/10/15) at
a fixed TC$=$120s$+$1s, with the full match protocol in
\Cref{app:anchor-eval}. This is the same kind of gauntlet the agents
run against \stockfish{}, only against calibrated, externally-rated
opponents. \Cref{tab:rq4-elo} reports it alongside the Elo each
engine claimed for itself (from its own bench scripts or
transcript). A dozen weaker engines lost (or won) every game against
this panel, so it can only place them ``below $\sim$\elo{1300}.'' To
pin those down we also let the engines play \emph{each other} and
fit one Bradley-Terry model over all the games at once, which turns
the bound into an actual number such as \texttt{chess-ruby-cc} at
\elo{1753}. Such refined cells are marked $^\dagger$. This
corpus-wide pooling is a step our analysis takes after the fact. A
single agent session, which only ever plays the references, cannot
do it, so the panel itself, not the refinement, is the practical
lesson for self-evaluation. The agents' own in-tree \pgn{} gauntlets,
where they exist, track the self-claim and we cite them only for
triangulation. They are not a cross-engine ruler, since each used a
different opponent set, time control, and game count. We omit a
perft (move-generation) column. That lower layer is audited under
\RQ{3} (\Cref{tab:eval-infra}), and the correctness signal that
matters here, namely that every gauntlet game is a legal, parseable
chess game, is already guaranteed by the scoring.

\begin{table}[!htbp]\centering\scriptsize
\caption{Evidence of strength per engine: self-reported
vs.\ external assessment. \textbf{Self}: the Elo number(s) the
engine itself reports (single value or min--max band, from
repository bench files plus agent-transcript claims).
\textbf{External Elo Assessment}: our re-evaluation of each
engine against five externally-rated reference engines
(Rustic~\elo{1820}, Asymptote~\elo{2150}, and \stockfish{}
Skill 5/10/15, calibrated) at a fixed TC$=$120s$+$1s, the
same kind of measurement the agents make against \stockfish{},
but against calibrated, externally-rated opponents
(\Cref{sec:rq4:anchor}). Where an engine lost (or won) every
game against the panel, the anchors only bound it. Those cells
(marked $^\dagger$) are resolved by a unified Bradley-Terry
refinement that additionally pools the corpus's internal
head-to-head games (\Cref{app:anchor-eval}). Unmarked values
are therefore Phase~B anchor measurements and $^\dagger$
values are Bradley-Terry refinements. The two rulers coincide
where both exist, and both are tabulated side by side in
\Cref{tab:rr-final}, generated from
\texttt{eval2/results/combined\_bt\_summary.md}.
$\Delta$: External minus Self (negative
$=$ self-overestimate ``Pattern A'', positive $=$
self-underestimate ``Pattern B'', midpoint of band used for
$\Delta$). ``---'' $=$ no comparable data (no UCI handshake,
underconverged, or lost every game with no refinement bound).
Self values marked $^{\S}$ are pre-measurement planning
estimates from the agent's session (the agent never ran a
Stockfish gauntlet to commit a number). All unmarked Self
values are post-measurement claims from a published bench file
or session match. The agents' own in-tree \pgn{} gauntlet
numbers, where present, are cited in the text for triangulation
only, not as a ruler. Move-generation correctness (which
engines ship a perft harness vs.\ pass an external perft
battery) is reported under the validation-infrastructure audit
of \Cref{tab:eval-infra}~(\Cref{sec:rq3:stack}). We do not
duplicate it here. Rows sorted by External Elo Assessment,
descending.}
\label{tab:rq4-elo}
\resizebox{\textwidth}{!}{%
\begin{tabular}{l l r r r}
\toprule
Engine & Language & Self & External Elo Assessment & $\Delta$ \\
\midrule
\texttt{chess-java-cc} & Java & 2212 & 2096 $\pm$ 132 & $-116$ \\
\texttt{chess-rust-cc-redo} & Rust & 1879 & 1989 $\pm$ 105 & $+110$ \\
\texttt{chess-revisit-java-toRust-codex} & Java$\to$Rust & — & 1922 $\pm$ 106 & — \\
\texttt{chess-rust-cc} & Rust & 2110 & 1841 $\pm$ 99 & $-269$ \\
\texttt{chess-ruby-cc} & Ruby & 1840 & 1753 $\pm$ 55$^\dagger$ & $-87$ \\
\texttt{chess-rust-codex} & Rust & 2043 & 1723 $\pm$ 103 & $-320$ \\
\texttt{cplusplus-chess} & C++ & 2087 & 1709 $\pm$ 111 & $-378$ \\
\texttt{chess-cplusplus-claude} & C++ & 1897 & 1680 $\pm$ 103 & $-217$ \\
\texttt{chess-newlang-codex} & DSL/Mini & 2170 & 1622 $\pm$ 68$^\dagger$ & $-548$ \\
\texttt{chess-why3-cc} & Why3$\to$OCaml & 1882 & 1618 $\pm$ 163 & $-264$ \\
\texttt{chess-py} & Python & 1800 & 1616 $\pm$ 55$^\dagger$ & $-184$ \\
\texttt{chess-java} & Java & 1806 & 1509 $\pm$ 131 & $-297$ \\
\texttt{chess-py-cc} & Python & 1500--1800$^{\S}$ & 1499 $\pm$ 55$^\dagger$ & $-151$ \\
\texttt{chess-Rocq} & Rocq$\to$OCaml & 1500 & 1499 $\pm$ 171 & $-1$ \\
\texttt{chess-purec} & C & 1997 & 1440 $\pm$ 193 & $-557$ \\
\texttt{chess-revisit-java-toCOBOL-codex} & Java$\to$COBOL & — & 1404 $\pm$ 69$^\dagger$ & — \\
\texttt{chess-assembly-codex} & x86-64 asm & 2481 & 1403 $\pm$ 196 & $-1078$ \\
\texttt{chess-purec-codex} & C & 1670--1972 & 1402 $\pm$ 73$^\dagger$ & $-419$ \\
\texttt{COBOL-chess} & COBOL & 1600--1700 & 1365 $\pm$ 80$^\dagger$ & $-285$ \\
\texttt{chess-ruby-codex} & Ruby & 920 & 1346 $\pm$ 72$^\dagger$ & $+426$ \\
\texttt{chess-brainfuck} & Brainfuck & 1582 & 1299 $\pm$ 94$^\dagger$ & $-283$ \\
\texttt{chess-cobol-cc} & COBOL & 1630 & 1274 $\pm$ 85$^\dagger$ & $-356$ \\
\texttt{lean-chess} & Lean 4 & — & 1271 $\pm$ 75$^\dagger$ & — \\
\texttt{chess-why3} & Why3$\to$OCaml & 1008 & 1210 $\pm$ 74$^\dagger$ & $+202$ \\
\texttt{chess-icon-codex} & Icon & — & 1168 $\pm$ 88$^\dagger$ & — \\
\texttt{chess-apl-codex54} & APL & 1040 & 686 $\pm$ 205$^\dagger$ & $-354$ \\
\texttt{chess-sql} & SQL & 900--1020 & 523 $\pm$ 297$^\dagger$ & $-437$ \\
\texttt{latex-chess-engine} & TeX & 1300 & — & — \\
\texttt{chess-mojo}$^{\S}$ & Mojo & 1100 & — & — \\
\texttt{chess-brainfuck-cc} & Brainfuck & 578 & — & — \\
\texttt{chess-latex-codex-replication} & LaTeX & 546 & — & — \\
\texttt{chess-css-cc} & CSS/HTML & — & — & — \\
\texttt{chess-css-codex} & CSS/HTML & — & — & — \\
\texttt{chess-css-codex-guided} & CSS/HTML & — & — & — \\
\bottomrule
\end{tabular}}
\end{table}

\paragraph{What \Cref{tab:rq4-elo} supports.}
The top of the table is mainstream-compiled. \texttt{chess-java-cc}
(\elo{2096}) and \texttt{chess-rust-cc-redo} (\elo{1989}) lead, with
the Java$\to$Rust port (\elo{1922}), \texttt{chess-rust-cc}
(\elo{1841}), and the C\texttt{++} and Ruby \claudecode{} engines
filling out a \elo{1700}--\elo{2100} top band that no specialised,
legacy, or esoteric engine reaches. Self-reports diverge from this
external number by hundreds of Elo, almost always upward. The
$\Delta$ column (external minus self) clusters around $-200$ to
$-400$ and runs to $-1078$ for \texttt{chess-assembly-codex}, the
record overestimation. Only three engines under-report (Pattern~B:
\texttt{chess-ruby-codex} $+426$, \texttt{chess-why3} $+202$, and
\texttt{chess-rust-cc-redo} $+110$, where the agent deliberately
published a conservative number), and \texttt{chess-Rocq}
($\Delta=-1$ against its Phase~B anchor measurement, though
the Bradley-Terry refinement is less kind,
\Cref{sec:rq3:stack}) is the lone session whose self-claim
lands on the external number. The ranking is not an artefact of one measurement:
where the panel resolves an engine, an independent corpus-wide
round-robin reproduces the same order (Pearson $r=0.94$,
\Cref{app:rr-scatter}). And language is a ceiling rather than a
point. Within Java the external estimate spans \elo{1509}
(\texttt{chess-java}) to \elo{2096} (\texttt{chess-java-cc}), within
Rust \elo{1723} to \elo{1989}, and within Ruby \elo{1346} to
\elo{1753}, while the two \LaTeX{} engines (self-reporting
$\approx$\elo{550} and $\approx$\elo{1300}, both unresolved by the
external panel) sit a full band apart in the same language.

\label{sec:rq4:anchor}%
The headline ranking settled, the rest of \RQ{4} pulls out the
observations the distribution makes possible: whether a strong
design survives a change of language, how Elo tracks the feature set
and the agent's oracle rigor, how the self-Elo claim moves within a
session, and where the corpus floor really sits.

\paragraph{Does a strong design survive a change of language?}
The two Java ports answer this almost in isolation. The agent
translated one engine, \texttt{chess-java-cc}, into Rust and into
COBOL under the \emph{same source design, same agent, same prompt
budget}. The Rust port keeps nearly all of the strength
(\elo{1922~$\pm$~106}, within reach of the scratch-built Rust
engines), while the COBOL port, which never resolved against the
panel, lands at \elo{1404~$\pm$~69} once refined. The design did not
change, only the language did, and a $\sim$\elo{520} gap opened,
with the slower language paying for it (\Cref{app:special-role}).
The two leaders confirm the other half: a CCRL-comparable
solid-club to weak-expert ceiling (\elo{2096~$\pm$~132} and
\elo{1989~$\pm$~105}, the Rust one corroborated by an independent
80-game head-to-head against its Java parent,
\Cref{app:deepdive-rust-redo}) that only compiled mainstream
languages reach.

\paragraph{Does the feature set explain the spread?}
Holding language fixed, engines with richer search pipelines
(aspiration windows, null-move, quiescence, TT) sit above those
without, and the feature-count from \Cref{tab:rq2-searcheval}
correlates positively with the external Elo in the mainstream rows.
We read this as a chain rather than a direct cause: the language
\emph{frames} which features a session can realistically target, and
the advanced ones that lift the mainstream engines (null-move, LMR,
a working transposition table) are exactly what an esoteric or slow
target can rarely afford. PL would then shape Elo partly through the
feature set it makes practical. The correlation stays descriptive,
since feature-count is also correlated with session effort and we
cannot isolate a causal feature effect.

\paragraph{Why does oracle rigor track Elo?}
Engines with higher external Elo also have more \pgn{} records and
more distinct \stockfish{} opponents in their history, up to nine
calibrated levels for \texttt{chess-ruby-cc}. Two effects compound.
More games are more observations, which both tighten the Elo
estimate and give the agent a cleaner gradient to tune against, so
rigor and strength co-produce each other as the agent moves to
harder gauntlets once it clears the easy ones. But running many
games is itself a function of speed: a weak or slow engine plays far
fewer games per hour, so the esoteric and slow PLs that already sit
low collect the least oracle signal, which widens the very gap they
are trying to close.

\paragraph{How does the self-Elo claim move during a session?}
\label{sec:rq4:curve}
Mining the agent's own messages for Elo claims yields five shapes
(\Cref{fig:elo-curves} in \Cref{app:elo-curves}): a steady climb
(4 engines), a plateau (${\sim}4$), a roller coaster (4), a single
claim (5), and silence (9). The roller coaster is the most telling.
The agent posts an Elo, a later run swings \emph{down} after an
optimisation quietly regressed an earlier gain, and the agent
usually catches the dip within one or two prompts and climbs back.
These swings cluster in the stagnation-heaviest sessions
(\texttt{COBOL-chess}, 7 flagged steps), so they double as a
supervision signal: a self-Elo that lurches up and down marks a
self-improvement loop that is actually working, catching and
reverting its own regressions, at the cost of the time and tokens
that surface in \RQ{5}.

\paragraph{Where the floor really sits.}
\label{sec:rq4:gaps}
Seven engines carry no external number, for reasons of interface or
raw speed rather than any failure to play chess. The three CSS/HTML
engines run in a browser and never expose a UCI loop, so the
gauntlet cannot drive them, and \texttt{chess-mojo} is the one
underconverged session, plateaued near \elo{900}. The remaining
three sit at or below the panel's floor: the \LaTeX{} replication
engine is genuinely weak (self-reporting $\approx$\elo{550} and
losing every game it played), while TeXCCChess and the slower
Brainfuck engine are too slow for a tournament clock (${>}30$~s per
move) and forfeit on time. Our search-throughput bench makes the gap
concrete: the fastest C engine searches ${\approx}4.2$\,M nodes/s,
Python ${\approx}42$\,k, and Ruby ${\approx}20$\,k, with the
\LaTeX{} and Brainfuck engines slower still. An esoteric-language
engine can play perfectly legal chess yet search orders of magnitude
too slowly to score against a calibrated opponent, which caps
measured strength without touching the ability to build a working
engine (\RQ{1}).

\paragraph{Model-mastery confounder.} Strength differences between
two engines in the same language and the same session style
(e.g., \texttt{chess-ruby-cc} at $\approx$\elo{1750} vs
\texttt{chess-ruby-codex} at $\approx$\elo{1350} on the external
ruler) may reflect differences in the underlying model's fluency
with the language rather than a property of the language itself. We
flag this in the discussion (\Cref{sec:discussion}) and in the
threats (\Cref{sec:threats}).

\begin{rqanswer}{4}
Re-played against five calibrated external references, the corpus
spans from a few hundred Elo at the floor (\texttt{chess-sql} at
$\approx$\elo{520}, with the no-win \LaTeX{} and CSS engines below
even that) to $\approx$\elo{2096} for \texttt{chess-java-cc}. The
strongest engines are solid-club to weak-expert players in Java and
Rust, and the median engine sits in the \elo{1400}--\elo{1700}
intermediate-club band. The dozen engines too weak to separate
against the panel are resolved by the Bradley-Terry refinement
($^\dagger$ in \Cref{tab:rq4-elo}), which agrees with the panel on
ordering (Pearson $r=0.94$ on the measurable-anchor engines,
\Cref{app:rr-scatter}).

The corpus answer to the title question, on this RQ, is sharp: the
top tier is mainstream-compiled only. Every engine in the
\elo{1900}--\elo{2100} band is Java or Rust, and no specialised,
domain-specific, legacy, or esoteric engine reaches it under
typical-budget conditions. Within a language, two sessions can still
diverge by hundreds of Elo (Java from \elo{1509} to \elo{2096}, Rust
from \elo{1723} to \elo{1989}, LaTeX from \elo{550} to \elo{1300} on
the engines' own scales), so language sets a loose upper bound and
the feature-set and oracle-rigor decisions move the engine under it.
Overall, \textbf{programming languages still matter for \emph{how
strong} an \agentbuilt{} engine can become, even when they no longer
matter for \emph{whether} one can be built at all (\RQ{1}).}
\end{rqanswer}
\FloatBarrier
\section{RQ5 (Cost and effort): does the PL matter for what
the AI teammate has to do?}
\label{sec:rq5}

\paragraph{Question.} What does it take, in user prompts,
tokens, dollars of agent spend, and in the breadth and
intensity of SE activities the agent has to exercise, to reach
a given \agentbuilt{} engine? How difficult was it to reach a
first working engine, \emph{per engine}?

Where \RQ{4} measured \emph{ceiling} differences across PLs,
\RQ{5} measures \emph{cost} differences. The two together are
the impact half of the title question. A coarse summary of the
evidence below is that a mainstream engine takes a
few prompts (median 7) and ${\sim}\$2$--\$115 canonical USD
(median ${\sim}\$40$), whereas a
Brainfuck or COBOL engine takes ${\sim}$25--50 prompts and
${\sim}\$60$--\$480, with debug fractions above 0.4 and
SE-activity intensity covering all 10 categories. PL still
shapes how much work the teammate actually does.


\Cref{tab:rq1-engines} (\cpageref{tab:rq1-engines}) reports
prompts, input and output
tokens, and a canonical USD figure (\emph{Norm.\ USD}) computed
uniformly across the corpus at a fixed reference rate of $\$1$
per million input tokens and $\$5$ per million output tokens
(rationale in \Cref{sec:corpus:evidence}). We use this
canonical figure throughout this section. Vendor-specific
list-price USD, the amount that \claudecode{} and \codex{}
would each invoice under their own cache-economics models, is
preserved in the replication artefact and intentionally kept
out of the body, because the two vendors' list prices do not
align well enough to support cross-engine comparison.

Effort is long-tailed and is driven by oracle iteration
rather than by LOC or by the language alone.
 The engines that
climb to top-tier Elo ran the most gauntlets, the engines that
stagnated against a hard PL accumulated debug prompts, and the
shape holds in tokens and in USD independently as well as
within each vendor.
 We also compute a \emph{debug fraction} per
session: the share of user prompts that ask the agent to fix
something broken rather than to build something new. It
separates smooth sessions from laborious ones. At the high
end, the Ruby \claudecode{} session spent almost half of its
prompts ($\approx$0.44) repairing move-generation bugs that
its own perft runs kept exposing, and the COBOL \claudecode{}
session spent $\approx$0.39 of its prompts fixing crashes and
illegal moves that surfaced during its gauntlet games. At the
low end, engines that compiled and played correctly on the
first attempt needed almost no fixing at all (0.00 for the
Rust \codex{} engine, $\approx$0.05 for the Python \codex{}
one).

\paragraph{Difficulty per language.}
Mainstream general-purpose languages have the cheapest median
engines (a ${\sim}\$40$ median over the nine mainstream
main-corpus rows with cost data, with the cheapest Rust, Java,
and C rows under \$30 and converging in under 30 wallclock
hours).
Legacy and constrained-execution-model rows are dominated by
long sessions with mid-to-high debug fractions, and Brainfuck
(\texttt{chess-brainfuck-cc}) spanned weeks of wallclock time
at a debug fraction around 0.4, the hardest case in the
corpus.
Specialised and academic rows (APL, Icon, Why3, Rocq, Lean) sit
in the middle, with modest hours, modest prompt counts, and low
debug fractions, but they plateau at lower Elo. The validation
infrastructure of \RQ{3}~(\Cref{tab:eval-infra}) is the
underlying cost driver: cost correlates more tightly with the
number of oracle iterations than with LOC, language, or final
strength.

\subsection{Where did the cost come from? The SE activities
behind each engine}
\label{sec:rq5:se-activities}

The cost table reports how much each session cost and the
eval-infrastructure table reports \emph{what kind} of oracle the
agent set up. Both stop short of saying \emph{what the agent
actually did step by step}: how much of the session was coding
versus debugging, how much was reading versus running perft,
whether build, test, and benchmark tooling was exercised at
all. We therefore decompose each session's trajectory against
the 10-capability SE-activity frame described in
\Cref{sec:corpus:se-activities} and report the per-engine
intensity profile. This is the finer-grained answer to
\emph{``where did the cost come from?''}. It also cross-cuts the
earlier RQs. The per-engine \emph{Impl} column is the activity
footprint behind the features of \Cref{sec:rq2}. Broad coverage
of \emph{Impl}, \emph{Debug}, \emph{Test}, and \emph{Read} is
direct evidence against any copy-and-translate reading of
\Cref{sec:rq3}. Intensity in \emph{Debug} and \emph{Bench}
foreshadows the strength spread of \Cref{sec:rq4}.

Each engine yields a 10-dimensional profile summarised on a
4-level Likert scale (absent, light, moderate, or heavy, based
on the share of session steps tagged with each capability,
with the methodology in \Cref{app:method:se-activities}). The
full per-engine matrix and a per-category summary are in
\Cref{tab:rq1-se-matrix} (\Cref{app:se-matrix}).

\paragraph{Six capabilities are near-universal.}
Across the 33 engines with recoverable trajectories,
\emph{implementation}, \emph{code comprehension}
(33/33), \emph{debugging} (31), \emph{testing} (29),
\emph{benchmarking \& evaluation} (28), and
\emph{build \& tooling} (23) all appear in a majority of
engines in \emph{every} language category. A chess-engine
session is not a pure coding task even in the easiest
targets: the agent writes or edits code, runs and inspects
existing code, runs perft and UCI smoke tests, plays
Stockfish gauntlets, builds the project, and commits,
regardless of whether the target is Python or Brainfuck. The
oracle-first iteration style surfaced under \RQ{3}
(first perft at step~2, first gauntlet at step~1) is the
early temporal signature of this population-wide pattern.

Reading the matrix row-wise, the high-intensity rows almost
always correspond to the high-USD rows of
\Cref{tab:rq1-engines}, which is the per-activity version of
the cost-as-oracle-iteration pattern. Excluding the
underconverged \texttt{chess-mojo} case (kept in the matrix for
transparency but excluded from main-corpus comparators,
\Cref{app:special-role}), \texttt{chess-rust-codex} tops the
mainstream category with 8 of 10 capabilities exercised
heavily, and \texttt{chess-apl-codex54} leads the specialised
category with 9 of 10 and a summed intensity of 25. The spread within a
single category is wide, with mainstream engines ranging from a
summed intensity of 3 (archived-transcript engines, see
\Cref{app:archived-floors}) to 24. Column-wise, \emph{Impl},
\emph{Test}, \emph{Debug}, \emph{Read}, \emph{Build}, and
\emph{Bench} carry non-absent cells in essentially every row.
\emph{Perf} is sparse, since performance vocabulary fires
mainly in legacy and esoteric targets. \emph{Design} and
\emph{Docs} are the least saturated columns, because the
capability-level prompt already serves as the plan. Across
categories, mainstream engines average 6.6 distinct
capabilities and esoteric engines 10.0, so what distinguishes
the bottom categories is less \emph{which} capabilities fire
than how intensely each one does. Building a working chess
engine across this corpus therefore requires software
engineering rather than coding alone. The language category
does not change \emph{whether} the agent interrogates its own
code, runs oracles, fixes its own bugs, and re-runs the build
and test cycle. It only changes \emph{how much} of that work is
needed.

\subsection{Where did the effort go, and was it well spent?}
\label{sec:rq5:anatomy}

\paragraph{How the effort unfolds.}
We segment each transcript into steps (one user prompt
together with the agent work it triggers, as in
\Cref{sec:corpus:se-activities}), label every step with a task
class such as feature work, debugging, or exploration, and
group consecutive steps of the same class into phases (the
annotation pipeline is part of the replication artefact). The
median session is compact, with 10 user prompts, 2.3 wallclock
hours, and 56~M pooled tokens. Per-engine totals run higher,
since an engine often spans several sessions: the median
main-corpus engine takes 14 prompts, and the totals
stretch to 55 prompts for \texttt{chess-rust-cc-redo} and
almost half a billion pooled tokens for \texttt{COBOL-chess}
and the Java$\to$Rust port (\Cref{tab:rq1-engines}). Feature
work accounts
for roughly 31\% of the median session's steps and open-ended
exploration for roughly 35\%, and the most common phase
sequence alternates between the two: the agent explores,
builds, and explores again. Much of the remaining effort, and
most of the effort in the expensive sessions, goes into the
validation loop. The median engine runs perft and a gauntlet
within its first two prompts (\Cref{sec:rq3:first-call}), and
the five most expensive engines of \Cref{tab:rq1-engines}
(\texttt{COBOL-chess}, the two Java ports,
\texttt{chess-brainfuck}, and \texttt{chess-cobol-cc}, from
$\approx$\$180 to \$475 canonical USD each) are precisely the
ones that kept re-entering that loop: play games, read the
outcome, decide what to change.

\paragraph{The reward signal that steered the spend.}
\label{sec:rq5:reward}
Every pass through that loop was scored by the agent's own
strength estimate, the self-Elo of \RQ{3}. That estimate is
biased upward by \elo{200} to \elo{1100}, for the
methodological reasons dissected in \Cref{sec:rq3:first-call}
and quantified corpus-wide by the self-reproduction sweep of
\Cref{app:anchor-eval:q2}. The agent used the same estimate to
decide what to keep, what to revert, and when to stop, so the
self-Elo is not a reporting detail but the criterion that
allocated the session budget. This has two cost consequences.

First, sessions stop too early. \texttt{chess-purec-codex}
halted after 4 prompts and \$22.25 of canonical spend because
the agent reported \elo{1670}--\elo{1972}, while the external
assessment is $\approx$\elo{1400}, so the user accepted an
engine several hundred Elo weaker than reported in the belief
that the target had been reached. Stopping is the single largest
cost decision in a session, and it belongs to the developer,
not to the agent. It is well-founded only when the evidence
that the goal has been reached is itself trustworthy.

Second, value-for-money comparisons mislead. The bias differs
by hundreds of Elo from engine to engine ($+\elo{141}$ to
$+\elo{681}$ across the sweep), so a ranking computed from
self-claims does not survive re-measurement. Under the agents'
own numbers, \texttt{chess-purec} (claim \elo{1997}) outranks
\texttt{chess-rust-cc-redo} (claim \elo{1879}). Under the
external assessment (unified Bradley-Terry ruler), the order
reverses by almost \elo{600}
(\elo{1415} versus \elo{1990}), and any Elo-per-dollar
comparison across engines, languages, or agents inherits such
reversals.

The misallocation is also cheap to undo. A single controlled
retry of the highest-bias engine (\texttt{chess-purec},
restarted from the same commit with the same model and a
30-prompt budget, with only the self-Elo loop swapped for a
gauntlet against two externally calibrated anchors) moved the
engine from \elo{1415~$\pm$~65} to \elo{1798~$\pm$~82} for
about \$30, roughly \$0.08 per Elo point. We treat this single
run ($n{=}1$) as supporting evidence rather than a
corpus-level result, and the full protocol is in
\Cref{app:anchor-eval:retry}. The broader lesson is that
developers are not powerless against wasted spend. Equipping
the session with a trustworthy signal, as the retry did, costs
little, but it presupposes that the developer masters the
correctness oracles, the benchmarking methodology, and enough
of the PL and its tooling to challenge implausible claims
before more budget is burned.

\begin{rqanswer}{5}
Individual engines take between a handful of prompts at a few
dollars and dozens of prompts at several hundred dollars, with
cost driven primarily by oracle-loop iteration rather than
language category alone. Difficulty per language is best read
as a distribution of per-engine values, and the most
consequential cost finding concerns the reward signal
(\Cref{sec:rq5:reward}): the Elo the agents measured for
themselves overstates the Elo an external evaluation measures.
Because the agents used that inflated signal to decide which
changes to keep and when to stop, part of the spend went into
changes that did not improve real strength, and some sessions
ended while large and cheap gains were still on the table.
Keeping the cost justified therefore remains a developer
responsibility. Choosing the PL sets the size of the bill, and
validating the agent's success claims decides whether the
spend stops for the right reason. Both call for
software-engineering judgement, about the oracles, the
benchmarking methodology, and the language itself, that the
agents in this corpus did not supply on their own.

Overall, \textbf{programming language still moves the cost and the SE-workload of \agentbuilt{}
construction by orders of magnitude. A mainstream
engine costs ${\sim}\$2$--\$115 canonical USD (median
${\sim}\$40$) and takes a
handful of prompts, whereas a Brainfuck or COBOL engine costs
${\sim}\$60$--\$480, takes ${\sim}$25--50 prompts, and carries a large fraction of
debugging and editing steps. PL choice is no longer
an existence question, but it is firmly an effort-and-cost
question}.
\end{rqanswer}
\FloatBarrier
\section{Threats to validity}
\label{sec:threats}

This is an exploratory empirical field study. Field studies observe
and characterise, and they do not support causal claims. The
confounders documented below collectively make causal claims about
programming-language effectiveness impossible with this single-run,
observational corpus, and together they specify what a follow-up
controlled experiment would have to hold fixed. The existence proofs,
synthesis evidence, and session-behaviour findings of
\Crefrange{sec:rq1}{sec:rq5} stand independently of the causal
questions this section leaves open. We follow the standard taxonomy
and discuss construct, internal, external, and conclusion validity in
turn.

\paragraph{Construct validity.}
Elo as a construct assumes a reasonably stable comparison against
\stockfish{}. The PGN-mined estimates each session produces vary in
time control (from 10s$+$0.1s to 120s$+$1s), opponent calibration
(\texttt{UCI\_LimitStrength} vs.\ actual node limits), and per-game
length, so they are not strictly comparable across engines and tend
to overestimate strength. The \texttt{chess-java-cc} session is
illustrative: the agent self-claimed $\approx$\elo{2200} from its
own 40-game gauntlets, whereas the engine settles at
\elo{2096~$\pm$~132} under the canonical Phase~B harness and
\elo{2094~$\pm$~56} under the unified Bradley--Terry model, still
the strongest engine in the corpus but roughly \elo{120} below the
self-claim. We mitigate this threat by taking every headline Elo
number in the body from the unified canonical harness
(\Cref{sec:rq4:anchor}, methodology in \Cref{app:anchor-eval}), which
plays all engines at a single time control of 120s$+$1s against
externally rated CCRL anchors, and by citing the agents' in-tree
estimates only for triangulation, never as a primary ruler.

Feature detection is regex-based, so unusual languages may
under-report features and depress the apparent feature counts of
precisely the most exotic corpus rows (\Cref{sec:rq2}). We mitigate
this by manually auditing the outliers.

Cost reporting depends on a rate card. The body uses a canonical
Norm.\ USD computed uniformly from observed token volumes at a fixed
reference rate of \$1 per million input tokens and \$5 per million
output tokens, ignoring vendor-specific cache discounts
(\Cref{sec:corpus:evidence}). The consequence is that absolute
dollar values must not be read as invoiced spend. The mitigation is
that the figure is intra-corpus comparable by construction, so the
relative ordering across engines is reliable, and the
vendor-specific list-price USD is available in the replication
artefact for transparency. We deliberately do not use list price for
cross-engine comparison because Anthropic's reduced cache-read rate
dominates \claudecode{} sessions while \codex{} sessions score
tokens primarily at the standard input rate, so the two invoices do
not represent the same underlying compute.

A handful of repositories contain transcripts from both
\claudecode{} and \codex{}, which would inflate per-engine prompt,
token, cost, and SE-activity aggregates if both sessions were
counted as development effort. Inspection shows the secondary agent
was always invoked post-mortem to review or annotate the repository,
with no engine-core writes. We mitigate the threat by excluding
post-mortem sessions from all per-engine aggregates, reporting only
the primary developer agent in \Cref{tab:rq1-engines} and flagging
affected rows with $\flat$. The detection rule (no engine-core
writes, at most five prompts, chronologically following the primary
session) is conservative, and reviewers can re-classify borderline
sessions via the replication artefact.

\paragraph{Internal validity.}
Elo numbers reflect the performance of a specific binary compiled
with specific flags by a specific compiler on specific hardware. All
development and benchmarking took place on macOS (Apple Silicon),
and at least one C engine was built with platform-specific
optimisations that increase search throughput on that hardware. The
consequence is that cross-engine Elo comparisons implicitly compare
full (language $\times$ compiler $\times$ flags $\times$ OS $\times$
CPU architecture) stacks, and we cannot separate ``this language
enables high-throughput search'' from ``this binary was compiled
with aggressive platform-native optimisations''. The canonical
re-evaluation (\Cref{app:anchor-eval}) mitigated the
measurement half of this threat: every engine played under the same
time control, on the same machine, against the same Docker-pinned
anchor opponents. The build half remains, since each binary keeps
the compiler and flags its session produced.

The volume and tone of follow-up prompts also varied per session,
which threatens internal validity because a human in the loop could
have steered some sessions more than others and thereby injected
methodology rather than mere supervision. The variation is in part
intrinsic to the paper's open-ended brief, since ``what does an AI
teammate produce given a capability-level goal?'' admits different
prompt budgets, and in part a consequence of basic interventions
needed across PLs of unequal ecosystem maturity, such as installing
a missing toolchain, pasting an error trace, or nudging with
``improve the Elo''. To bound the threat we audited all 374
follow-up prompts across 24 corpus engines and found zero
protocol-deviating prompts (named-algorithm or named-feature
directives) once turn-context provenance is taken into account, so
the meta-protocol held even though the volume varied. The audit
script, the per-prompt classification, and the turn-context audit
notes are released with the replication artefact
(\hyperref[sec:data-availability]{Data and code availability}).
The residual gap we cannot fully close is qualitative tone, such as
encouragement versus scepticism within the same prompt class. A
fully automated, human-out-of-the-loop replication, in which an LLM
supervisor deterministically emits the canonical prompt schedule
(\Cref{fig:protocol}, released as \texttt{PROTOCOL.md} with the
artefact), would close this gap structurally. The same document
catalogues the pitfalls such a supervisor must avoid.

The SE-activity frame, the feature catalogue, and the
qualitative audits were coded by a single human rater, the
first author, and we report no agreement statistic between
independent human raters (\Cref{app:method:qualitative-pilot}).
 However, reliable coding here requires joint
expertise in chess-engine internals and in reading long agent
transcripts, which made recruiting a comparable second rater
impractical within this study. The cross-checks we do have are
of two kinds. The per-step tagging rules are deterministic
predicates over released structured records
(\Cref{sec:corpus:se-activities}), so any reader can re-code
the corpus and diff the result. In addition, the analysis
agent independently drafted a report for every engine, and
the first author adjudicated every divergence between the
report and his own reading (\Cref{app:llm-usage}), so a
human-versus-LLM cross-check was exercised corpus-wide even
though a human-versus-human statistic was not.

\paragraph{External validity.}
The corpus is a convenience sample. 34 \agentbuilt{} engines
authored by one developer over about two and a half months (2026-02
to 2026-04) are not a random sample of agent--language pairs, so any
generalisation risks overreach. We mitigate this by scoping claims
to ``what a motivated developer gets with a few thousand dollars of
frontier agent spend'' (${\approx}\$3{,}100$ canonical USD summed
over the 27 engines of \Cref{tab:rq1-engines} with recoverable
cost data, special-role experiments included). Data was collected on \opus{4.6}, \opus{4.7},
\gpt{5-codex}, \gpt{5.3-codex}, and \gpt{5.4}, and later versions may shift feature
choices and achievable Elo. All results also come from
closed-source commercial agents on proprietary frontier models, and
effective context-window size, prompt-cache infrastructure, and
polyglot fluency at the system-construction scale are reasons
open-weight parity may not hold, so findings should not be
extrapolated to open-weight models and open harnesses without direct
measurement (\Cref{sec:discussion} frames this as a measurable
follow-up for which our artefacts provide ground truth). We
deliberately did not run a one-engine open-weight baseline inside
this study, because a credible baseline needs the full multi-week
protocol in at least one language, and a shallow attempt would
invite exactly the cross-ruler comparisons this section warns
against. Relatedly,
agents may have different fluencies across languages depending on
training-data volume, and this design cannot separate language
effects from training-data-fluency effects (\Cref{sec:discussion}).
The output-side counterpart of this confound is memorisation. Chess
engines are abundant in public code, so an agent could reproduce a
known engine rather than synthesise one, which would contaminate
both novelty and strength claims. The novelty audit
(\Cref{sec:rq2:audit}) mitigates this with four independent
signals, and across the corpus it flagged exactly one engine-core
fingerprint match and one library-assisted engine
(\Cref{sec:rq3:evasion}).
Finally, every session terminated at a human-judged plateau and
per-engine investment varied widely (${\sim}\$1.5$ to
${\sim}\$474$ canonical USD), so reported
Elo values are lower bounds and cross-language comparisons conflate
language properties with investment level. The canonical
re-evaluation removed the measurement-strategy part of this
conflation by replaying every engine under one protocol, but the
budget part remains.

\paragraph{Conclusion validity.}
We compute weighted-average Elo with inverse-variance weighting and
report 95\% confidence intervals. For small-$N$ engines with fewer
than 20 games per opponent the intervals are wide enough that
per-engine ordering should not be over-read. Feature-vs-Elo
correlation analyses are descriptive, and we do not apply causal
inference. Each engine is also \emph{one draw} from a high-variance
distribution over the plans the agent might have produced (LLM
sampling, plan-selection randomness, sequential tool-call history).
With one to three sessions per language, cross-language orderings
are suggestive at best, and language-level claims in this paper are
about tendencies across the observable distribution rather than
about individual sessions. We also did not re-run sessions under a
repeated-measures design. A session costs tens to hundreds of
dollars plus a deep qualitative pass, so the budget bought breadth
across 34 agent-language pairs rather than repetition within each
pair. That trade-off matches the form of the headline claims, which
are existence claims: agent $A$ produced a working engine with
feature set $F$ at strength $E$ in language $L$, and a single
successful session establishes such a claim.
 Besides, the repetition we
do have operates at two levels. Every canonical Elo value
aggregates hundreds of repeated games with confidence intervals,
and the one controlled session-level retry, \texttt{chess-purec}
(\Cref{app:anchor-eval:retry}), quantifies how much a single
session-level variable can matter, since changing only the reward
signal moved externally measured strength from \elo{1415~$\pm$~65}
to \elo{1798~$\pm$~82}.

\section{Related work}
\label{sec:related}

\paragraph{Positioning at a glance.}
\Cref{tab:related-positioning} places the present study against the
major evaluation paradigms for LLM-generated code. Three dimensions
matter for our question. (i)~\emph{Task granularity}:
whether the agent synthesises a function, repairs a single issue,
edits an existing file, or designs and builds a complete system from
a blank slate. (ii)~\emph{Language spectrum}: how many languages,
and whether coverage extends beyond mainstream general-purpose
targets. (iii)~\emph{Oracle family}: whether correctness is
checked by a pre-written test suite, a binary build-passes criterion,
or an external language-agnostic performance oracle. To our knowledge,
the combination of full-system construction, a 15-language spectrum
that reaches strictly esoteric targets, and a cardinal
external performance oracle is not matched by any prior work.
Individual dimensions are each well studied, but not together. We
read \Cref{tab:related-positioning} as a map of adjacent regimes, not
a ranking.

\begin{table*}[t]
\centering
\small
\caption{Positioning of the present study among code-evaluation
benchmarks and coding-agent demonstrations. \textbf{Task}: Func.\ =
function-level synthesis, Issue = single-issue repair in an existing
repo, Edit = code editing, System = design-and-build a complete
system from scratch. \textbf{Spectrum}: M = mainstream only, M$^+$ =
mainstream plus a few adjacent targets, Wide = reaches esoteric or
constrained-execution targets. \textbf{Oracle}: unit tests, repo test
suite, binary build-passes, or cardinal external performance. ``This
paper'' is the only row combining System-scope construction with a
Wide spectrum and an external cardinal oracle.}
\label{tab:related-positioning}
\resizebox{\textwidth}{!}{%
\begin{tabular}{lllll}
\toprule
Benchmark / study & Task & \#Langs & Spectrum & Oracle \\
\midrule
HumanEval~\cite{chen2021humaneval}                       & Func.  & 1   & M    & unit tests \\
MBPP~\cite{austin2021mbpp}                               & Func.  & 1   & M    & unit tests \\
LiveCodeBench~\cite{jain2024livecodebench}               & Func.  & 1   & M    & unit tests \\
MultiPL-E~\cite{cassano2023multipl}                      & Func.  & 22  & M$^+$ & unit tests \\
HumanEval-XL~\cite{peng2024humanevalxl}                  & Func.  & 12  & M$^+$ & unit tests \\
CRUXEval-X~\cite{xu2025cruxevalx}                        & Reason & 19  & M$^+$ & reference behaviour \\
SWE-bench family~\cite{jimenez2024swebench,yang2025swebenchmm,badertdinov2025swebrebench}
                                                         & Issue  & 1--2 & M    & repo tests \\
Multi-SWE-bench~\cite{zan2025multiswebench}              & Issue  & 7   & M    & repo tests \\
Aider Polyglot~\cite{gauthier2024aiderpolyglot}          & Edit   & 6   & M    & Exercism tests \\
DevBench~\cite{li2024devbench}                           & Lifecycle & few & M & mixed \\
Commit0~\cite{zhao2024commit0}                           & Repo (fill) & 1 (Python) & M & repo tests \\
DevEval~\cite{li2024deveval}                             & Repo (gen) & 1 (Python) & M & repo tests \\
Endoh~\cite{endoh2026bench}                              & Small task & 13 & M$^+$ & unit tests \\
Carlini/Anthropic compiler~\cite{carlini2026compiler}    & System & 1 (Rust) & M & binary (Linux boots) \\
METR time-horizon study~\cite{kwa2026measuringaiabilitycomplete}
                                                         & Mixed  & ---  & M$^+$ & auto + human time-baseline \\
ProgramBench~\cite{yang2026programbench}
                                                         & System (recover) & any & M & 248k auto-fuzzed tests \\
\midrule
\textbf{This paper}                                      & \textbf{System} & \textbf{15} & \textbf{Wide} & \textbf{cardinal Elo + perft} \\
\bottomrule
\end{tabular}%
}
\end{table*}

\paragraph{Code evaluation benchmarks.}
The dominant paradigm uses bounded, unit-test-graded tasks:
function-level synthesis (HumanEval~\cite{chen2021humaneval},
MBPP~\cite{austin2021mbpp},
LiveCodeBench~\cite{jain2024livecodebench}), repository-level
issue repair on Python (SWE-bench~\cite{jimenez2024swebench}
and its family
SWE-bench Multimodal~\cite{yang2025swebenchmm},
SWE-smith~\cite{yang2025swesmith},
SWE-rebench~\cite{badertdinov2025swebrebench}), and the
lifecycle-broadest entry DevBench~\cite{li2024devbench}. The
task is bounded, the oracle is a pre-written test suite, and
the language is fixed.

\paragraph{Coding-agent architectures.}
Li et al.~\cite{li2025riseaiteammatessoftware} frame this
generation of tools as SE~3.0, in which autonomous coding
agents work as teammates under human supervision.
SWE-agent~\cite{yang2024sweagent} established the importance
of agent-computer interfaces.
AutoCodeRover~\cite{zhang2024autocoderover} adds AST-aware
search and fault-localisation.
Agentless~\cite{xia2024agentless} shows simpler
localise$\to$repair$\to$validate pipelines can match richer
orchestration. RepoCoder~\cite{zhang2023repocoder} introduces
iterative retrieval-generation loops for repository
completion. USEagent~\cite{applis2025useagent} examines when
patching scaffolds become ``AI software engineers''.
Riddell et al.~\cite{riddell2026multihop} characterise the
reasoning failures of {LLM} agents performing root-cause
analysis in cloud systems.

\paragraph{Multi-agent frameworks and long-horizon
construction.}
Role-decomposed multi-agent frameworks
(ChatDev~\cite{qian2024chatdev},
MetaGPT~\cite{hong2024metagpt},
OpenHands~\cite{openhands2024}) demonstrate prototype-scale
end-to-end software artefacts. The closest from-scratch
system-construction precedent is Carlini and Anthropic's
parallel-Claudes
experiment~\cite{carlini2026compiler}: 16~Claude instances
collaborating through git built a dependency-free Rust C
compiler bootstrapping Linux~6.9.

\paragraph{Long-horizon agent capability.}
The METR time-horizon study~\cite{kwa2026measuringaiabilitycomplete}
measures the same frontier-agent capability we exploit, on an
orthogonal axis: \emph{how long} a realistic software-style task
an autonomous agent can complete with 50\% reliability against
800$+$ human baselines, reporting an ${\approx}$7-month doubling
of the frontier 50\%-time-horizon. The parallel
MirrorCode~\cite{adamczewski2026mirrorcode} effort from Epoch~AI
extends the long-horizon claim from hours to weeks-scale coding
tasks. They calibrate the long-horizon autonomy that makes our
question well-posed: once agents reliably handle multi-hour
software tasks, asking \emph{which} programming language the agent
should use becomes empirically tractable. We hold the domain fixed
(chess engines) and vary the language. METR/MirrorCode hold the
language paradigm fixed and vary task duration, making the two
lines of work complementary rather than competing.

\paragraph{From-scratch system reconstruction (complementary).}
A natural complement to our paper is
ProgramBench~\cite{yang2026programbench} (Meta/Stanford/Harvard,
May~2026): agents must rebuild 200 existing programs
(\texttt{FFmpeg}, \texttt{SQLite}, \texttt{PHP}, \dots) from
binary~$+$ documentation alone (internet and original source
blocked, execute-only permissions on the gold binary), scored by
${\sim}248{,}000$ auto-fuzzed behavioural tests. The best frontier
model fully resolves 0/200 tasks and clears 95\% of tests on only
3\% of tasks. The two studies attack the same from-scratch
construction problem from opposite ends. ProgramBench
\emph{hides} the specification (the agent must recover behavior
from a binary), \emph{leaves the implementation language to the
agent} (which uniformly picks one mainstream compiled target),
and runs under an \emph{adversarial sandbox} (no internet, no
source repository, judge-watched, \$9 per-task budget) engineered
to control source leakage. Our paper inverts all three: the
specification is \emph{publicly canonical and backed by strong
external oracles} (the rules of chess, plus \perft{} and Elo), the program is held fixed and the \emph{language is
varied across 17 PLs} so that polyglot from-scratch construction
becomes the dependent variable, and the corpus is built in a
\emph{naturalistic} setting (open internet, multi-prompt
sessions, organic human supervision) that mirrors how developers
actually use coding agents today. The source leakage
ProgramBench's sandbox engineers away is therefore exactly what our
four-signal novelty audit (\Cref{sec:rq2:audit}) characterises in
the wild. ProgramBench and our paper both extend the earlier
single-language from-scratch benchmarks,
Commit0~\cite{zhao2024commit0} (54~Python library skeletons,
fill-in-the-blanks against the original test suite) and
DevEval~\cite{li2024deveval} (manually-annotated Python
repository-level generation), in different directions: hidden-spec
generality vs.\ known-spec polyglot reach.

\paragraph{Polyglot capability of coding agents.}
Code-LLMs are trained on hundreds of languages
(StarCoder2~\cite{lozhkov2024starcoder2} via The Stack~v2 at
600+) but evaluations show large language-dependent gaps:
MultiPL-E~\cite{cassano2023multipl} (22~languages),
HumanEval-X / CodeGeeX~\cite{zheng2023codegeex},
HumanEval-XL~\cite{peng2024humanevalxl} (12~PLs).
\citet{cassano2025multiplt} report transfer-learning gains on
low-resource languages. Cross-lingual
transfer~\cite{baltaji2025crosslingual} and
CRUXEval-X~\cite{xu2025cruxevalx} document Python bias even in
code reasoning, and
\citet{boruchgruszecki2026agnostics} use language-agnostic
verifiers as a low-resource answer.
Multi-SWE-bench~\cite{zan2025multiswebench} and the parallel
SWE-PolyBench~\cite{rashid2025swepolybench} extend issue repair
to several mainstream languages.
Aider Polyglot~\cite{gauthier2024aiderpolyglot} evaluates
editing across six, and
Endoh~\cite{endoh2026bench} measures agent iteration cost
across 13~languages on unit-test pass/fail.

\paragraph{Benchmark quality and grey-literature critiques.}
Koohestani et al.~\cite{koohestani2025benchmarking}
consolidate 100$+$ SE benchmarks into a quality framework
(task-fidelity, oracle quality, language coverage,
contamination risk, reproducibility) and report systematic
gaps. KatanaQuant's grey-literature
critique~\cite{katanaquant2025correctness} contrasts an
LLM-built Rust SQLite re-implementation against reference~C
SQLite, reports ${\sim}20{,}000\times$ slowdowns, and argues
that frontier LLMs are sycophantic plausibility-matchers
(citing METR, GitClear, Mercury, BrokenMath, DORA~2024).

\paragraph{Chess as a testbed for AI.}
Chess has been an AI testbed since Shannon~\cite{shannon1950}.
Modern engines are hybrids of handcrafted evaluation and
neural networks (AlphaZero~\cite{silver2017alphazero},
NNUE~\cite{nasu2018nnue}).
This maturity cuts both ways: reference engines and chess
libraries abound, so a from-scratch claim cannot be taken on
faith. We therefore audited the novelty of every engine in
the corpus, with the four-signal audit method described in
\Cref{sec:corpus:evidence}, the accompanying supervision
protocol in \Cref{sec:corpus:supervision}, and the results
reported in \Cref{sec:rq2:audit}.

\Cref{tab:related-positioning} summarises our study positioning.
To our knowledge, no prior work combines
from-scratch multi-component system construction with a
language spectrum that reaches strictly esoteric targets together
with an external performance oracle.

\section{Discussion, conclusion, and future work}
\label{sec:discussion}
\label{sec:conclusion}

Programming languages (PLs) still matter to your AI coding agent
teammate, but the way they matter has shifted, and that
shift is the empirical contribution of this paper. With
frontier coding-agent teammates, every PL
in our taxonomy turned out to host a working chess engine,
including languages no human had ever publicly used to build
one. The corpus contains, to our knowledge, the first
chess engines of comparable scope ever written in pure-TeX,
pure-CSS, APL, Icon, Lean~4, and Why3/Rocq, plus an x86-64
\uci{} assembly engine without an open-source predecessor we
could identify, and
Brainfuck and COBOL engines well beyond any prior hobby
artefact in those languages. This is more than a quantitative
extension of the polyglot evaluation
literature~\cite{cassano2023multipl,zan2025multiswebench,gauthier2024aiderpolyglot}.
It is a separate empirical statement that long-tail and
supposedly unsuited PLs have stopped being a coverage
barrier for AI-teammate-built non-trivial systems.
Niche-language and legacy-rewrite scenarios previously
dismissed as uneconomic become now an option. The box below recapitulates the
findings behind this answer.

\begin{tcolorbox}[
  breakable,
  colback=gray!5, colframe=gray!60!black, boxrule=0.4pt,
  title=\textbf{Findings at a glance},
  fonttitle=\bfseries\footnotesize,
  sharp corners, left=6pt, right=6pt, top=4pt, bottom=4pt,
]
\small
\begin{itemize}[leftmargin=1.2em,itemsep=2pt,topsep=2pt]
  \item \textbf{Polyglot coverage.} Frontier agents produce
    working engines across all five language categories
    (mainstream, specialised, domain-specific, legacy, esoteric).
  \item \textbf{Ceiling, not destiny.} Within the same PL,
    engines produced by different agents or sessions land
    hundreds of Elo rating apart. Language sets a loose upper bound and
    feature-set/oracle-rigor decisions move the engine under it
    (\RQ{2}, \RQ{4}).
  \item \textbf{Genuine synthesis, no copy.} 27 of the 29 main-corpus
    engines pass a four-signal novelty audit (no engine-core
    library import, no canonical fingerprint, no self-reported
    copy claim). Feature-level analysis on quiescence, transposition
    tables, and evaluation shows a shared canonical skeleton
    with substantial sub-feature and PL-idiomatic
    divergence rather than textual copy (\RQ{2}).
  \item \textbf{Agents wire up their own validation, without
    being asked.} Move-generation correctness checks (\perft{})
    and strength-test games against \stockfish{} appear within
    the first one or two developer--agent exchanges of the
    typical session (\RQ{3}).
  \item \textbf{Self-evaluation is biased upward.} Most agents'
    in-loop self-Elo overestimates the externally measured Elo
    by \elo{200} to \elo{1100}, a weakness of the benchmarking
    setup the agents converge to on their own, with a magnitude
    that varies from engines and PLs. Every per-PL strength
    and cost figure in this paper therefore rests on our
    external, language-agnostic re-evaluation, and bringing
    that supervision into the agent's own loop is where developers' domain knowledge can pay off (\RQ{3}--\RQ{5}).
  \item \textbf{Cheating is real and language-shaped.} A CSS engine
    that silently used \texttt{python-chess} backed by qualitative
    supervision, and a Brainfuck task with a tempting Python detour
    that the agent refused in the \claudecode{} session (the
    \codex{} Brainfuck session kept part of its logic in Python,
    \Cref{app:github-repos}). Evasion is a language-constraint
    failure mode that needs explicit catching (\RQ{3}).
  \item \textbf{Strength scales with language category.} The top
    tier (\elo{1900}--\elo{2100}) is mainstream-compiled only,
    the mid-band (\elo{1300}--\elo{1700}) mixes interpreted
    mainstream, formal-methods, legacy, and DSL engines, and
    the floor (\elo{<\!700} or no measurable rating) is held by
    APL, SQL, and the markup engines. Academic/specialised
    PLs (Lean~4, Why3, Rocq) follow a different curve: fast
    to a working engine, low ceiling on Elo (\RQ{4}).
  \item \textbf{Cost scales with oracle iteration, not LOC.}
    Mainstream engines settle in a few prompts and tens of
    millions of tokens. Specialised, legacy, and esoteric
    engines need tens of prompts and hundreds of millions of
    tokens, with debugging dominating the tail (\RQ{5}).
\end{itemize}
\end{tcolorbox}

\subsection{What the language effect is, and is not}
\label{sec:discussion:answer}

Once every attempted PL proves reachable, the question becomes
\emph{what matters when choosing one?}. The corpus exposes four
PL-correlated axes (ceiling Elo, cost multiplier,
software-engineering workload shape, and verification
affordance) and supplies worked examples of each. Two readings
the data does not support deserve to be ruled
out. \textbf{Language does not determine Elo} on its own:
within the same language we see hundreds of Elo of variance
(\elo{1509}--\elo{2096} in Java and \elo{1723}--\elo{1989}
in Rust on the external ruler, \elo{550}--\elo{1300} in
\LaTeX{} on the engines' own scales), which we cannot decompose into deliberate feature
selection versus session-level stochasticity (LLM sampling,
sequential plan divergence). \textbf{Nor are language and
platform separable} from one another: ``language'' here is
shorthand for the full (language, compiler, optimisation
flags, OS, CPU architecture) stack, part of what reads as a
language effect may be a training-data fluency effect of the
specific LLM (\Cref{sec:threats}), and budget and
benchmarking strategy add two more confounders that the
corpus does not control. The better reading is
\emph{output quality $\approx$ f(language $\times$ platform,
feature set, oracle rigor, human steering, budget, session
randomness)}, with the factors correlated rather than
independent.

Within those caveats, the language ceiling is real and large.
The search-throughput bench of \RQ{4} spans about two orders of
magnitude, yet the Python engine still reaches the
intermediate-club band (${\approx}$\elo{1600}) while searching
roughly $100\times$ slower than the fastest compiled engines,
showing that algorithmic quality (aspiration windows, null-move
pruning, transposition tables) substantially compensates for raw
throughput within limits. The three-feature analysis of
\Cref{sec:rq2:subfeature,sec:rq2:qsearch} adds the
structural lens: a convergent algorithmic skeleton, divergent
per-PL idiom, and freely varying numerical constants, the
strongest evidence that the agents reproduce algorithmic
structure from a shared specification~\cite{CPW} while
generating the data layer independently.

\subsection{What the corpus does \emph{not} reach: the gap to
top-level engines}
\label{sec:discussion:ceiling-gap}

\emph{None} of the engines in this corpus reach the strength
of top-level chess engines. The corpus ceiling sits at
${\sim}$\elo{2100}, whereas a CCRL top-tier engine plays at
${\sim}$\elo{3500+} and Stockfish at full strength several
hundred Elo above that. We see four compounding reasons and
read this gap as a concrete future-work agenda.
\textbf{(i)~Stopping policy.} Sessions ended at a user-judged
plateau under a reasonable budget, and we did not push for
GM-strength play because that was not the experiment's
target. \textbf{(ii)~Prompting strategy.} The canonical first
prompt names ``Elo rating'' as a goal but not a target, and
the typical follow-up is ``improve the Elo'' rather than
``reach \elo{2800}''. Stronger incentives plausibly extract
more strength per dollar, but we did not measure this.
\textbf{(iii)~PL ceiling.} Several languages (Brainfuck,
LaTeX, CSS, SQL) sit well below top-tier strength under any
budget. \textbf{(iv)~Top-tier engineering needs sustained
expert interaction.} A modern ${\sim}$\elo{3500+} engine
reflects decades of chess-engine knowledge (bitboard tricks,
NNUE training, Syzygy tablebases, search-tree-shaping
heuristics tuned over millions of self-play games) plus
expert steering. Whether an AI teammate plus an expert can
close the gap to top-tier engines is a worthy follow-up that
our corpus does not answer.
An important counterpoint bounds the gap from below: the
Phase-B anchor ladder includes \texttt{Rustic Alpha 3.0.4}
(CCRL \elo{1820}), a deliberately chosen \emph{non-strawman}
solid human-authored Rust engine. Two of our engines beat it
head-to-head and several others close to within 100 Elo. The
gap to GM-strength engines is real, but the gap to a
competent human-built club engine is roughly closed.

\subsection{Implications for practice}

For SE teams adopting AI coding teammates, three lessons
stand out. Agents will wire up strong oracles
autonomously when the domain admits them, so expect it and
verify it: the in-loop self-Elo signal the agents optimise
against can be methodologically biased, and that bias is the
human teammate's correction to refine. Capability-level
prompts plus a handful of
``improve'' follow-ups go a long way, as the median
main-corpus engine is built in 14 prompts
(\Cref{tab:rq1-engines}). And PL choice
still earns its budget line: the agent does the work, but the
human still chooses the medium, and a poor target multiplies
cost and effort by orders of magnitude while capping
reachable quality. A fourth lesson generalises the
verification-affordance axis beyond the PL itself: in this
corpus, the agents' autonomous build-test-revise loop made
verifiable progress because the domain supplied strong
external oracles (\Cref{sec:rq3}). When deciding where to
deploy a coding agent, asking ``what can my domain check
automatically?'' is therefore at least as informative as
asking how familiar the underlying LLM is with the domain
and its languages.

\subsection{Future work}

Three follow-up directions extend naturally, with the
artefacts in place for each. First, \emph{controlled
replications} to separate PL effects from model mastery and
session randomness: the six confounders catalogued in
\Cref{sec:threats} specify exactly what such an experiment
must hold fixed. The full factorial over those confounders is
combinatorially large, and every cell is costly, since a
single variant means a complete agent session plus a
canonical re-evaluation. Designing replications that stay
informative under a realistic budget is therefore a research
challenge in itself, and our exploratory study opens it by
indicating which factors appear to move outcomes
most. Second, an \emph{open-weight and
open-harness comparison}. Our results were produced with
closed-source frontier models (\opus{4.6}/\opus{4.7},
\gpt{5-codex}/\gpt{5.3-codex}/\gpt{5.4})
at ${\approx}\$2{,}100$ canonical USD summed over the 23
main-corpus engines with recoverable cost data
(\Cref{tab:rq1-engines}, with vendor list prices
discussed in \Cref{sec:corpus:evidence}). Do open-weight
models (DeepSeek-Coder, Qwen-Coder, Llama-Coder) and open
harnesses (OpenHands~\cite{openhands2024}, Aider, SWE-agent)
reach comparable polyglot coverage? Do they exploit oracles
the same way, under comparable budgets?
Our corpus and methodology provide a ready ground truth for
that comparison. Third, \emph{cross-domain generalisation}.
Chess gave us three things that make the evidence comparable
across languages: a non-trivial feature-rich system, a
hierarchy of language-agnostic oracles, and a series of
demanding software engineering tasks both agents consistently
attempt. The same traits make chess a realistic complex case
rather than a sanitised benchmark: Elo is noisy, reference
engines drift across versions and hardware, domain knowledge
is in play, and games cost compute. The next study need not
be chess, however. Compilers, SQL query engines, regex
engines, type checkers, cryptographic primitives, and
constraint solvers all admit strong external oracles, would
support the same style of study, and are natural candidates
for the next generation of polyglot system-construction
benchmarks, testing whether the ``PL is one factor among
several'' reading holds beyond chess.

\subsection{Concluding remarks}

We release \textbf{34 \agentbuilt{} chess engines}
(29 main-corpus $+$ 5 special-role) across 17 primary
languages, per-engine evidence reports, the canonical
re-evaluation harness (\Cref{app:anchor-eval}), and
a reproducible Python pipeline at
\url{https://github.com/acherm/agentic-chessengine-metaanalysis}
(per-engine repositories listed in \Cref{app:github-repos}).

AI teammates have moved software construction one level up,
from line-by-line authoring to capability-level steering, and
the medium (the programming language) still shapes the
message. This paper establishes that these language effects
exist: the PL bounds the strength an agent-built system can
reach and multiplies what it costs to build.
We hope the corpus, protocol, and research questions of this paper
will be revisited with diverse programming languages, open and next-generation coding agents, and
from chess to other domains with strong external oracles.

\section*{Data and code availability}
\label{sec:data-availability}
\addcontentsline{toc}{section}{Data and code availability}

All data, scripts, and per-project reports used in this paper are
released under a permissive license at:

\begin{center}
\url{https://github.com/acherm/agentic-chessengine-metaanalysis}
\end{center}

The 34 per-engine repositories are also released individually under
the same organisation
(\href{https://github.com/acherm}{\texttt{github.com/acherm}}). The
full mapping from engine to repository URL is given in
\Cref{app:github-repos}. The artifact contains:
\begin{itemize}
  \item \texttt{scripts/}: the Python pipeline
    (\texttt{common.py}, \texttt{discover.py},
    \texttt{extract\_sessions.py}, \texttt{per\_project\_report.py},
    \texttt{elo\_and\_perft.py}, \texttt{extract\_elo.py},
    \texttt{synthesis.py}),
  \item \texttt{data/}: machine-readable appendices
    (\texttt{overview.json}, \texttt{synthesis.json},
    \texttt{elo.json}, \texttt{novelty.json},
    \texttt{prompt\_protocol.json}, \texttt{projects/<name>.json}),
  \item \texttt{reports/}: the per-project reports (45 Markdown
    files), the cross-project synthesis, and four qualitative deep
    dives.
\end{itemize}

\paragraph{Session transcripts.}
Interaction data is released in two forms. The per-engine
measurements (prompts, tokens, costs, tool use, and trajectories)
are released as the committed \texttt{data/} JSON snapshot,
extracted on 2026-04-20 by the released scripts, which is the
canonical record of \claudecode{} interaction for this corpus. The
raw \codex{} session rollouts are released as well, with personally
identifying information redacted. Per-session provenance notes are
included in the artefact.

\section*{Use of large language models}
\addcontentsline{toc}{section}{Use of large language models}

LLMs play three roles in this work. They are the object of
study (\Cref{sec:background:agents}). They assisted the
analysis: the extraction pipeline, the evaluation harness,
and the statistical scripts were written with \claudecode{}
assistance and reviewed by the authors, and every published
number is computed by that released code rather than by LLM
judgement (\Cref{sec:corpus:evidence}, \Cref{app:llm-usage}).
They assisted the writing: the manuscript was drafted and
edited with \claudecode{} (Anthropic \opus{4.x}-family and
\textsc{Fable~5} models, 2026-04 to 2026-06) under detailed
author direction. The authors reviewed and revised all
generated text, verified the factual claims against the
released data, and take full responsibility for the content.
A self-assessment against the reporting checklist of the
community guidelines for LLM
studies~\citep{baltes2025llmguidelines} is given in
\Cref{app:guidelines}.

\section*{Acknowledgments}
\addcontentsline{toc}{section}{Acknowledgments}

This work is supported by the Inria Défi
\emph{LLM4Code}~(\url{https://project.inria.fr/llm4code/}).

\section*{Conflicts of interest}
\addcontentsline{toc}{section}{Conflicts of interest}

The authors declare no conflicts of interest. In particular, the
authors have no financial, employment, or contractual relationship
with OpenAI or Anthropic, the providers of the coding-agent models
(Codex / GPT-5-codex family and Claude Code / Opus 4.6/4.7) used to
build the corpus.

\bibliographystyle{abbrvnat}
\bibliography{references}

\appendix
\section{Corpus supplement: additional figures and the
novelty audit}
\label{app:corpus-supplement}

This appendix collects views referenced from the main text
that we kept out of the body to control figure and table
count.
\Cref{app:elo-curves} reproduces the agent-claimed Elo
trajectories cited in \RQ{4},
\Cref{app:novelty-audit} reproduces the per-engine novelty
audit cited in \RQ{2}, and
\Cref{app:rr-scatter} reproduces the round-robin-vs-anchor Elo
scatterplot cited in \RQ{4}.

\subsection{Agent-claimed Elo trajectories during a session}
\label{app:elo-curves}

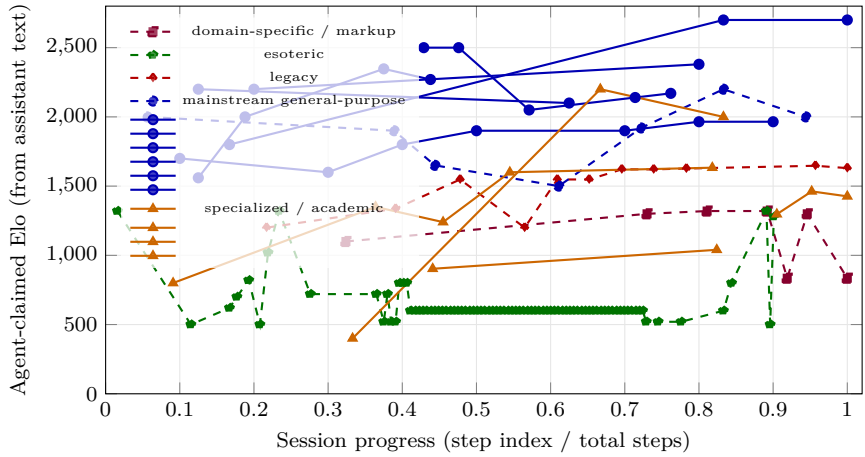
\begin{figure}[!htbp]
\centering
\begin{tikzpicture}
\begin{axis}[
  width=0.95\linewidth, height=0.55\linewidth,
  xlabel={Session progress (step index / total steps)},
  ylabel={Agent-claimed Elo (from assistant text)},
  xmin=0, xmax=1.02,
  ymin=0, ymax=2800,
  grid=both, grid style={gray!20},
  legend style={font=\tiny, at={(0.02,0.98)}, anchor=north west, draw=none, fill=white, fill opacity=0.75, text opacity=1},
  legend columns=1,
  tick label style={font=\footnotesize},
  label style={font=\footnotesize},
  mark size=1.7pt,
]
  \addplot[color=purple!70!black, mark=square*,dashed, mark options={fill=purple!70!black}, thick] coordinates { (0.324,1100) (0.730,1300) (0.811,1320) (0.892,1320) (0.919,835) (0.946,1300) (1.000,835) };
  \addlegendentry{domain-specific / markup}
  \addplot[color=green!45!black, mark=pentagon*,dashed, mark options={fill=green!45!black}, thick] coordinates { (0.016,1320) (0.115,500) (0.167,620) (0.177,700) (0.193,820) (0.208,500) (0.219,1019) (0.234,1320) (0.276,720) (0.365,720) (0.375,520) (0.380,720) (0.385,520) (0.391,520) (0.396,800) (0.401,800) (0.406,800) (0.411,600) (0.417,600) (0.422,600) (0.427,600) (0.432,600) (0.438,600) (0.443,600) (0.448,600) (0.453,600) (0.458,600) (0.464,600) (0.469,600) (0.474,600) (0.479,600) (0.484,600) (0.490,600) (0.495,600) (0.500,600) (0.505,600) (0.510,600) (0.516,600) (0.521,600) (0.526,600) (0.531,600) (0.536,600) (0.542,600) (0.547,600) (0.552,600) (0.557,600) (0.562,600) (0.568,600) (0.573,600) (0.578,600) (0.583,600) (0.589,600) (0.594,600) (0.599,600) (0.604,600) (0.609,600) (0.615,600) (0.620,600) (0.625,600) (0.630,600) (0.635,600) (0.641,600) (0.646,600) (0.651,600) (0.656,600) (0.661,600) (0.667,600) (0.672,600) (0.677,600) (0.682,600) (0.688,600) (0.693,600) (0.698,600) (0.703,600) (0.708,600) (0.714,600) (0.719,600) (0.724,600) (0.729,520) (0.745,520) (0.776,520) (0.833,600) (0.844,800) (0.891,1320) (0.896,500) (0.901,1280) };
  \addlegendentry{esoteric}
  \addplot[color=red!70!black, mark=diamond*,dashed, mark options={fill=red!70!black}, thick] coordinates { (0.217,1200) (0.391,1337) (0.478,1548) (0.565,1200) (0.609,1548) (0.652,1548) (0.696,1621) (0.739,1621) (0.783,1627) (0.957,1647) (1.000,1630) };
  \addlegendentry{legacy}
  \addplot[color=blue!70!black, mark=*,dashed, mark options={fill=blue!70!black}, thick] coordinates { (0.056,2000) (0.389,1900) (0.444,1650) (0.611,1500) (0.722,1920) (0.833,2200) (0.944,2000) };
  \addlegendentry{mainstream general-purpose}
  \addplot[color=blue!70!black, mark=*, mark options={fill=blue!70!black}, thick] coordinates { (0.167,1800) (0.833,2700) (1.000,2700) };
  \addlegendentry{}
  \addplot[color=blue!70!black, mark=*, mark options={fill=blue!70!black}, thick] coordinates { (0.429,2500) (0.476,2500) (0.571,2050) (0.714,2140) (0.762,2170) };
  \addlegendentry{}
  \addplot[color=blue!70!black, mark=*, mark options={fill=blue!70!black}, thick] coordinates { (0.200,2200) (0.800,2380) };
  \addlegendentry{}
  \addplot[color=blue!70!black, mark=*, mark options={fill=blue!70!black}, thick] coordinates { (0.100,1700) (0.300,1600) (0.400,1800) (0.500,1900) (0.700,1900) (0.800,1965) (0.900,1965) };
  \addlegendentry{}
  \addplot[color=blue!70!black, mark=*, mark options={fill=blue!70!black}, thick] coordinates { (0.125,2200) (0.625,2100) };
  \addlegendentry{}
  \addplot[color=blue!70!black, mark=*, mark options={fill=blue!70!black}, thick] coordinates { (0.125,1560) (0.188,2000) (0.375,2347) (0.438,2270) };
  \addlegendentry{}
  \addplot[color=orange!80!black, mark=triangle*, mark options={fill=orange!80!black}, thick] coordinates { (0.091,800) (0.364,1350) (0.455,1240) (0.545,1600) (0.818,1633) };
  \addlegendentry{specialized / academic}
  \addplot[color=orange!80!black, mark=triangle*, mark options={fill=orange!80!black}, thick] coordinates { (0.441,903) (0.824,1040) };
  \addlegendentry{}
  \addplot[color=orange!80!black, mark=triangle*, mark options={fill=orange!80!black}, thick] coordinates { (0.333,400) (0.667,2200) (0.833,2000) };
  \addlegendentry{}
  \addplot[color=orange!80!black, mark=triangle*, mark options={fill=orange!80!black}, thick] coordinates { (0.905,1297) (0.952,1462) (1.000,1426) };
  \addlegendentry{}
\end{axis}
\end{tikzpicture}
\caption{Agent-claimed Elo over normalised session progress, for engines with at least two Elo mentions in their transcripts. One polyline per engine, with colour and marker shape by language category. \emph{Dashed} polylines mark zigzag trajectories (an Elo claim that dropped then recovered within the session, \Cref{sec:rq4:curve}). The mainstream general-purpose engines cluster near the top, rise monotonically, and cap around \elo{2000}--\elo{2200}. The legacy and domain-specific cells plateau at \elo{1000}--\elo{1600}, and the esoteric row sits at the bottom. The zigzag engines are the ones where the agent caught and rolled back a regression mid-session, a supervision signal that costs tokens but prevents strength loss.}
\label{fig:elo-curves}
\end{figure}

\noindent\Cref{fig:elo-curves} shows the five qualitative
shapes of agent-claimed Elo curves discussed in
\Cref{sec:rq4:curve}. Mainstream general-purpose engines
cluster near the top, legacy and domain-specific cells
plateau in the middle, and the esoteric row sits at the bottom.
Dashed polylines mark the zigzag trajectories that signal a
within-session regression-and-rollback, the supervision
signal we discuss in \RQ{4}~(\Cref{sec:rq4:curve}) and
\RQ{5}~(\Cref{sec:rq5:reward}).

\subsection{Per-engine novelty audit}
\label{app:novelty-audit}

\begin{table}[!htbp]\centering\scriptsize
\caption{\RQ{3} Novelty evidence, per engine. `Class.' is our classification. ``Scratch'' means the engine-core files contain no chess-library import, no canonical-engine fingerprint, and no self-reported copy claim. ``Lib-asst'' means the engine core imports an external chess library. ``Fp-match'' means a distinctive fingerprint (e.g., \texttt{python-chess} API surface) appears inside the engine core. ``Copy-claim'' means the transcript contains explicit authorship language (``ported from X'', ``adapt from X'', \ldots). `Core imp.' and `Core fp.' are the engine-core-only evidence streams (test/harness files are scored separately). `Tool dep.' flags a chess library used only in the evaluation tooling. `Strong' and `Weak' are counts of transcript mentions of canonical engines (``adapt from''/``port from'' vs.\ ``like''/``based on'').}
\label{tab:rq3-novelty}
\resizebox{\textwidth}{!}{%
\begin{tabular}{l l l l l c c c}
\toprule
Engine & Lang. & Class. & Core imp. & Core fp. & Tool dep. & Strong & Weak \\
\midrule
chess-apl-codex54 & APL & scratch & — & — &  & 0 & 2 \\
chess-assembly-codex & Assembly & scratch & — & — & \checkmark & 0 & 0 \\
chess-brainfuck & Brainfuck & scratch & — & — & \checkmark & 0 & 0 \\
chess-brainfuck-cc & Brainfuck & scratch & — & — & \checkmark & 0 & 2 \\
chess-purec & C & scratch & — & — &  & 0 & 1 \\
chess-purec-codex & C & scratch & — & — &  & 0 & 0 \\
chess-cplusplus-claude & C++ & scratch & — & — &  & 0 & 1 \\
cplusplus-chess & C++ & scratch & — & — &  & 0 & 8 \\
COBOL-chess & COBOL & scratch & — & — &  & 0 & 11 \\
chess-cobol-cc & COBOL & scratch & — & — &  & 0 & 0 \\
chess-css-codex & CSS/HTML & fingerprint-match & python: import chess & python-chess API surface & \checkmark & 0 & 0 \\
chess-css-codex-guided & CSS/HTML & scratch & — & — &  & 0 & 0 \\
chess-icon-codex & Icon & scratch & — & — &  & 0 & 0 \\
chess-java & Java & scratch & — & — &  & 0 & 0 \\
chess-java-cc & Java & scratch & — & — &  & 0 & 0 \\
chess-latex-codex-replication & LaTeX & scratch & — & — &  & 0 & 0 \\
latex-chess-engine & LaTeX (TeX) & scratch & — & — & \checkmark & 0 & 0 \\
lean-chess & Lean 4 & scratch & — & — &  & 0 & 0 \\
chess-py & Python & scratch & — & — &  & 0 & 3 \\
chess-py-cc & Python & scratch & — & — &  & 0 & 1 \\
chess-Rocq & Rocq$\to$OCaml & scratch & — & — &  & 0 & 1 \\
chess-ruby-cc & Ruby & scratch & — & — &  & 0 & 0 \\
chess-ruby-codex & Ruby & scratch & — & — &  & 0 & 0 \\
chess-rust-cc & Rust & scratch & — & — &  & 0 & 2 \\
chess-rust-cc-redo & Rust & scratch & — & — &  & 0 & 1 \\
chess-rust-codex & Rust & library-assisted & rust: chess crate & — & \checkmark & 0 & 0 \\
chess-sql & SQL & scratch & — & — &  & 0 & 0 \\
chess-why3-cc & Why3$\to$OCaml & scratch & — & — &  & 0 & 0 \\
chess-why3 & Why3$\to$OCaml & scratch & — & — &  & 0 & 0 \\
\bottomrule
\end{tabular}
}
\end{table}

\noindent\Cref{tab:rq3-novelty} reports the four-signal
novelty audit per engine cited from \Cref{sec:rq2:audit}:
class (\emph{scratch}, \emph{lib-asst}, \emph{fp-match},
\emph{copy-claim}), engine-core import and fingerprint hits,
tooling-only library dependencies, and counts of strong
(``ported from''/``adapt from'') and weak (``inspired by'',
``based on'') transcript references to canonical engines.
Headline buckets are summarised inline at \Cref{sec:rq2:audit}.
The table here lets a reviewer audit any engine cell directly.

\paragraph{The four signals, in detail.}
(1)~\emph{Manifest dependencies}: parse
\texttt{requirements.txt}, \texttt{pyproject.toml},
\texttt{Cargo.toml}, \texttt{Gemfile}, \texttt{package.json},
\texttt{pom.xml}, \texttt{build.gradle}, \texttt{go.mod} for
known chess libraries (\texttt{python-chess},
\texttt{sunfish}, \texttt{chess.js}, \texttt{shakmaty},
\texttt{cozy-chess}, the Rust \texttt{chess} crate,
\texttt{chesspresso}, \ldots).
(2)~\emph{Source-level imports}: grep source files for
imports of any of those libraries.
(3)~\emph{Canonical-engine fingerprints}: a hand-curated
catalogue of distinctive constants (Sunfish piece values,
\texttt{python-chess} API surface, \texttt{chess.js}
\texttt{PIECE\_SYMBOLS}, the TSCP \texttt{\#define} block).
(4)~\emph{Transcript authorship claims}: mine \claudecode{}
and \codex{} transcripts for strong patterns
(``ported from'', ``translated from'') and weak patterns
(``inspired by'', ``based on'') followed by a
canonical-engine name.

\paragraph{Engine-core vs.\ tooling files.}
A recurrent false positive in our first pass was
\texttt{python-chess} imported by engines whose primary
language is not Python: the agent uses it in a test harness
or PGN converter (\texttt{play\_stockfish.py},
\texttt{pgn2latex.py}). A dependency used only for evaluation
tooling is not evidence the engine is library-assisted. We
therefore split every import or fingerprint hit by whether
its file sits in an engine-core path (\texttt{src/},
\texttt{lib/}, project root) or a tooling path, and only
engine-core hits determine the bucket.

\paragraph{Fingerprint catalogue limits.}
The fingerprint catalogue is deliberately small and
hand-curated. We do not claim exhaustive coverage. A future
pass will add locality-sensitive hashing over distinctive
integer constants and identifier-ngram Jaccard similarity
against a vendored corpus of canonical engines.

\paragraph{Regex limits for unusual syntaxes.}
The feature scan in \Cref{sec:rq2:features} relies on regex
patterns over source text. Brainfuck is a known weak spot:
the compiled 5.6\,MB \texttt{chess.bf} does not expose
identifiers like ``castling'' or ``en passant'' (those live
only in the Python code generator), so the Brainfuck rows
under-report. LaTeX engines, where chess concepts are spelled
out in macro names, report cleanly. We treat zeros in
unusual-syntax rows as ``not detected'' rather than ``not
implemented'' and cross-reference the per-engine reports
before making a claim.

\subsection{Round-robin vs anchor Elo scatterplot}
\label{app:rr-scatter}

\begin{figure}[!htbp]
\centering
\includegraphics[width=0.85\linewidth]{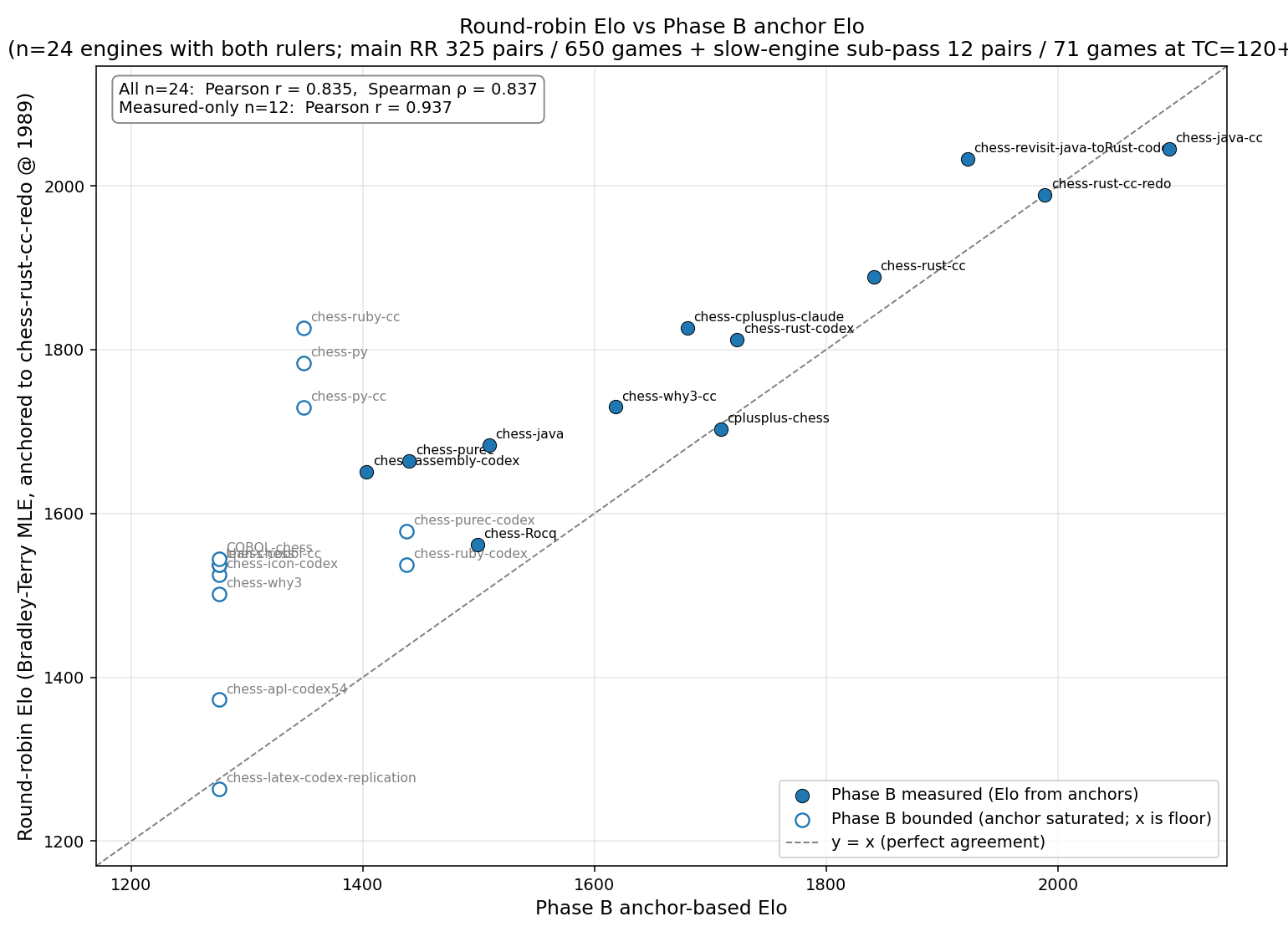}
\caption{Round-robin Elo (Bradley-Terry MLE) versus Phase~B
anchor-based Elo (inverse-variance combination of the calibrated
external references). The round-robin pool is 650 head-to-head
games over 325 pairs among the 26 UCI-conformant main-corpus
engines plus a slow-engine sub-pass (71 further games over 12
pairs vs.\ \texttt{chess-brainfuck}, \texttt{chess-sql}, and
\texttt{latex-chess-engine}). The scatter plots the $n=24$
engines for which both rulers are available. Filled markers:
engines with at least one unsaturated anchor pair. Open markers:
engines whose anchor-based Elo is a saturation floor only (the
agent's measurement methodology provides no resolution below
${\sim}$\elo{1500} so they pile up at the floor). The dashed
line is $y=x$. Pearson $r=0.94$ on the measured-only subset
($n=12$). Pearson $r=0.84$, Spearman $\rho=0.84$ across all
$n=24$ engines plotted. Discussion of the agreement and
disagreement patterns is in \Cref{app:rr}.}
\label{fig:rr-vs-phaseb}
\end{figure}

\section{Method details}
\label{app:method-details}

This appendix collects the methodology paragraphs that would
otherwise inflate \S\ref{sec:corpus} and break the reading
flow. Cross-references from the body point here when a reader
needs the exact pipeline behind a number cited in the
\RQ{} sections.

\subsection{Repository classification}
\label{app:method:repo-classification}

Language attribution uses file-extension counts: the
\emph{primary language} of an engine is the code-language
(excluding Markdown, JSON, YAML, and TOML) with the highest
LOC. Vendored third-party trees (\texttt{cutechess-cli/},
\texttt{stockfish/}, \texttt{fairy-stockfish/}, bundled
opponent-engine source trees under \texttt{tests/engines/})
are excluded from LOC and language counts to avoid attributing
benchmarking-infrastructure code to the engine itself.

\subsection{Feature-level cross-engine analysis pipeline}
\label{app:method:feature-pipeline}

The file-level feature scan (\Cref{sec:rq2}) records whether a
pattern such as \texttt{quiescence|qsearch} appears
\emph{anywhere} in a project, giving one bit per feature per
engine. To move from presence to depth (which optimisations
were implemented, how the algorithm was adapted to the target
language), we built a deeper pipeline around two scripts.

\paragraph{\texttt{feature\_locator.py --hotspot}.} For a
named feature (e.g.\ ``Transposition table''), the locator
searches all source files for matching lines and scores each
match by: keyword-in-function-name (+20),
line-is-a-function-definition (+12), search-adjacent file
(+6), keyword-in-line (+3), test-file penalty ($-$25). The
highest-scoring match becomes the \emph{anchor}. A
language-aware boundary extractor walks outward to recover the
enclosing function body. Eleven language families are
handled: brace-counting (C/C++/Java/Rust), indentation
(Python), \texttt{keyword/end} (Ruby/Icon), \texttt{let\,rec}
(OCaml/Why3), \texttt{Fixpoint}/\texttt{Definition}
(Coq/Rocq), \texttt{def}/\texttt{partial\,def} (Lean~4),
paragraph names (COBOL), assembly labels, and
$\nabla$\ldots$\nabla$ delimiters (APL). A secondary stem scan
(e.g.\ \texttt{quiesc} from \texttt{quiescence}) catches
function names that match algorithmically but not lexically.

\paragraph{\texttt{feature\_compare.py}.} Given the per-engine
hotspot JSON, the comparator applies 13 algorithmically
motivated sub-feature regexes to each extracted snippet and
produces (i) a sub-feature presence matrix
(\checkmark/\textperiodcentered{} per engine $\times$
sub-feature), (ii) pairwise \emph{token-Jaccard}
(bag-of-normalised-tokens, language keywords filtered), and
(iii) pairwise \emph{feature-Jaccard} (Jaccard over the
13-dimensional binary sub-feature vector). Token-Jaccard
measures lexical proximity. Feature-Jaccard measures
algorithmic proximity, independent of surface syntax.

\paragraph{Three features, one triad.} We applied this
pipeline to three features chosen to form a complete
cross-section of a chess engine. \emph{Quiescence search} is a
search algorithm concentrated in one function (the pattern
fires in 28 of the scan's 36 targets, the corpus engines plus
three canonical reference engines, and the hotspot extractor
recovers a valid function body for 26 of those 28). \emph{Transposition
table} is a data structure distributed across initialisation,
class definition, and probe/store call sites (21 valid
extractions out of 24 pattern matches).
\emph{Evaluation/PST} is further fragmented across constants
file, \texttt{evaluate()} function, and incremental update
hooks (the highest extraction failure rate, $\approx$47\%, with
10 valid extractions out of 19 matches). Raw artefacts are in
\texttt{reports/features/}. Per-feature narratives are in
\texttt{QUIESCE/TT/EVAL\_FEATURE\_ANALYSIS.md} and synthesised
in \texttt{FEATURE\_ANALYSIS\_REPORT.md} in the replication
artefact.

\subsection{SE-activity coding frame}
\label{app:method:se-activities}

The 10 activity categories used throughout the paper were
\emph{induced} from the trajectory data rather than imposed
a priori. Starting point was the per-step record assembled by
the two-pass protocol (\Cref{sec:corpus:evidence}): user
prompt, tool calls, bash commands, files written, files
edited, and (for \claudecode{}) user-intent heuristics. We
inspected a stratified sample of sessions and iteratively
merged and split candidate categories until the resulting set
satisfied four constraints: (i)~grounded in concrete
trajectory evidence, (ii)~language-agnostic, (iii)~orthogonal
enough to be informative while permitting genuine overlap
(a debugging step that edits a file is both \emph{debugging}
and \emph{implementation}), and (iv)~covering, with fewer than
3\% of steps in the stratified sample carrying no tag. An
initial 13-category draft was revised after auditing showed
that \emph{implementation} was too narrowly tagged
(``new source file with a recognised extension'' missed most
coding activity, since agents do most work through edits) and
that protocol-engineering, refactoring, and
requirements-spec splits were not reliably separable. The
revised frame merges them into \emph{implementation} and
\emph{design}. The full rule set lives in the replication
artefact as \texttt{scripts/se\_capabilities.py}.

\paragraph{Per-step tagging rules.} The coder is
deterministic, runs against the structured per-step record,
and assigns each step zero or more tags. Each category has
one to three evidence sources: a filename regex (e.g.\ the
\emph{testing} category fires on \texttt{/tests?/} paths or
\texttt{test\_*}/\texttt{*\_test.*} stems), a bash-kind
filter (e.g.\ \emph{benchmarking \& evaluation} fires on
\texttt{cutechess}/\texttt{stockfish} invocations), a
user-intent heuristic, or a user-prompt vocabulary regex. A
step can carry multiple tags. The coder does not count LOC
or weight tags. Each firing contributes a single event.

\paragraph{Likert intensity scale.} Per-engine session length
spans two orders of magnitude (one-shot prompt-and-go to
192-step Brainfuck iterations), so raw event counts are not
directly comparable. We summarise each engine as a
10-dimensional intensity profile using a 4-level ordinal
scale on the share of the session tagged with each
capability: \textbf{absent} (0 events), \textbf{light}
($\bullet$: one event regardless of session length, or share
$\leq$~5\%), \textbf{moderate} ($\bullet\bullet$: share
in~(5\%,~20\%]), and \textbf{heavy}
($\bullet\bullet\bullet$: share~$>$~20\%). The thresholds are
coarse by design.

\paragraph{Scope and limits.} The frame gives three derived
measures: coverage (whether an engine ever exercised a
capability), breadth (how many of the 10 categories an engine
exercised), and intensity (the Likert score per category per
engine). None measures correctness or value. Tagging is also
heuristic and under-counts engines whose primary transcripts
are archived (\Cref{app:archived-floors}).

\subsection{Qualitative triangulation pilot}
\label{app:method:qualitative-pilot}

To complement the quantitative evidence with sourced
qualitative material, we ran a small structured coding pilot
over two author-authored blog posts on specific corpus
engines~\cite{acher2026texccchess,acher2026bfchess}, using a
nine-code frame tied to the paper's RQs (three codes per
RQ-cluster). 33 excerpts (15 TeX, 18 Brainfuck) were
extracted, and 8 of 9 codes saturate across both sources. The
frame, coded excerpts, and code$\times$source frequency
matrix are in the replication artefact
(\hyperref[sec:data-availability]{Data and code availability}). Inter-rater reliability is
not yet established (single coder), and we position the pilot
as complementary evidence rather than a primary finding.

\subsection{The Chess Engine Feature Explorer (web tool)}
\label{app:method:explorer}

To support the qualitative side of the feature analysis we
built a small web tool, \emph{Chess Engine Feature Explorer},
that browses the per-engine hotspots and compares variants
side by side. The tool was used during the qualitative
analysis to inspect sub-feature differences across PLs,
specifically to check, per feature, where two engines share a
sub-feature and where they diverge. The live tool is hosted
at
\url{https://blog.mathieuacher.com/agentic-chessengine-metaanalysis/reports/features/feature_explorer.html}
and the dedicated quiescence-art view (constellation laid out
around the canonical CPW pseudocode) at
\url{https://blog.mathieuacher.com/agentic-chessengine-metaanalysis/reports/features/quiescence_art_advanced.html}.
Both are also released with the replication artefact alongside
the per-feature JSON.

\begin{landscape}
\begin{figure}[p]
\centering
\includegraphics[width=0.97\linewidth,height=0.85\textheight,keepaspectratio]{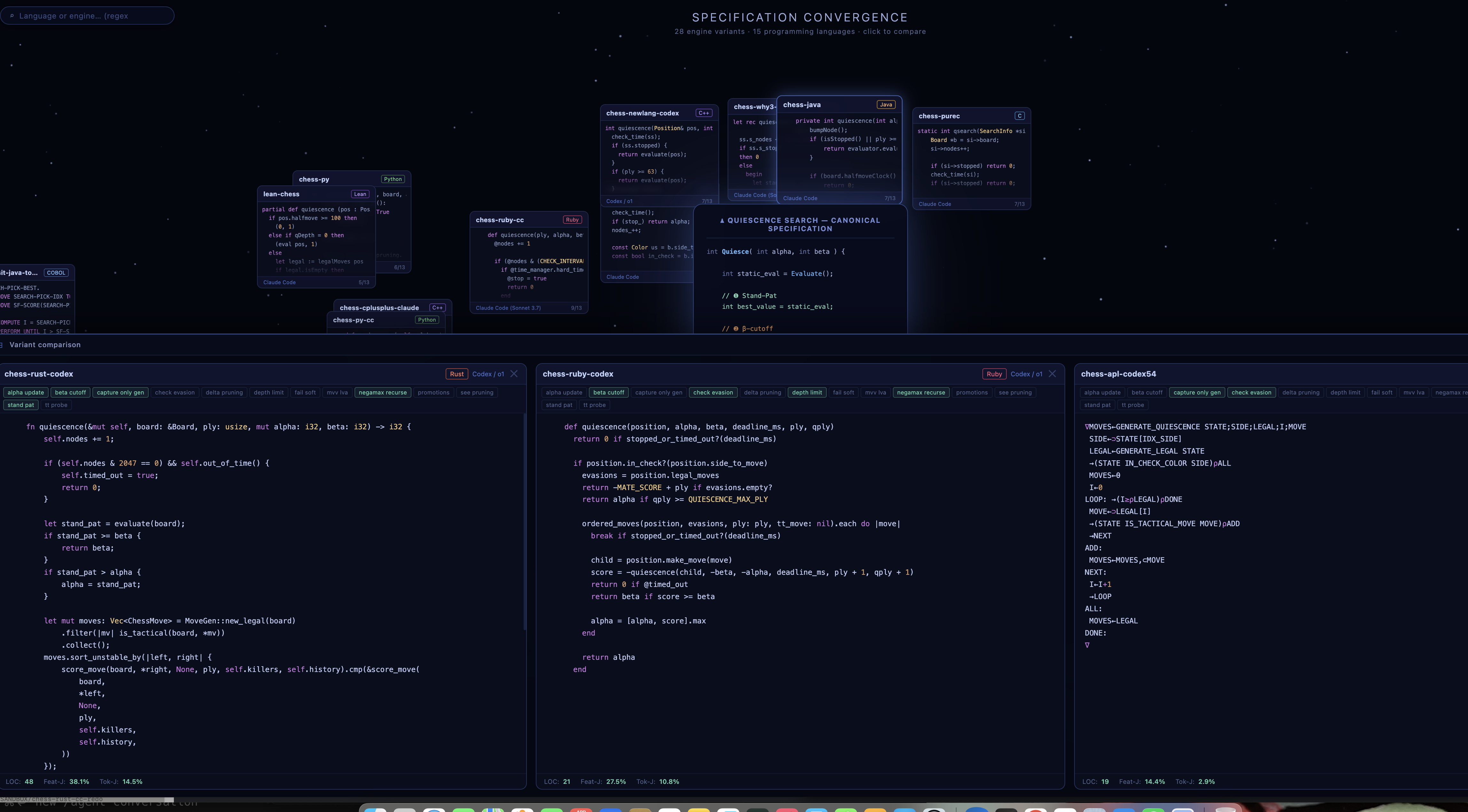}
\caption{Chess Engine Feature Explorer, \emph{specification
view}: 28 engine variants arrayed around the
canonical CPW pseudocode skeleton (centre). The visualisation
makes the size of the variant space immediately legible: any
two cards can be selected for side-by-side comparison
(\Cref{fig:feature-explorer-compare}). The bottom strip shows
three concrete variants (\texttt{chess-rust-codex} in Rust,
\texttt{chess-ruby-codex} in Ruby,
\texttt{chess-apl-codex54} in APL) selected for the
side-by-side view, with feat-J / tok-J / LOC summaries in
each card. Live at
\url{https://blog.mathieuacher.com/agentic-chessengine-metaanalysis/reports/features/quiescence_art_advanced.html}.}
\label{fig:feature-explorer-galaxy}
\end{figure}
\end{landscape}

\begin{figure*}[!htbp]
\centering
\includegraphics[width=0.95\linewidth]{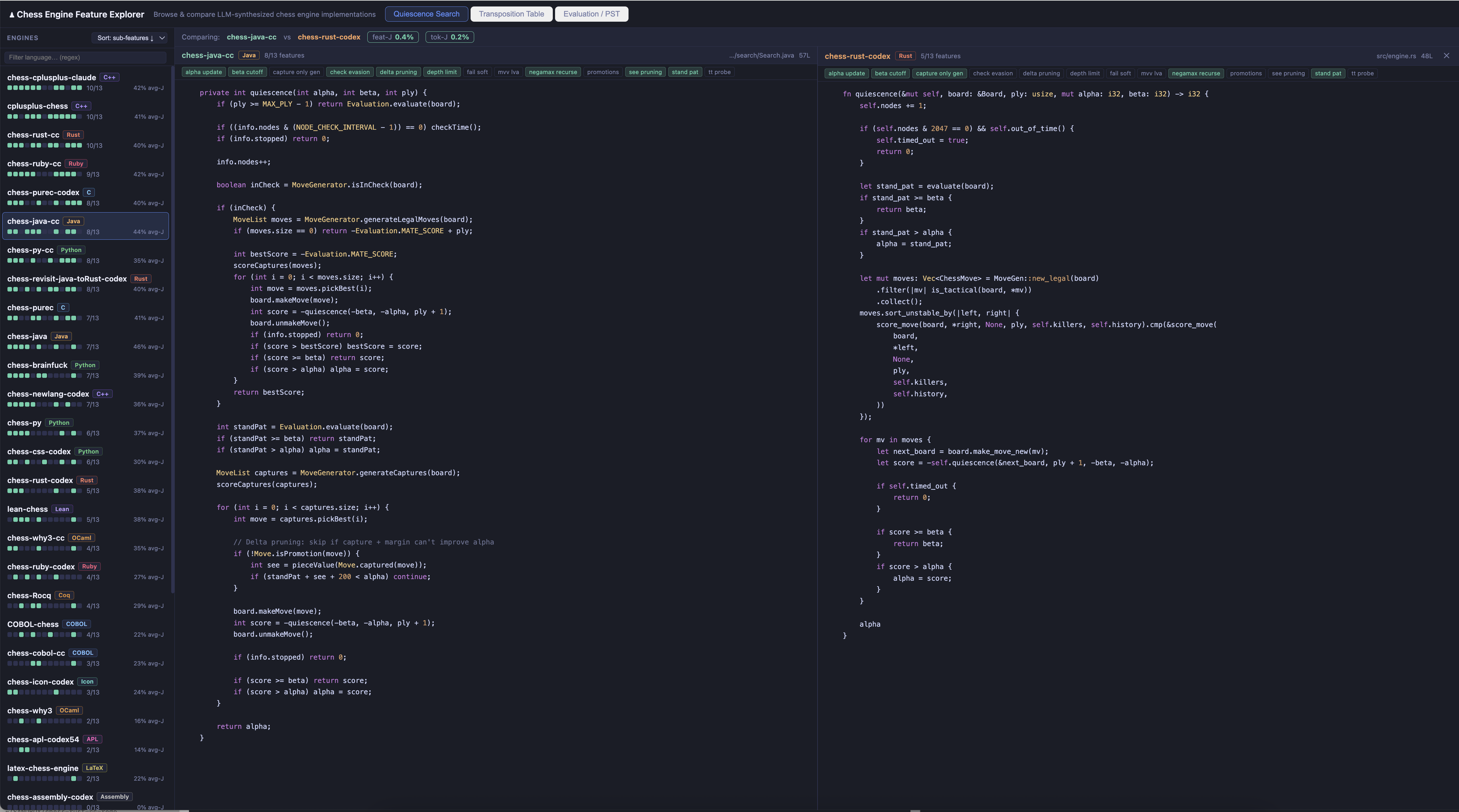}
\caption{Chess Engine Feature Explorer,
\emph{compare view}: a Java vs.\ Rust pair
(\texttt{chess-java-cc} vs.\ \texttt{chess-rust-codex}) on
quiescence search. Sub-feature tags at the top of each panel
mark which of the 13 algorithmic sub-features fired in that
engine (highlighted = present, dim = absent). The source
code is rendered below. The header bar reports
feat-J/tok-J similarity (here 44\%/22\%: fewer than half the
sub-features and about a fifth of the normalised tokens are
shared) and the
sub-feature counts (8/13 vs.\ 5/13). The left-hand engine
list lets the reviewer pick any two variants from the 28
engine cards. Sorting by sub-feature count, language, or
average feat-J similarity is supported.}
\label{fig:feature-explorer-compare}
\end{figure*}

\Cref{fig:feature-explorer-galaxy} (left panel) shows the
\emph{specification view}: 28 engine variants laid out on a
constellation, with the
canonical CPW-style quiescence pseudocode in the centre and
the agent-built variants floating around it. Clicking any
two cards opens a side-by-side comparison.
\Cref{fig:feature-explorer-compare} shows the
\emph{compare view}: a Java vs.\ Rust pair
(\texttt{chess-java-cc} vs.\ \texttt{chess-rust-codex}) with
the 13 sub-feature tags rendered as chips at the top of each
panel and the underlying source code below. The two
sub-feature rows make divergence visible at a glance:
shared chips highlighted, missing chips greyed out, and the
feat-J / tok-J similarity scores reported in the header bar
(here feat-J $=$ 44\% and tok-J $=$ 22\%: the pair shares
fewer than half of the 13 sub-features and about a fifth of
the normalised tokens despite the common canonical skeleton).

We use the tool throughout the paper as the qualitative
counterpart to the \texttt{feature\_compare.py} pipeline of
\Cref{app:method:feature-pipeline}: where the pipeline
reports aggregate distances, the explorer makes the
per-engine code visible so a reviewer can audit \emph{which}
sub-features differ between any two variants and \emph{how
the divergence manifests} in source. This is also the
primary reason we report sub-feature divergence rather than
pure ``convergence'': the paper's algorithmic-similarity
findings come from the pipeline's distance numbers, but the
qualitative read of those numbers in the explorer makes
clear that variants share a common high-level spec while
diverging substantially in sub-feature selection and
language-specific form.

\subsection{SE-activity per-engine intensity matrix}
\label{app:se-matrix}

\begin{table*}[!htbp]\centering\scriptsize
\caption{SE-activity intensity profile per engine, grouped by the RQ1 language taxonomy. Each cell is the Likert level on the share of the session's steps tagged with that capability: $-$ absent, $\bullet$ light (1 event or $\leq$5\%), $\bullet\bullet$ moderate (5--20\%), $\bullet\bullet\bullet$ heavy ($>$20\%). \textsc{Steps} = total trajectory steps, computed over the full recoverable record, a different accounting from the prompt counts of \Cref{tab:rq1-engines} (see \Cref{sec:corpus:se-activities}). \textsc{\#Caps} = number of capabilities exercised at any level. \textsc{Int.} = summed Likert (0--30). Horizontal rules separate language categories, and rows within each category are sorted by intensity descending. Column abbreviations: Design (design \& spec), Impl (implementation: writing or editing source files), Test (testing: perft, UCI smoke, test harness), Debug (debugging), Build (build \& tooling), VCS (version control), Bench (benchmarking: strength measurement), Read (code comprehension), Perf (performance engineering), Docs (documentation). A dagger ($^\dagger$) marks engines whose main session transcripts were compacted by \claudecode{} before capture: the row reflects only the surviving subagent sidecars, which is a strict lower bound on the session's actual SE activity (\Cref{app:archived-floors}). A double-dagger ($^\ddagger$) marks the special-role \emph{underconverged} case (\texttt{chess-mojo}, \Cref{app:special-role}). Its row is shown for transparency and excluded from main-corpus row-wise comparisons.}
\label{tab:rq1-se-matrix}
\resizebox{\textwidth}{!}{%
\begin{tabular}{l l cccccccccc r r r}
\toprule
Engine & Lang. & Design & Impl & Test & Debug & Build & VCS & Bench & Read & Perf & Docs & Steps & \#Caps & Int. \\
\midrule
\texttt{chess-mojo}$^\ddagger$ & Mojo & $\bullet\bullet\bullet$ & $\bullet\bullet\bullet$ & $\bullet\bullet\bullet$ & $\bullet\bullet\bullet$ & $\bullet\bullet\bullet$ & $\bullet\bullet\bullet$ & $\bullet\bullet\bullet$ & $\bullet\bullet\bullet$ & $-$ & $-$ & 6 & 8 & 24 \\
\texttt{chess-rust-cc-redo} & Rust & $\bullet\bullet$ & $\bullet\bullet\bullet$ & $\bullet\bullet\bullet$ & $\bullet\bullet$ & $\bullet\bullet\bullet$ & $\bullet$ & $\bullet\bullet\bullet$ & $\bullet\bullet\bullet$ & $\bullet\bullet$ & $\bullet\bullet$ & 49 & 10 & 24 \\
\texttt{chess-rust-codex} & Rust & $\bullet\bullet\bullet$ & $\bullet\bullet\bullet$ & $\bullet\bullet\bullet$ & $\bullet\bullet\bullet$ & $\bullet\bullet\bullet$ & $\bullet\bullet\bullet$ & $\bullet\bullet\bullet$ & $\bullet\bullet\bullet$ & $-$ & $-$ & 3 & 8 & 24 \\
\texttt{chess-purec-codex} & C & $\bullet\bullet\bullet$ & $\bullet\bullet\bullet$ & $\bullet\bullet\bullet$ & $\bullet\bullet\bullet$ & $\bullet\bullet\bullet$ & $\bullet$ & $\bullet\bullet\bullet$ & $\bullet\bullet\bullet$ & $\bullet$ & $-$ & 4 & 9 & 23 \\
\texttt{cplusplus-chess} & C++ & $\bullet\bullet\bullet$ & $\bullet\bullet\bullet$ & $\bullet\bullet\bullet$ & $\bullet\bullet\bullet$ & $\bullet\bullet\bullet$ & $\bullet$ & $\bullet\bullet\bullet$ & $\bullet\bullet$ & $-$ & $\bullet\bullet$ & 16 & 9 & 23 \\
\texttt{chess-java} & Java & $\bullet$ & $\bullet\bullet\bullet$ & $\bullet\bullet\bullet$ & $\bullet\bullet\bullet$ & $\bullet$ & $\bullet$ & $\bullet\bullet\bullet$ & $\bullet\bullet\bullet$ & $-$ & $-$ & 7 & 8 & 18 \\
\texttt{chess-revisit-java-toRust-codex} & Rust (port) & $-$ & $\bullet$ & $\bullet\bullet\bullet$ & $\bullet$ & $\bullet\bullet\bullet$ & $\bullet\bullet\bullet$ & $\bullet\bullet\bullet$ & $\bullet\bullet\bullet$ & $\bullet$ & $-$ & 5 & 8 & 18 \\
\texttt{chess-ruby-cc} & Ruby & $\bullet$ & $\bullet\bullet\bullet$ & $\bullet\bullet\bullet$ & $\bullet\bullet\bullet$ & $\bullet\bullet$ & $-$ & $\bullet\bullet\bullet$ & $\bullet\bullet\bullet$ & $-$ & $-$ & 10 & 7 & 18 \\
\texttt{chess-ruby-codex} & Ruby & $\bullet\bullet\bullet$ & $\bullet\bullet\bullet$ & $\bullet\bullet\bullet$ & $-$ & $-$ & $\bullet\bullet\bullet$ & $\bullet\bullet\bullet$ & $\bullet$ & $-$ & $-$ & 7 & 6 & 16 \\
\texttt{chess-py} & Python & $\bullet\bullet\bullet$ & $\bullet\bullet\bullet$ & $\bullet\bullet\bullet$ & $\bullet$ & $-$ & $-$ & $\bullet\bullet\bullet$ & $\bullet$ & $-$ & $-$ & 10 & 6 & 14 \\
\texttt{chess-rust-cc}$^\dagger$ & Rust & $-$ & $\bullet$ & $-$ & $\bullet\bullet\bullet$ & $\bullet\bullet\bullet$ & $-$ & $\bullet$ & $\bullet\bullet\bullet$ & $\bullet\bullet\bullet$ & $-$ & 8 & 6 & 14 \\
\texttt{chess-cplusplus-claude}$^\dagger$ & C++ & $-$ & $\bullet$ & $-$ & $\bullet\bullet\bullet$ & $\bullet$ & $-$ & $\bullet$ & $\bullet\bullet\bullet$ & $-$ & $-$ & 18 & 5 & 9 \\
\texttt{chess-java-cc} & Java & $\bullet\bullet\bullet$ & $\bullet$ & $-$ & $\bullet$ & $-$ & $-$ & $-$ & $\bullet\bullet\bullet$ & $-$ & $-$ & 6 & 4 & 8 \\
\texttt{chess-purec}$^\dagger$ & C & $-$ & $\bullet$ & $\bullet$ & $\bullet$ & $\bullet$ & $-$ & $-$ & $\bullet\bullet\bullet$ & $-$ & $-$ & 5 & 5 & 7 \\
\texttt{chess-py-cc}$^\dagger$ & Python & $-$ & $-$ & $-$ & $-$ & $\bullet$ & $-$ & $\bullet$ & $\bullet$ & $-$ & $-$ & 1 & 3 & 3 \\
\midrule
\texttt{chess-apl-codex54} & APL & $\bullet\bullet\bullet$ & $\bullet\bullet\bullet$ & $\bullet\bullet\bullet$ & $\bullet\bullet\bullet$ & $\bullet\bullet$ & $\bullet\bullet$ & $\bullet\bullet\bullet$ & $\bullet\bullet\bullet$ & $\bullet\bullet\bullet$ & $-$ & 34 & 9 & 25 \\
\texttt{chess-newlang-codex} & DSL & $\bullet\bullet\bullet$ & $\bullet\bullet\bullet$ & $\bullet\bullet\bullet$ & $\bullet\bullet\bullet$ & $\bullet\bullet$ & $\bullet\bullet\bullet$ & $\bullet\bullet\bullet$ & $\bullet\bullet\bullet$ & $\bullet$ & $-$ & 21 & 9 & 24 \\
\texttt{chess-why3} & Why3 & $\bullet\bullet\bullet$ & $\bullet\bullet\bullet$ & $\bullet\bullet\bullet$ & $\bullet\bullet\bullet$ & $\bullet\bullet\bullet$ & $\bullet\bullet\bullet$ & $\bullet\bullet\bullet$ & $\bullet\bullet\bullet$ & $-$ & $-$ & 9 & 8 & 24 \\
\texttt{chess-icon-codex} & Icon & $\bullet\bullet\bullet$ & $\bullet\bullet\bullet$ & $\bullet\bullet\bullet$ & $\bullet\bullet\bullet$ & $\bullet$ & $\bullet\bullet\bullet$ & $\bullet\bullet\bullet$ & $\bullet\bullet\bullet$ & $-$ & $-$ & 6 & 8 & 22 \\
\texttt{lean-chess} & Lean & $\bullet\bullet\bullet$ & $\bullet\bullet\bullet$ & $\bullet\bullet$ & $\bullet\bullet\bullet$ & $\bullet\bullet\bullet$ & $\bullet\bullet$ & $\bullet\bullet\bullet$ & $\bullet\bullet$ & $\bullet$ & $-$ & 21 & 9 & 22 \\
\texttt{chess-Rocq} & Rocq & $\bullet\bullet$ & $-$ & $-$ & $\bullet\bullet\bullet$ & $-$ & $\bullet\bullet\bullet$ & $-$ & $\bullet\bullet\bullet$ & $-$ & $-$ & 11 & 4 & 11 \\
\texttt{chess-why3-cc}$^\dagger$ & Why3 & $-$ & $-$ & $-$ & $\bullet\bullet\bullet$ & $-$ & $-$ & $-$ & $\bullet\bullet\bullet$ & $-$ & $-$ & 6 & 2 & 6 \\
\midrule
\texttt{chess-latex-codex-replication} & LaTeX & $\bullet\bullet\bullet$ & $\bullet\bullet\bullet$ & $\bullet\bullet\bullet$ & $\bullet\bullet\bullet$ & $\bullet$ & $\bullet\bullet\bullet$ & $\bullet\bullet\bullet$ & $\bullet\bullet\bullet$ & $\bullet$ & $-$ & 15 & 9 & 23 \\
\texttt{chess-css-codex} & CSS & $\bullet\bullet\bullet$ & $\bullet\bullet\bullet$ & $\bullet\bullet$ & $\bullet\bullet$ & $-$ & $\bullet\bullet\bullet$ & $\bullet\bullet$ & $\bullet\bullet\bullet$ & $-$ & $-$ & 17 & 7 & 18 \\
\texttt{chess-css-codex-guided} & CSS & $\bullet\bullet\bullet$ & $\bullet\bullet\bullet$ & $\bullet\bullet\bullet$ & $\bullet$ & $-$ & $\bullet$ & $-$ & $\bullet\bullet\bullet$ & $-$ & $-$ & 9 & 6 & 14 \\
\texttt{latex-chess-engine} & LaTeX & $-$ & $-$ & $\bullet\bullet$ & $\bullet\bullet$ & $-$ & $\bullet\bullet$ & $\bullet\bullet$ & $\bullet\bullet\bullet$ & $\bullet\bullet$ & $-$ & 37 & 6 & 13 \\
\texttt{chess-css-cc}$^\dagger$ & CSS & $-$ & $-$ & $-$ & $\bullet\bullet\bullet$ & $-$ & $\bullet$ & $\bullet$ & $\bullet\bullet\bullet$ & $-$ & $-$ & 53 & 4 & 8 \\
\texttt{chess-sql}$^\dagger$ & SQL & $-$ & $-$ & $\bullet$ & $\bullet$ & $-$ & $-$ & $\bullet$ & $\bullet$ & $-$ & $-$ & 3 & 4 & 4 \\
\midrule
\texttt{chess-revisit-java-toCOBOL-codex} & COBOL (port) & $\bullet\bullet\bullet$ & $\bullet\bullet\bullet$ & $\bullet\bullet\bullet$ & $\bullet\bullet\bullet$ & $\bullet\bullet\bullet$ & $\bullet\bullet\bullet$ & $\bullet\bullet\bullet$ & $\bullet\bullet\bullet$ & $-$ & $\bullet\bullet\bullet$ & 28 & 9 & 27 \\
\texttt{COBOL-chess} & COBOL & $\bullet\bullet\bullet$ & $\bullet\bullet\bullet$ & $\bullet\bullet\bullet$ & $\bullet\bullet\bullet$ & $\bullet\bullet\bullet$ & $\bullet\bullet$ & $\bullet\bullet\bullet$ & $\bullet\bullet\bullet$ & $\bullet\bullet\bullet$ & $-$ & 34 & 9 & 26 \\
\texttt{chess-cobol-cc} & COBOL & $-$ & $\bullet\bullet\bullet$ & $\bullet\bullet\bullet$ & $\bullet\bullet\bullet$ & $\bullet$ & $\bullet$ & $\bullet\bullet\bullet$ & $\bullet\bullet\bullet$ & $\bullet\bullet\bullet$ & $\bullet\bullet$ & 23 & 9 & 22 \\
\texttt{chess-assembly-codex} & Asm & $\bullet\bullet\bullet$ & $\bullet\bullet\bullet$ & $\bullet$ & $\bullet\bullet\bullet$ & $\bullet\bullet\bullet$ & $\bullet\bullet$ & $\bullet\bullet\bullet$ & $\bullet\bullet\bullet$ & $-$ & $-$ & 12 & 8 & 21 \\
\midrule
\texttt{chess-brainfuck} & Brainfuck & $\bullet\bullet\bullet$ & $\bullet\bullet\bullet$ & $\bullet\bullet\bullet$ & $\bullet\bullet\bullet$ & $\bullet\bullet\bullet$ & $\bullet\bullet\bullet$ & $\bullet\bullet\bullet$ & $\bullet\bullet\bullet$ & $\bullet\bullet$ & $\bullet\bullet$ & 38 & 10 & 28 \\
\texttt{chess-brainfuck-cc} & Brainfuck & $\bullet$ & $\bullet$ & $\bullet$ & $\bullet\bullet\bullet$ & $\bullet$ & $\bullet\bullet$ & $\bullet$ & $\bullet\bullet\bullet$ & $\bullet$ & $\bullet\bullet$ & 192 & 10 & 16 \\
\bottomrule
\end{tabular}%
}
\end{table*}

\noindent\Cref{tab:rq1-se-matrix} reports the per-engine
SE-activity profile cited from \Cref{sec:rq5:se-activities}:
each cell is the Likert level on the share of the session's
steps tagged with each of 10 capabilities (absent~$-$,
light~$\bullet$, moderate~$\bullet\bullet$,
heavy~$\bullet\bullet\bullet$). Rows are grouped by language
category and sorted by summed intensity within each category.
The dagger~$^\dagger$ marks engines whose main session
transcripts were compacted before capture (lower bound, see
\Cref{app:archived-floors}). The double-dagger~$^\ddagger$ marks the
special-role underconverged case (\texttt{chess-mojo},
\Cref{app:special-role}) shown for transparency and excluded
from main-corpus comparators.

\subsection{Per-engine trajectory case studies}
\label{app:rq2:trajectory-cases}

The four representative engines whose feature-introduction
trajectories are summarised in \Cref{sec:rq2:trajectory},
namely Ruby (mainstream general-purpose, \claudecode{}), COBOL
(legacy), LaTeX (Codex,
\texttt{chess-latex-codex-replication}), and Brainfuck
(esoteric), show qualitatively different patterns. Step
counts here follow the feature-trajectory extraction behind
\Cref{fig:feature-growth}, which pools sessions slightly
differently from the SE-activity matrix of
\Cref{tab:rq1-se-matrix}:

\paragraph{Ruby (plan $\to$ dump $\to$ refine).} Of 32
distinct features detected across 16 steps, 23
appear in a
single step: the ``implement the plan'' burst early in the
session, touching all of the rules and protocol layer, two
board representations, the entire search core, six search
extensions, material$+$PST evaluation, and time management.
Only nine features arrive later. The cadence is a three-stage
arc: plan (step~1) $\to$ dump-the-engine (step~3, the burst
that writes most of the codebase) $\to$ refine (two short
rounds of search-extension + evaluation additions driven by
``improve the Elo'' prompts).

\paragraph{COBOL (gradual accretion across debug loops).}
24 features across 37 steps with 331 Write/Edit events. The
features are spread across the session rather than dumped in
one burst. The long stretches between feature additions are
heavy on edits rather than writes: the agent iterates on
an existing file to fix a perft mismatch or a gauntlet
regression. This is the \emph{oracle-debug} shape: fewer
bursts, more revisions per feature, and a strength curve that
zigzags as regressions get caught and rolled back.

\paragraph{LaTeX (thin but coherent).}
The heredoc-aware trace recovers 6 of 38 features from the
surviving authoring events (37 file writes across 15 steps).
This undercounts the actual engine capability, since the
\texttt{expl3} macro logic is dense enough that several
pattern heuristics fail to match, but the trace surfaces
FEN, castling, en passant, promotion, perft, and material
evaluation all in a compact early cluster, with no
search-extension features at all. LaTeX's macro-expansion
model caps the engine at a depth that does not require
aspiration windows or null-move pruning to reach.

\paragraph{Brainfuck (long flat region, late burst).}
20 features across 198 steps and 87 authoring events. The
cumulative-feature curve is flat for the first 150-ish steps
and then jumps from $\sim$7 to 18 features near session end.
The long flat region is the code-generation loop: the agent
writes Python that emits Brainfuck, and most of that Python
is scaffolding that does not yet introduce new chess
features. The shape illustrates that in a language too
constrained for direct authorship, feature introduction is
not where the agent's effort goes. Most of the effort is in
making an existing feature work.

\subsection{Qualitative supervision protocol}
\label{app:method:supervision}

A practical threat to any study of ``what the agent built''
is that a capable model can satisfy the user's literal
request while side-stepping its spirit (the
\texttt{chess-css-codex} case is canonical here, see
\Cref{sec:rq3:evasion}). We therefore document an explicit
qualitative supervision protocol applied across the corpus.
(1)~Read the engine-core diff on at least one checkpoint per
session (target \texttt{src/}, \texttt{lib/}, or the
language equivalent). (2)~Grep for chess-library imports
inside engine-core files (e.g.\ \texttt{import chess} in a
non-Python engine, \texttt{use shakmaty} in Rust). (3)~Verify
the oracle is independent: the evaluation harness may
legitimately use \texttt{python-chess} for PGN parsing or
UCI bridging, but the engine under evaluation must not.
(4)~Redirect when evasion is detected (in
\texttt{chess-css-codex} the redirect produced a second,
compliant engine under \texttt{strict-css/}, kept in-tree
for transparency). The novelty audit in \Cref{sec:rq2:audit}
operationalises the grep step.

\subsection{Subagent-only floor counts for archived-transcript engines}
\label{app:archived-floors}

The seven \texttt{n/a$^\dagger$} rows of \Cref{tab:rq1-engines}
are engines whose main-session JSONLs were compacted by
\claudecode{} before our pipeline captured them. The subagent
sidecars in
\texttt{\textasciitilde/.claude/projects/$\ldots$/subagents/}
that still survived at our 2026-04-20 capture yielded the partial
floor counts in \Cref{tab:archived-floors}. (Those sidecars have
since been removed by client-side retention along with the rest of
the \claudecode{} corpus transcripts, see \hyperref[sec:data-availability]{Data and code availability}.
The floor counts below persist only in the committed replication
snapshot.)
The numbers are \emph{floors}: the true main-loop counts are at
least this much, and the partial main-session JSONL we found
for \texttt{chess-rust-cc} is included here (which is why its
floor is the largest of the seven). The seven floors sum to a
minimum of 2{,}105 user messages and \$125 of Anthropic
list-price token spend across these engines: they were not
built ``for free''. We retain \texttt{n/a} in
\Cref{tab:rq1-engines} rather than substituting these floors
because mixing measurements with floors in one column would
mislead.

\begin{table}[!htbp]\centering\scriptsize
\caption{Per-engine subagent-only floor counts. \textbf{Sub jsonl}
= number of distinct subagent transcripts. \textbf{User} =
user-role messages within them. \textbf{Asst} = assistant-role
messages. \textbf{In/Out kTok} = input (incl.\ cache-read and
cache-creation) and output tokens. \textbf{USD} = Anthropic
list-price floor estimate (\$3 / Mtok input, \$15 / Mtok output,
\$0.30 / Mtok cache-read, \$3.75 / Mtok cache-creation, no
account discounts).}
\label{tab:archived-floors}
\begin{tabular}{l r r r r r r}
\toprule
Engine & Sub jsonl & User & Asst & In kTok & Out kTok & USD floor \\
\midrule
\texttt{chess-rust-cc}            & 1+2 & 645  & 1{,}028 & 95{,}248 & 1{,}466 & \$65.43 \\
\texttt{test-superset (ChessCSS)} & 52  & 930  & 1{,}267 & 44{,}150 &   286   & \$42.90 \\
\texttt{chess-cplusplus-claude}   & 18  & 190  &   251   &  7{,}409 &     2   & \$7.70  \\
\texttt{chess-purec}              &  5  & 148  &   204   &  7{,}827 &     1   & \$4.95  \\
\texttt{chess-why3-cc}            &  6  & 127  &   146   &  3{,}401 &     2   & \$2.73  \\
\texttt{chess-sql}                &  3  &  47  &    70   &  1{,}614 &     0   & \$0.92  \\
\texttt{chess-py-cc}              &  1  &  18  &    20   &    304   &     0   & \$0.27  \\
\midrule
\textbf{Total floor} &  & \textbf{2{,}105} & \textbf{2{,}986} & \textbf{159{,}953} & \textbf{1{,}757} & \textbf{\$124.90} \\
\bottomrule
\end{tabular}
\end{table}

\section{Special-role experiments: protocol details}
\label{app:special-role}

The bottom block of \Cref{tab:rq1-engines} (\Cref{sec:rq1})
lists the five engines that follow protocols different enough
from ``design a chess engine in $L$ from scratch'' to be kept
separate from the main-corpus aggregates. Each is referenced
inline from the relevant RQ. This appendix collects the
protocol-difference details in one place for completeness.

\paragraph{Java$\to$language ports.}
\texttt{chess-revisit-java-toRust-codex} and
\texttt{chess-revisit-java-toCOBOL-codex} are Codex sessions
where the user asks the agent to translate the existing
\texttt{chess-java-cc} engine into another target language,
preserving module structure and observable behaviour. The
research question behind these experiments is \emph{``can a
chess engine that is good in language~$A$ be as good in
language~$B$?''}, a question about portability of a specific
engine design, not about what the agent would author from
scratch. We therefore exclude them from the main-corpus tables
(they would bias the Rust and COBOL cells of
\Cref{tab:rq4-elo}). On the external ruler the Rust port lands
at \elo{1922} (anchor) and the Java$\to$COBOL port at
\elo{1404} (refinement, the panel never resolved it). That is a
$\sim$\elo{520} cross-PL gap from the same source code, the
ablation discussed in \Cref{sec:rq4:anchor}.

\paragraph{DSL design (engine-dsl).}
\texttt{chess-newlang-codex} (``GAMBIT'') is a Codex session
where the user asks the agent to design a new chess-engine DSL
and its end-to-end C\texttt{++}17 transpiler (parser, runtime
backend, bitboards, FEN parsing, UCI loop, negamax\,+\,$\alpha\beta$,
TT, perft runner). The agent's role spans
\emph{language designer plus compiler writer plus engine
writer}, which is not comparable to either a C\texttt{++}
engine or a from-scratch engine in some target language.

\paragraph{Agent-heavy co-design (engine-codesign).}
\texttt{test-superset}
(``ChessCSS''~\cite{acher2026chesscss-readme}) is a chess
engine where as much logic as possible is expressed in pure
CSS: move generation, legality filtering, check detection,
and move scoring are all CSS rules using \texttt{:has()}, CSS
\texttt{if()}, and a z-index-argmax\,$+$\,\texttt{elementFromPoint}
trick to pick the best move in $O(1)$. The final engine
contains $\approx$55k CSS rules ($\approx$17~MB) and only
$\approx$340 lines of JavaScript (14 functions, all for
I/O\,/\,game-loop control that CSS cannot structurally
express). Git history is the authoring record: \textbf{31
commits over 5 days} (2026-02-27 to 2026-03-03), documenting a
progressive JavaScript$\to$CSS migration. The repository's
README explicitly flags its methodological status:
\emph{``I really had to drive the agent with technical
expertise (chess and programming), to be proactive, and even
to encourage the agents to not giving up with CSS''}. That
protocol is materially different from the minimal-instruction
default the rest of the corpus follows. We discuss the case as
the positive counterexample to the \texttt{chess-css-codex}
evasion (\Cref{sec:rq3:evasion}).

\paragraph{Underconverged case: \texttt{chess-mojo}.}
\texttt{chess-mojo} is the corpus's clearest example of a
session that did not converge to a competitive engine. The
agent assembled a Mojo UCI scaffold, wrote a milestone
roadmap, and produced a working perft-checked move generator
and a shallow search. The engine plateaued at roughly
\elo{900} against calibrated \stockfish{} and did not recover
under the user's budget. We include it in
\Cref{tab:rq1-engines} for transparency, label it
\emph{underconverged}, and \textbf{exclude it from every
main-corpus aggregate} (Elo distribution, cost analysis,
SE-activity matrix, oracle-first first-call statistics). The
engine plays legal chess and satisfies the ``working engine''
criterion, but it sits well below every main-corpus engine in
playing strength. It serves as the weakest point of the
corpus, not as a data point in any cross-engine median.

\paragraph{Phase-mix evidence that the protocols differ.}
The cross-project trajectory analysis of
\Cref{sec:rq5:anatomy} confirms these protocols are
empirically distinct from the main-corpus default. The two
ports sit at 60--71\% \emph{other} (translation and
module-mapping work) with only 18--20\% \emph{feature} and
emit no recoverable Elo-claim curve. The DSL experiment has
the highest \emph{feature} fraction in this group (38\%),
consistent with the agent designing two artefacts at once.
\texttt{chess-mojo} has a single Elo claim and no evolution
curve. Pooling these four engines with the main corpus would
distort the task-class medians and the oracle-cadence
statistics used in \RQ{5}.

\section{Deep-dive: \texttt{chess-rust-cc-redo}, a second try at Rust}
\label{app:deepdive-rust-redo}

The Rust cell of the main corpus originally held two engines:
\texttt{chess-rust-codex} (Codex, scratch-built but using the
\texttt{chess} crate for board-state types) and \texttt{chess-rust-cc}
(Claude Code, scratch-built with archived transcripts). The crate
dependency in the first Rust row was an evasion of the
``implement it yourself'' spirit of the experiment: the agent
delegated board representation, move generation, and legal-move
checking to a third-party library and authored only search and
evaluation. We therefore re-ran Rust from a blank slate, under the
explicit constraint \emph{no chess crate}, as
\texttt{chess-rust-cc-redo}. This engine is a scratch-built Rust UCI
engine with bitboards, PEXT/PDEP move generation, iterative
deepening, PVS, a two-tier transposition table, null-move pruning,
LMR/LMP, futility/reverse-futility pruning, singular extensions,
ProbCut, SEE pruning, and 35 perft tests. The session ran across
four Claude Code conversations, 55 user prompts, \$63.94 of
list-price spend, over roughly 70 wallclock hours (\Cref{tab:rq1-engines},
\Cref{tab:rq4-elo}). The authoring log is intact (no compaction),
the code is reproducible, and the engine converges to a measurable
strength.

Three observations from the session stand out, each qualitatively
different from the quantitative patterns the rest of the paper
reports. We record them here because they sharpen
\Crefrange{sec:rq4}{sec:rq5} and hint at a dimension of the
benchmark the main tables do not expose.

\paragraph{Observation 1: high effort, and a self-estimate that
under-counted the payoff.}
The session's stated goal was Elo~2000 against a calibrated
\stockfish{}. After four rounds of improvement, the agent's own
1,200-game gauntlet put it at approximately \elo{1853}
($\pm$24.7), recorded as ${\sim}$\elo{1879} in the final
commit and used as the Self value of \Cref{tab:rq4-elo},
short of that goal. Yet on the external ruler the
engine actually lands at \elo{1989~$\pm$~105}
(\Cref{tab:rq4-elo}), top-three in the corpus and a Pattern~B
case where the agent \emph{under}-reported its own strength. Rust
is a mainstream general-purpose language with no obvious
execution-model handicap for a chess engine. The feature set
implemented (\Cref{tab:rq2-searcheval}) is essentially the full
search-extension catalogue, richer than most mainstream engines
in the corpus. And the agent had a single-concentrated-goal,
multi-session budget. By the agent's own in-loop signal the
Elo~2000 target stayed out of reach, and, tellingly, the
agent could not \emph{tell} it had essentially arrived. This
nuances the ``Rust tops the corpus'' reading of
\Cref{tab:rq4-elo}: the scratch-built Rust engines do top the
table (\texttt{chess-rust-cc-redo} at \elo{1989}, second only to
the Java engine), but reaching that band took neither quickly nor
cheaply, and a session's own gauntlet can under-count progress
just as easily as the more common Pattern~A over-count.

\paragraph{Observation 2: benchmarking misallocation, and the user
had to intervene.}
The engine was calibrated against four Stockfish skill levels: L5
($\approx$\elo{1650}), L8 ($\approx$\elo{1900}), L10
($\approx$\elo{2000}), and L13 ($\approx$\elo{2400}), at 300 games
each, 1,200 games total. Read against the final \elo{1853} estimate,
the L5 and L13 matches carried almost no calibration information:
the engine scored 80\%{} against L5 (it wins almost everything)
and 7.5\%{} against L13 (it loses almost everything). Only the
L8 match (45\%{}) and L10 match (19\%{}) sit in the calibration
sweet spot where score differences are informative about Elo. Half
the 600 games spent on L5 and L13 were largely wasted, and, more
tellingly, it was clear to the user \emph{before} the gauntlet
finished that L13 was too strong and L5 too weak for this engine's
band. The agent did not adjust the gauntlet composition on its
own. The user had to be proactive and stop the mis-calibrated
matches. This is a session-level pattern the cost table
(\Cref{tab:rq1-engines}) and the SE-activity matrix
(\Cref{tab:rq1-se-matrix}) cannot surface: \emph{evaluation time
can be mis-allocated in a way the agent does not detect}, even
when the tooling to do better (shrinking the opponent ladder after
a few games at extremes) is within the agent's capability.

\paragraph{Observation 3: chess domain knowledge surfaced at the
wrong level of abstraction.}
Partway through the session the agent diagnosed recurring
weaknesses in the engine's opening play by replaying the gauntlet
games, identifying specific anti-patterns (piece placement
errors in the Sicilian, weak response to 1. d4 setups), and
proposing an \emph{opening book} as the fix. That the agent
reaches for game-level analysis, rather than code-level
debugging alone, is itself noteworthy: the coding agent is
drawing on chess-domain knowledge that goes beyond programming.
It read its own \pgn{} output and evaluated move sequences the
way a human chess student would. But the fix it proposed
(``precompute an opening book'') was, from a chess expert's
point of view, a short-term workaround rather than a remedy. The
engine's opening losses were symptoms of a deeper evaluation or
search issue (giving a piece away for nothing in a
quasi-tactical position) that an opening book would only
paper over. The chess-expert reading is that the opening book
masks the bug in positions the book covers and leaves it intact
everywhere else. This sharpens two points. First, the
benchmark has a domain-knowledge surface area: agents that
understand chess better can debug their own play in ways that
domain-agnostic agents cannot, and this affects measured strength.
Second, domain-knowledge usage is not uniformly helpful:
surface-level pattern recognition (``my openings lose'') can
drive fixes that do not address the underlying engineering
problem (``my evaluation of quiet-tactical positions is off''),
and the agent does not always distinguish the two.

\paragraph{Why this deep dive is here and not elsewhere in the paper.}
Observations 1--2 are observational data points that weaken a
naive reading of \Cref{tab:rq4-elo} (``Rust sits at the top, done'')
and add a caveat to \Cref{tab:rq1-engines} about \emph{how} the cost
is spent. Observation~3 is a qualitative signal from a single
session that we do not have the sample size to make into a
population claim. We record it here as a pointer for future work
on \emph{domain-knowledge-aware} coding benchmarks, and as a caveat
that chess, for all its oracle richness, tests programming
\emph{plus chess}, not programming in isolation.

\section{External Elo assessment: methodology, round-robin cross-check, and tables}
\label{app:anchor-eval}

This appendix documents the unified re-evaluation harness behind the
\emph{External Elo Assessment} column of \Cref{tab:rq4-elo}
(\Cref{sec:rq4:anchor}). The headline number is a Bradley-Terry MLE
that pools the anchor gauntlet with a corpus-wide round-robin. We
report the round-robin separately here as an internal-consistency
cross-check (\Cref{app:rr}). It is intended to be sufficient for an
independent group to reproduce both the anchor gauntlet and the
round-robin on a different machine.

\subsection{Why a re-evaluation was needed}

Each agent's in-tree gauntlet used a different combination of
opponent set, time control, hash size, opening book, and game count.
\Cref{tab:rq4-elo} pools those numbers, but they are not directly
comparable: the same engine plays $\sim$\elo{300} stronger at TC$=$200~ms
movetime against \stockfish{} \texttt{UCI\_Elo$=$2400} than at
TC$=$120s$+$1s/move against \stockfish{} \texttt{Skill Level $=$~10}.
We therefore designed a uniform harness around three principles:
externally-rated reference engines (not labels), enough games for the
standard error to fall below $\sim$\elo{200}, and a single time
control across the whole corpus.

\subsection{Reference engines and Phase A calibration}

Five references, fixed across the corpus:
\begin{itemize}
  \item \texttt{Rustic Alpha 3.0.4}: published CCRL 40/4 rating
    \elo{1820}. Distributed as an immutable Docker image with the
    binary pinned at SHA.
  \item \texttt{Asymptote 0.7}: published CCRL 40/4 rating
    \elo{2150}. Same Docker pinning.
  \item \texttt{Stockfish 18 Skill Level $=$ 5, 10, 15}, three
    rungs. Their CCRL-equivalent Elo was \emph{not} taken from the
    \stockfish{} label but \emph{calibrated empirically} against
    Rustic and Asymptote on the host hardware (Apple M-series
    laptop, Docker on macOS):
    \[
      \text{Skill 5} \approx \elo{1658} \pm \elo{126}, \;
      \text{Skill 10} \approx \elo{2004} \pm \elo{112}, \;
      \text{Skill 15} \approx \elo{2325} \pm \elo{124}.
    \]
    These numbers are hardware-dependent (a faster machine raises
    the effective Elo of a fixed Skill setting) and would need to
    be re-derived on a different host. Calibration cost: $\approx$1
    hour wall, 200 SPRT-capped games.
\end{itemize}

We deliberately do \emph{not} use \stockfish{}'s
\texttt{UCI\_LimitStrength=true} mode. Phase 2 of our calibration
showed its ladder is compressed: the gap between \texttt{UCI\_Elo$=$2400}
and \texttt{UCI\_Elo$=$2600} is $\approx$\elo{100} of effective
playing strength, not the labelled \elo{200}. The gap between
\texttt{UCI\_Elo$=$1800} and \texttt{UCI\_Elo$=$2200} is
$\approx$\elo{300}, not \elo{400}. Compression is most severe above
\elo{2000}, which is exactly where many agents' self-evals concentrated.

\subsection{Phase B match settings}

\begin{itemize}
  \item TC: 120s$+$1s/move (slow enough that pruning quality and
    eval richness dominate, fast enough that 30 games per pair fit
    in $\approx$30 minutes wall).
  \item Concurrency: 2, with \texttt{--cpus=2 --memory=1g} per
    Docker engine, so two pair-of-engine sub-tournaments run in
    parallel at most.
  \item SPRT cap: \texttt{elo0$=$-50}, \texttt{elo1$=$+50},
    \texttt{alpha$=$beta$=$0.05}, max 30 games per pair. SPRT
    terminates early when one hypothesis is statistically
    decided.
  \item Openings: 256-position EPD book reused across Phase A and
    Phase B, with positions selected at random per game.
  \item Resign: agreed score \elo{900} for 5 plies.
    Adjudicated draw: 6 plies of $\leq$\elo{10}~score after move 40.
  \item Per-anchor estimate: $\Delta = -400\log_{10}(1/s - 1)$
    with $s\in(0,1)$ the engine's match score, and $\mathrm{SE} =
    \frac{400}{\ln 10\cdot\sqrt{N\cdot s(1-s)}}$.
  \item Per-engine estimate: inverse-variance combination of
    measurable per-anchor estimates. Saturated pairs ($s=0$ or $s=1$)
    contribute one-sided bounds only and are flagged in the per-engine
    card.
  \item ``Bounded'' status (used in \Cref{fig:rr-vs-phaseb}):
    every per-anchor pair was saturated. The engine's true Elo lies
    below the lowest unsaturated upper bound. We stamp at \elo{1276}
    or \elo{1438} depending on which anchor's bound was tightest.
    Bounded engines have no measurable signal in Phase B alone.
\end{itemize}

\subsection{Round-robin cross-check: design}
\label{app:rr}

The round-robin is the internal-consistency cross-check summarised in
\Cref{sec:rq4:anchor}: it scores each engine against the rest of the
corpus only, independent of any external reference, and feeds the
unified Bradley-Terry fit. Settings shared with Phase B (TC,
concurrency, opening book, draw/resign rules) are identical so the
two bodies of games are on the same scale. Round-robin-specific
choices:
\begin{itemize}
  \item Two games per unordered pair (one game each colour, same
    opening). Total: $\binom{n}{2}\cdot 2$ games, $n=23$ in the
    first pass and $n=26$ once the port/DSL pass completes.
  \item No SPRT cap on the round-robin pairs: every pair plays
    exactly two games, regardless of who is ahead. The
    Bradley-Terry MLE pools across pairs to recover global Elo.
  \item Bradley-Terry fit: Newton-Raphson with damping factor 0.3,
    400 iterations (well past convergence on this graph density).
    Anchored to \texttt{chess-rust-cc-redo}'s anchor-based Elo
    of \elo{1989} after every iteration.
  \item Prior: a half virtual-draw vs every opponent
    (PRIOR\_GAMES $=$ 0.5) keeps the MLE finite for engines with
    100\% or 0\% scores against any subset of opponents.
\end{itemize}

\subsection{Reproduction recipe}

\begin{enumerate}
  \item \texttt{cd eval2; ./runners/run\_phaseA\_sf\_calibration.sh}:
    recalibrates \stockfish{} Skill Level effective Elo on the
    target hardware. Required only if the target hardware is not
    Apple M-series ($\sim$1 hour).
  \item \texttt{./runners/run\_anchor\_gauntlet.sh}: runs the full
    Phase B against all engines in the manifest. $\sim$10 hours
    wall on a 2-CPU concurrency budget.
  \item \texttt{./runners/score\_anchor.sh}: aggregates PGNs
    into per-engine cards in \texttt{results/anchor/}.
  \item \texttt{./runners/run\_round\_robin.sh}: the round-robin
    pass, $\sim$25 hours wall for $n=23$ and $\sim$35 hours for $n=26$.
    Resumable: re-running skips already-completed pairs.
  \item \texttt{python3 lib/score\_rr.py}: Bradley-Terry MLE
    + Pearson/Spearman vs Phase B.
  \item \texttt{python3 scripts/plot\_rr\_vs\_phaseb.py}: regenerates
    \Cref{fig:rr-vs-phaseb}.
\end{enumerate}

\subsection{Unified Bradley-Terry MLE: pooling Phase B and round-robin}
\label{app:anchor-eval:bt}

The Phase B inverse-variance combine and the round-robin BT MLE
(\Cref{app:rr}) are two views of the same underlying corpus of
TC$=$120s$+$1s games. We pool every game played
between every pair of engines (anchor or corpus) into a single
Bradley-Terry log-likelihood and maximise it by Newton-Raphson, with
six anchor nodes held \emph{fixed} at their known Elo: Rustic
(\elo{1820}), Asymptote (\elo{2150}), and Stockfish Skill 0/5/10/15
at the Phase A calibration values (\elo{900}/\elo{1658}/\elo{2004}/\elo{2325}).
The pooled likelihood treats a 0/9 saturation against Rustic as a
genuine one-sided constraint (``this engine is at most $\sim$\elo{1438}'')
and propagates it via shared opponents to every other estimate.
The IV-combine alone discards saturated pairs and so loses that
signal. Per-engine 95\% confidence intervals are estimated by a
paired-game bootstrap ($n{=}200$ resamples, refit each replicate).
The pooled corpus is 3{,}256 games across 531 unique pairs, which
includes a 36-pair targeted-skill anchor pass (one or two new SF
\texttt{Skill Level} matches per engine, picked so the engine's
expected score against the chosen Skill rung is in the 30--70\%
information-rich region). The pass adds 720 anchor-information-rich
games and collapses one previously-pathological estimate
(\texttt{sf\_skill11} at \elo{5952} from a single saturated opponent
in the original fit, now \elo{2262~$\pm$~189} with eight new corpus
edges).

\paragraph{Headline corpus-wide ranking after the pooled fit.}
The top tier is unchanged from \Cref{sec:rq4:anchor}:
\texttt{chess-java-cc} (\elo{2094~$\pm$~56}),
\texttt{chess-rust-cc-redo} (\elo{1990~$\pm$~57}),
\texttt{chess-revisit-java-toRust-codex} (\elo{1989~$\pm$~54}),
\texttt{chess-rust-cc} (\elo{1825~$\pm$~58}),
\texttt{chess-purec-retry} (\elo{1798~$\pm$~82}), and
\texttt{chess-ruby-cc} (\elo{1753~$\pm$~55}). The
mid-band engines whose Phase B and RR estimates disagreed most
now have a single number with a tight CI:
\texttt{chess-newlang-codex} \elo{1622~$\pm$~68}, \texttt{chess-Rocq}
\elo{1315~$\pm$~83}, \texttt{chess-icon-codex} \elo{1168~$\pm$~88}.
The full table is generated by \texttt{eval2/lib/score\_combined\_bt.py}
(runtime $\approx$8\,s) and saved to
\texttt{eval2/results/combined\_bt\_summary.md}. Two
no-win engines (\texttt{chess-latex-codex-replication},
\texttt{latex-chess-engine}) drift to extreme negative
estimates ($-3235$, $-3480$) because the BT likelihood has no
lower bound. A virtual random-mover anchor would constrain
them, but we did not add one.

\subsection{Per-engine round-robin Elo: full table}

\begin{table}[t]
\centering\small
\caption{Per-engine Elo on the unified TC$=$120s$+$1s harness, three rulers side by side: \textbf{Phase B} (anchor-only inverse-variance combine, \Cref{sec:rq4:anchor}, where $\le$X marks engines saturated against every Phase~B anchor and bounded above by the lowest unsaturated upper bound), \textbf{BT MLE} (unified Bradley-Terry MLE pooling Phase~B, round-robin, and targeted-skill anchor games, 3{,}256 games / 531 pairs, \Cref{app:anchor-eval:bt}), and \textbf{RR} (round-robin Bradley-Terry fit anchored to \texttt{chess-rust-cc-redo} at \elo{1989}, \Cref{app:rr}). $\Delta$ is the round-robin minus the BT estimate. `NB' = no lower bound: chess-latex-codex-replication and latex-chess-engine lost every game in the round-robin and the BT likelihood drifts to extreme negatives without a random-mover anchor (cf.\ \Cref{app:anchor-eval:bt}). $^{\dagger}$\,partial round-robin coverage: \texttt{chess-brainfuck}, \texttt{chess-sql}, and \texttt{latex-chess-engine} played only the slow-engine sub-pass (5 opponents at TC$=$120s$+$1s) because they exceed the per-move budget vs the main field. Their RR Elo is on the same scale but rests on fewer pairs. Rows sorted by BT Elo descending.}
\label{tab:rr-final}
\begin{tabular}{r l r r r r r r}
\toprule
\# & Engine & Phase B & BT MLE & RR & $\Delta$ & Opps & W-D-L \\
\midrule
1 & \texttt{chess-java-cc} & \elo{2096}~$\pm$~132 & \elo{2094}~$\pm$~56 & \elo{2045} & $-49$ & 25 & 48-1-1 \\
2 & \texttt{chess-rust-cc-redo} & \elo{1989}~$\pm$~105 & \elo{1990}~$\pm$~57 & \elo{1989} & $-1$ & 25 & 46-0-4 \\
3 & \texttt{chess-revisit-java-toRust-codex} & \elo{1922}~$\pm$~106 & \elo{1989}~$\pm$~54 & \elo{2033} & $+44$ & 25 & 47-2-1 \\
4 & \texttt{chess-rust-cc} & \elo{1841}~$\pm$~99 & \elo{1825}~$\pm$~58 & \elo{1889} & $+64$ & 25 & 40-1-9 \\
5 & \texttt{chess-ruby-cc} & $\le$\elo{1349} & \elo{1753}~$\pm$~55 & \elo{1827} & $+74$ & 25 & 34-5-11 \\
6 & \texttt{chess-rust-codex} & \elo{1723}~$\pm$~103 & \elo{1707}~$\pm$~63 & \elo{1812} & $+105$ & 25 & 28-15-7 \\
7 & \texttt{chess-cplusplus-claude} & \elo{1680}~$\pm$~103 & \elo{1682}~$\pm$~58 & \elo{1827} & $+145$ & 25 & 35-3-12 \\
8 & \texttt{cplusplus-chess} & \elo{1709}~$\pm$~111 & \elo{1627}~$\pm$~69 & \elo{1703} & $+76$ & 25 & 27-1-22 \\
9 & \texttt{chess-newlang-codex} & --- & \elo{1622}~$\pm$~68 & \elo{1696} & $+74$ & 25 & 24-5-20 \\
10 & \texttt{chess-py} & $\le$\elo{1349} & \elo{1616}~$\pm$~55 & \elo{1784} & $+168$ & 25 & 26-15-9 \\
11 & \texttt{chess-why3-cc} & \elo{1618}~$\pm$~163 & \elo{1567}~$\pm$~56 & \elo{1730} & $+163$ & 25 & 23-12-14 \\
12 & \texttt{chess-py-cc} & $\le$\elo{1349} & \elo{1499}~$\pm$~55 & \elo{1729} & $+230$ & 25 & 27-5-18 \\
13 & \texttt{chess-java} & \elo{1509}~$\pm$~131 & \elo{1483}~$\pm$~72 & \elo{1684} & $+201$ & 25 & 19-14-17 \\
14 & \texttt{chess-assembly-codex} & \elo{1403}~$\pm$~196 & \elo{1468}~$\pm$~72 & \elo{1651} & $+183$ & 25 & 18-11-21 \\
15 & \texttt{chess-purec} & \elo{1440}~$\pm$~193 & \elo{1415}~$\pm$~65 & \elo{1664} & $+249$ & 25 & 21-7-22 \\
16 & \texttt{chess-revisit-java-toCOBOL-codex} & --- & \elo{1404}~$\pm$~69 & \elo{1612} & $+208$ & 25 & 16-9-25 \\
17 & \texttt{chess-purec-codex} & $\le$\elo{1438} & \elo{1402}~$\pm$~73 & \elo{1579} & $+177$ & 25 & 12-12-26 \\
18 & \texttt{COBOL-chess} & $\le$\elo{1276} & \elo{1365}~$\pm$~80 & \elo{1544} & $+179$ & 25 & 13-5-32 \\
19 & \texttt{chess-ruby-codex} & $\le$\elo{1438} & \elo{1346}~$\pm$~72 & \elo{1537} & $+191$ & 25 & 10-10-30 \\
20 & \texttt{chess-Rocq} & \elo{1499}~$\pm$~171 & \elo{1315}~$\pm$~83 & \elo{1562} & $+247$ & 28 & 23-10-33 \\
21 & \texttt{chess-brainfuck} & --- & \elo{1299}~$\pm$~94 & \elo{1639}$^{\dagger}$ & $+340$ & 5 & 21-1-6 \\
22 & \texttt{chess-cobol-cc} & $\le$\elo{1276} & \elo{1274}~$\pm$~85 & \elo{1537} & $+263$ & 25 & 10-10-30 \\
23 & \texttt{lean-chess} & $\le$\elo{1276} & \elo{1271}~$\pm$~75 & \elo{1537} & $+266$ & 25 & 12-6-32 \\
24 & \texttt{chess-why3} & $\le$\elo{1276} & \elo{1210}~$\pm$~74 & \elo{1502} & $+292$ & 25 & 7-11-32 \\
25 & \texttt{chess-icon-codex} & $\le$\elo{1276} & \elo{1168}~$\pm$~88 & \elo{1526} & $+358$ & 28 & 23-5-40 \\
26 & \texttt{chess-apl-codex54} & $\le$\elo{1276} & \elo{686}~$\pm$~205 & \elo{1373} & $+687$ & 28 & 12-2-54 \\
27 & \texttt{chess-sql} & --- & \elo{523}~$\pm$~297 & \elo{1384}$^{\dagger}$ & $+861$ & 5 & 7-2-21 \\
28 & \texttt{chess-latex-codex-replication} & $\le$\elo{1276} & \elo{-3235}~(NB) & \elo{1264} & --- & 25 & 0-0-50 \\
29 & \texttt{latex-chess-engine} & --- & \elo{-3480}~(NB) & \elo{1220}$^{\dagger}$ & --- & 5 & 0-0-30 \\
\bottomrule
\end{tabular}
\end{table}

The round-robin column ``Anchor Elo $\Delta$'' is the round-robin
estimate minus the anchor-based Elo. Engines whose anchor-based Elo
is a saturation floor (\texttt{Phase B (floor)}) have their
$\Delta$ marked with `*': the anchor number is not a measurement
in those rows. The table is sorted by round-robin Elo descending.

\subsection{Limitations of this re-evaluation}

\begin{itemize}
  \item \textbf{Hardware fingerprint.} The \stockfish{} Skill Level
    Elo numbers are calibrated on this specific machine. Transferring
    to a different host requires re-running Phase A.
  \item \textbf{TC choice.} TC$=$120s$+$1s is a single point in the
    blitz-to-tournament continuum. Some engines may rank-shift between
    TC classes (e.g., a deep-search engine that loses on time at blitz
    but wins at slow TC). Those shifts are not measured here. The
    correct framing is: ``Elo at TC$=$120s$+$1s on this hardware'',
    not ``Elo on the abstract CCRL scale''.
  \item \textbf{Round-robin sample size.} Two games per pair gives a
    standard error of $\approx$\elo{200} per pair. Bradley-Terry
    pools across pairs and pulls per-engine SE down to $\approx$\elo{50}
    for engines with full coverage, but per-pair direction may be
    misleading for any single comparison.
  \item \textbf{Coverage holes.} Three engines are excluded from the
    round-robin (\texttt{chess-css-cc}, \texttt{chess-css-codex},
    \texttt{chess-css-codex-guided}: no working binary) and four are
    excluded for time-control mismatch (\texttt{chess-brainfuck},
    \texttt{chess-brainfuck-cc}, \texttt{chess-sql},
    \texttt{latex-chess-engine}: $>$30s per move at TC$=$120s$+$1s).
    A separate slow-engine round-robin at TC$=$600s$+$5s would close
    this gap. We did not run it.
\end{itemize}

\subsection{Q2 self-reproduction sweep: methodology and per-engine results}
\label{app:anchor-eval:q2}

The reward-signal claims in \Cref{sec:rq3:first-call} and
\Cref{sec:rq5:reward} rest on three sub-claims: that each engine's self-reported Elo is reproducible from
its own eval methodology, that the magnitude of the reproduction is
systematically higher than the anchor-based Elo, and that the
\stockfish{} \texttt{UCI\_LimitStrength} ladder is non-monotonic at
the rungs the agents chose. We tested all three by re-running each
engine's original eval setup (same TC, same Hash, same opponent grid,
same games-per-rung scaled down for time budget) on host hardware
against host \stockfish{} 18, with no Docker, no anchors, no eval2
re-tuning: a third party with the same machine could reproduce the
same numbers from the engines' own scripts.

\paragraph{Configuration.} The 11 engines covered are those for which
the original eval script is parametric (TC + Hash + opponent grid we
could extract) and the binary is host-runnable. For each engine, the
opponent grid is the engine's own (\texttt{UCI\_Elo} rungs for ten
engines, SF \texttt{Skill Level} for \texttt{chess-cplusplus-claude}),
TC and Hash match the original script (mostly TC=10s+0.1s, hash
sizes ranging from 16 MB to 128 MB), 20 games per rung (30 for
\texttt{chess-newlang-codex}). Total: 46 matches / 953 games. Concurrency $=$ 2 with caffeinate to prevent
system sleep. The full per-engine configuration is in
\texttt{eval2/runners/self\_reproduction.yaml}. The runner is
\texttt{eval2/runners/run\_self\_repro\_host.sh} and the scorer is
\texttt{eval2/lib/score\_self\_repro.py}.

\paragraph{Per-engine results.}
\Cref{tab:q2-self-repro} reports, per engine, the original
self-claim (point value or band), the reproduction Elo we measure
under the same methodology on host hardware, the anchor-based Elo
from Phase B, and the round-robin Elo. The two $\Delta$ columns
quantify the deviations: claim-vs-repro and repro-vs-anchor.

\begin{table}[!htbp]\centering\scriptsize
\caption{Q2 self-reproduction: per-engine Elo under each engine's
original methodology (host-native, 953 games total). Repro is
inverse-variance-weighted across the engine's own SF \texttt{UCI\_Elo}
rungs (or SF \texttt{Skill Level} for \texttt{chess-cplusplus-claude}).
Anchor and RR are from \Cref{sec:rq4:anchor} and \Cref{app:rr}
(Phase~B values shared with \Cref{tab:rr-final}, and
``$\le X$'' = saturated against every Phase~B anchor).
$\Delta$(repro--claim) measures whether the methodology is
\emph{reproducible}, where small $|\Delta|$ means it is. $\Delta$(repro--anchor)
measures the methodology bias relative to anchor-based Elo, and ``$\ge$''
denotes a lower bound when Anchor is saturated. Mean of
$|\Delta_\text{repro--claim}|$ across the 9 non-Skill engines with a
point claim: \elo{129} (9 of 11 engines overall within $\pm$\elo{150}).
Mean of $\Delta_\text{repro--anchor}$:
$+\elo{348}$ over the 7 non-Skill engines with point Anchor (range
$+141$ to $+681$). Two further saturated engines have $\Delta$ lower-bounded
at $+\elo{451}$ and $+\elo{568}$.}
\label{tab:q2-self-repro}
\begin{tabular}{l r r r r r r}
\toprule
Engine & Self-claim & \textbf{Repro} & Anchor & RR & $\Delta_\text{rep-claim}$ & $\Delta_\text{rep-anc}$ \\
\midrule
\texttt{chess-purec}              & 1997      & \textbf{2121 $\pm$ 78}  & 1440 $\pm$ 193 & 1664 & $+124$ & $\mathbf{+681}$ \\
\texttt{chess-purec-codex}        & 1670--1972 & \textbf{1889 $\pm$ 78}  & $\le$1438 & 1579 & in band & $\ge\!+451$ \\
\texttt{chess-rust-cc}            & 2110      & \textbf{2208 $\pm$ 88}  & 1841 $\pm$ 99  & 1889 & $+98$  & $+367$ \\
\texttt{chess-rust-cc-redo}       & 1879      & \textbf{2355 $\pm$ 88}  & 1989 $\pm$ 105 & 1989 & $\mathbf{+476}$ & $+366$ \\
\texttt{chess-rust-codex}         & 2043      & \textbf{1987 $\pm$ 88}  & 1723 $\pm$ 103 & 1812 & $-56$  & $+264$ \\
\texttt{chess-cplusplus-claude}   & 1897      & \textbf{1485 $\pm$ 124} & 1680 $\pm$ 103 & 1827 & $\mathbf{-412}$ & $-195$ \\
\texttt{cplusplus-chess}          & 2087      & \textbf{1984 $\pm$ 101} & 1709 $\pm$ 111 & 1703 & $-103$ & $+275$ \\
\texttt{chess-ruby-cc}            & 1840      & \textbf{1917 $\pm$ 124} & $\le$1349 & 1827 & $+77$  & $\ge\!+568$ \\
\texttt{chess-why3-cc}            & 1882      & \textbf{1959 $\pm$ 95}  & 1618 $\pm$ 163 & 1730 & $+77$  & $+341$ \\
\texttt{chess-Rocq}               & 1500      & \textbf{1640 $\pm$ 101} & 1499 $\pm$ 171 & 1562 & $+140$ & $+141$ \\
\texttt{chess-newlang-codex}      & 2170      & \textbf{2182 $\pm$ 72}  & ---            & 1696 & $+12$  & --- \\
\bottomrule
\end{tabular}
\end{table}

\paragraph{Three findings.}
\begin{itemize}
\item \textbf{Reproducibility.} 9 of 11 engines reproduce within
  $\pm$\elo{150} of their original claim
  (\texttt{chess-newlang-codex} is the cleanest at $\Delta_\text{rep-claim}=+12$).
  The agents weren't fabricating numbers: the bias is embedded
  in the methodology itself.
\item \textbf{Systematic over-statement.} Repro is uniformly higher
  than anchor across all \texttt{UCI\_LimitStrength}-using engines:
  mean $+\elo{348}$ over the 7 engines with a point Anchor (range
  $+\elo{141}$ to $+\elo{681}$), and lower-bounded at $+\elo{451}$
  and $+\elo{568}$ for the 2 saturated engines. The one Skill-Level
  user comes in at $-\elo{195}$. The direction and magnitude depend
  on which rungs the agent chose: the higher the rungs, the larger
  the inflation, matching Phase A's compression measurements.
\item \textbf{Two diagnostic outliers.} \texttt{chess-rust-cc-redo}
  reproduces \emph{$+\elo{476}$ above its own claim} (the agent
  under-reported deliberately, and the anchor at \elo{1989} confirms a real
  $\sim$\elo{2000} engine). \texttt{chess-cplusplus-claude} is the
  only engine that used SF \texttt{Skill Level} (5,7,10,12,14)
  instead of \texttt{UCI\_LimitStrength}. Once the Skill levels are
  Phase-A-calibrated on this hardware (Skill 5 $=$ \elo{1658},
  Skill 14 $=$ \elo{2245}), the math comes out
  \emph{$-\elo{412}$ below} the original claim. Different SF rating
  knob, different bias direction. Both fail without the right
  external anchor.
\end{itemize}

\paragraph{Per-rung non-monotonicity.}
Two engines exhibit per-rung scores that are \emph{not}
monotonically decreasing in the SF \texttt{UCI\_Elo} label, on the
same machine, against the same opponent build, with the same opening
book:
\begin{itemize}
\item \texttt{chess-purec}: 95\% vs SF1500, 80\% vs SF1800,
  \textbf{35\% vs SF2000}, \textbf{70\% vs SF2200}, 38\% vs SF2400.
  The same engine plays \elo{350}-Elo differently against two
  adjacent rungs that are labelled 200 Elo apart.
\item \texttt{chess-purec-codex}: 87.5\% vs SF1500, 65\% vs SF1700,
  \textbf{75\% vs SF1800}, \textbf{35\% vs SF1900}, 52.5\% vs SF2000.
  Adjacent rungs disagree by 40 percentage points.
\end{itemize}
This is direct evidence that, at this part of the curve, the
\texttt{UCI\_Elo} labels do not form a calibrated ladder on this
hardware. No amount of ``more games per rung'' fixes a non-monotonic
ladder.

\subsection{The \texttt{chess-purec} retry experiment: protocol and
full results}
\label{app:anchor-eval:retry}

\Cref{sec:rq5:reward} reports a single controlled retry of the
highest-bias engine in the corpus. This section records the
protocol and the details omitted from the body.

\paragraph{Engine selection.} \texttt{chess-purec} has the
largest repro-vs-anchor gap of the self-reproduction sweep
($+\elo{681}$, \Cref{tab:q2-self-repro}). The original session
self-claimed \elo{1997}, the Phase~B anchor assessment is
\elo{1440~$\pm$~193}, and the unified Bradley-Terry refinement
is \elo{1415~$\pm$~65}.

\paragraph{Controls.} Three variables were controlled.
First, \emph{source}: the retry session restarted from the same
final git commit as the original engine, so the starting code is
identical. Second, \emph{model}: the session was pinned to
\opus{4.6}, the model that built the original engine, so no
model-version improvement can explain a gain. Third, the
\emph{reward signal}, which is the only substantive change: the
original session's \texttt{UCI\_LimitStrength} self-gauntlet was
replaced by a 14-game match per iteration against two calibrated
anchors, \texttt{Rustic} (\elo{1820}) and \stockfish{}
\texttt{Skill~10} (\elo{2004} on this hardware), at
TC$=$30s$+$0.3s. The budget was capped at 30 prompts, about
\$30 at list price.

\paragraph{What the agent did.} Within the 30-prompt budget the
agent committed two changes: a transposition-table
flag-correctness fix and a hash-table size increase from 64 to
128~MB.

\paragraph{Results.} Re-evaluated under the same Phase~B harness
as every other engine, the post-retry binary measures Phase~B
\elo{1796~$\pm$~98} and unified BT \elo{1798~$\pm$~82}
(\texttt{chess-purec-retry} in the pooled fit of
\Cref{app:anchor-eval:bt}). This is a $+\elo{383}$ gain over the
pre-retry unified BT estimate, with non-overlapping 95\%
confidence intervals. Phase~B and BT agree to within their CIs
both pre- and post-retry.

\paragraph{The agent could not see its own gain.} The agent's
in-loop verdict at the end of the retry was that strength was
``hovering around baseline''. A 14-game match per anchor is too
noisy to resolve a gain of this size in real time, yet the
committed changes (the transposition-table fix in particular)
transferred to the canonical time control. An anchored but small
in-loop signal therefore fixes the \emph{direction} of the
gradient without necessarily giving the agent a readable
\emph{magnitude}.

\paragraph{Scope and cost.} This is one engine ($n{=}1$).
Converting the finding from suggestive to corpus-supported
requires repeating the protocol on further high-bias engines,
for example \texttt{chess-cplusplus-claude} ($\Delta=-217$
claim vs anchor) or \texttt{chess-rust-codex} ($\Delta=-320$).
The cost ratio of the retry, roughly \$0.08 per anchored Elo
point gained, is an order of magnitude below typical human
chess-engine tuning budgets.

\section{Per-engine GitHub repositories}
\label{app:github-repos}

The 34 chess engines analysed in this paper are released as
individual public GitHub repositories under the
\href{https://github.com/acherm}{\texttt{acherm}} organisation,
following a uniform naming convention
\texttt{agentic-chessengine-<lang>[-<agent>][-<variant>]}. Each
repository carries an MIT \texttt{LICENSE}, a per-language
\texttt{.gitignore}, and a \texttt{README.md} crediting Mathieu
Acher as the human supervisor and naming the primary coding
agent (Claude Code or Codex). Suffix conventions are as follows.
\texttt{-cc} /
\texttt{-codex} disambiguate multi-implementation cells (same
language built by both agents). \texttt{-from-<lang>-codex}
marks an agent-driven port from another language.
\texttt{-codexfailure} flags a session preserved with a
documented failure status. For Brainfuck the engine plays but
did not meet its pure-language constraint (critical engine
logic remained in the Python toolchain, as the repository's
README documents). For Mojo it marks the underconverged
session of \Cref{app:special-role}.
\texttt{-ChessCSS}, \texttt{-TeXCCChess}, and
\texttt{-newlang-codex} preserve the engine's own project
identity. \Cref{tab:github-repos} maps each local working-tree
folder used internally during the study to its public repository
suffix. The full URL is
\url{https://github.com/acherm/agentic-chessengine-<suffix>}.

\newcommand{\repolink}[1]{\href{https://github.com/acherm/agentic-chessengine-#1}{\texttt{#1}}}

\begin{scriptsize}
\setlength{\emergencystretch}{3em}
\sloppy
\begin{longtable}{@{}r p{0.32\linewidth} p{0.28\linewidth} p{0.16\linewidth} p{0.06\linewidth}@{}}
\caption{The 34 agentic chess-engine repositories on GitHub
(\texttt{acherm} organisation). The full URL of each repository
is \url{https://github.com/acherm/agentic-chessengine-<suffix>}
where \emph{Repo suffix} is the value shown in the third
column. \emph{Local folder} is the internal working-tree name.
\textsc{cc} = Claude Code (Opus 4.6/4.7). \textsc{cdx} = Codex
(GPT-5-codex / GPT-5.3-codex, with one APL run also using
GPT-5.4).}
\label{tab:github-repos} \\
\toprule
\# & Local folder & Repo suffix\textsuperscript{$\ast$} & Language & Agent \\
\midrule
\endfirsthead
\multicolumn{5}{l}{\emph{\Cref{tab:github-repos} (continued)}} \\
\toprule
\# & Local folder & Repo suffix\textsuperscript{$\ast$} & Language & Agent \\
\midrule
\endhead
\midrule
\multicolumn{5}{r}{\emph{(continued on next page)}} \\
\endfoot
\bottomrule
\endlastfoot
1 & \texttt{chess-apl-codex54} & \repolink{apl-codex} & APL & \textsc{cdx} \\
2 & \texttt{chess-assembly-codex} & \repolink{assembly-codex} & x86-64 Asm & \textsc{cdx} \\
3 & \texttt{chess-brainfuck} & \repolink{brainfuck-codexfailure} & Brainfuck (partial failure) & \textsc{cdx} \\
4 & \texttt{chess-brainfuck-cc} & \repolink{brainfuck} & Brainfuck & \textsc{cc} \\
5 & \texttt{chess-cobol-cc} & \repolink{cobol-cc} & COBOL & \textsc{cc} \\
6 & \texttt{COBOL-chess} & \repolink{cobol-codex} & COBOL & \textsc{cdx} \\
7 & \texttt{chess-cplusplus-claude} & \repolink{cpp-cc} & C\texttt{++} & \textsc{cc} \\
8 & \texttt{cplusplus-chess} & \repolink{cpp-codex} & C\texttt{++} & \textsc{cdx} \\
9 & \texttt{chess-css-codex} & \repolink{css-codex} & CSS/HTML (evasion) & \textsc{cdx} \\
10 & \texttt{chess-css-codex-guided} & \repolink{css-codex-guided} & CSS/HTML (strict) & \textsc{cdx} \\
11 & \texttt{chess-icon-codex} & \repolink{icon-codex} & Icon & \textsc{cdx} \\
12 & \texttt{chess-java} & \repolink{java-codex} & Java & \textsc{cdx} \\
13 & \texttt{chess-java-cc} & \repolink{java-cc} & Java & \textsc{cc} \\
14 & \texttt{chess-latex-codex-replication} & \repolink{latex-codex-replication} & LaTeX (expl3) & \textsc{cdx} \\
15 & \texttt{latex-chess-engine} & \repolink{latex-TeXCCChess} & TeX & \textsc{cdx} \\
16 & \texttt{lean-chess} & \repolink{lean-codex} & Lean 4 & \textsc{cdx} \\
17 & \texttt{chess-mojo} & \repolink{mojo-codexfailure} & Mojo (underconverged) & \textsc{cdx} \\
18 & \texttt{chess-newlang-codex} & \repolink{dsl-newlang-codex} & C\texttt{++}/GAMBIT DSL & \textsc{cdx} \\
19 & \texttt{chess-purec} & \repolink{c-cc} & C & \textsc{cc} \\
20 & \texttt{chess-purec-codex} & \repolink{c-codex} & C & \textsc{cdx} \\
21 & \texttt{chess-py} & \repolink{python-codex} & Python & \textsc{cdx} \\
22 & \texttt{chess-py-cc} & \repolink{python-cc} & Python & \textsc{cc} \\
23 & \texttt{chess-revisit-java-toRust-codex} & \repolink{rust-from-java-codex} & Rust (port from Java) & \textsc{cdx} \\
24 & \texttt{chess-Rocq} & \repolink{rocq-cc} & Rocq $\to$ OCaml & \textsc{cc} \\
25 & \texttt{chess-ruby-cc} & \repolink{ruby-cc} & Ruby & \textsc{cc} \\
26 & \texttt{chess-ruby-codex} & \repolink{ruby-codex} & Ruby & \textsc{cdx} \\
27 & \texttt{chess-rust-cc} & \repolink{rust-cc} & Rust & \textsc{cc} \\
28 & \texttt{chess-rust-cc-redo} & \repolink{rust-cc-redo} & Rust & \textsc{cc} \\
29 & \texttt{chess-rust-codex} & \repolink{rust-codex} & Rust & \textsc{cdx} \\
30 & \texttt{chess-sql} & \repolink{sql-cc} & SQL & \textsc{cc} \\
31 & \texttt{chess-why3} & \repolink{why3-codex} & Why3 $\to$ OCaml & \textsc{cdx} \\
32 & \texttt{chess-why3-cc} & \repolink{why3-cc} & Why3 $\to$ OCaml & \textsc{cc} \\
33 & \texttt{test-superset} & \repolink{css-ChessCSS} & CSS + minimal JS & \textsc{cc} \\
34 & \texttt{chess-revisit-java-toCOBOL-codex}\textsuperscript{$\dagger$} & \repolink{cobol-from-java-codex} & COBOL (port from Java) & \textsc{cdx} \\
\end{longtable}
\end{scriptsize}

\noindent\textsuperscript{$\ast$}\,Full repository URL is
\url{https://github.com/acherm/agentic-chessengine-<suffix>}.\\
\noindent\textsuperscript{$\dagger$}\,Materialised from the
sub-folder \texttt{chess-revisit-java-toCOBOL/} inside the Rust
port working tree (\#23) into its own SANDBOX directory before
release.

\section{Use of LLMs in conducting this study}
\label{app:llm-usage}

Following the community guidelines for empirical studies
involving LLMs~\citep{baltes2025llmguidelines}, this appendix
documents where LLMs were used to conduct and report the
study, what they contributed, and how the authors verified
each contribution. LLMs appear in three roles. They are the
\emph{object of study}, namely the corpus agents of
\Cref{sec:background:agents}. They are an \emph{analysis
instrument}, assisting the construction of the measurement
code. They are a \emph{writing assistant} for the analysis
reports and for the manuscript. The instrument and writing
roles were filled mainly by \claudecode{} sessions run inside
the meta-analysis repository between 2026-04 and 2026-06, on
Anthropic \opus{4.x}-family and \textsc{Fable~5} models.

The guiding principle is \emph{code over judgement}. Wherever
the paper reports a number, that number is computed by a
deterministic script over released primary data (transcripts,
repositories, game records), and the script is released too.
LLM assistance therefore concentrated on producing code and
prose that the authors could inspect and verify, not on
producing measurements directly. No quantitative result in
the paper is an unverified LLM judgement.
\Cref{tab:llm-usage} lists each task with its LLM role, the
author role, and the artefact a reader can use to check it.

\begin{table}[!htbp]
\centering\small
\caption{Division of labour between LLMs and authors, per
analysis task.}
\label{tab:llm-usage}
\begin{tabularx}{\linewidth}{>{\raggedright\arraybackslash}p{0.20\linewidth} >{\raggedright\arraybackslash}X >{\raggedright\arraybackslash}X}
\toprule
Task & LLM role & Author role and verifiable artefact \\
\midrule
Corpus creation, 34 engines
(\Cref{sec:corpus:acquisition}) &
Object of study. The agents plan, code, build, debug, and
validate the engines. &
Capability-level steering only, audited over 374 follow-up
prompts (per-prompt audit released with the artefact).
Artefacts: engine repositories, \codex{} rollouts. \\
\addlinespace
Extraction pipeline
(\Cref{sec:corpus:evidence}) &
Drafted the Python code that derives LOC, features, session
ledgers, backlogs, and novelty signals. &
Specified each derivative, reviewed the code, spot-checked
outputs against raw transcripts and repositories. Artefacts:
\texttt{scripts/} and \texttt{data/} JSON, re-runnable. \\
\addlinespace
Feature catalogue and depth pipeline
(\Cref{sec:corpus:feature-analysis}) &
Drafted the extraction and hotspot-comparison code. &
Curated the 38-feature catalogue from the chess-programming
reference, hand-audited every per-variant verdict for the
three case-study features. Artefacts: per-feature JSON,
feature explorer. \\
\addlinespace
SE-activity coding frame
(\Cref{sec:corpus:se-activities}) &
Implemented the deterministic per-step tagging predicates. &
Induced, tested, and pruned the categories, fixed the tagging
rules. Artefact: structured per-step records that any reader
can re-code. \\
\addlinespace
Canonical evaluation platform
(\Cref{sec:corpus:canonical-eval}, \Cref{app:anchor-eval}) &
Drafted the harness code (Docker pinning, gauntlet
orchestration, Bradley--Terry fit). &
Designed anchors, time control, and adjudication, checked
calibration against externally rated opponents. Artefacts:
PGNs, configurations, scripts. \\
\addlinespace
Statistics and paper tables &
Drafted the scripts that compute Elo aggregates and emit the
LaTeX tables. &
Verified formulas and numbers. The generators are the single
source of the published tables. Artefact: generator scripts
in the replication package. \\
\addlinespace
Per-engine reports and synthesis
(\Cref{sec:corpus:evidence}) &
Drafted the narratives from tagged evidence. &
Reviewed each report against its evidence pointers,
re-verified every claim promoted into the paper. Artefacts:
45 per-engine reports and cross-project syntheses in
\texttt{reports/}. \\
\addlinespace
Manuscript &
Drafted and edited LaTeX text under author direction. &
Set structure, framing, and claims, revised all text, checked
all numbers against the generated tables. Artefacts: paper
sources and table generators. \\
\bottomrule
\end{tabularx}
\end{table}

\paragraph{Agent-drafted prose and its verification.}
The per-engine reports and the qualitative deep dives are
agent-drafted. Each was generated under an evidence-citation
discipline that requires every claim to carry a pointer to a
repository file, a git commit, a transcript line, a run log,
or a computed statistic, and the first author reviewed each
report against those pointers. The reports are intermediate
analysis notes rather than publication text, and any report
claim that reached the paper was re-verified against the
underlying primary data.

\paragraph{Analysis-session transcripts.}
The analysis sessions are stored client-side by the product
under a retention policy, so we treat them as ephemeral. The
durable record of the analysis is the repository itself: the
git history, the released scripts, the data snapshots, and
the reports they generate.

\paragraph{Resource demands and justification.}
The corpus consumed about \$3{,}100 in canonical USD across
34 engines, plus roughly two and a half months of
intermittent developer effort and the evaluation compute of
the canonical harness. We consider the spend justified on
four grounds. Whether programming languages still matter to
coding agents at system-construction scale is not answered by
the existing benchmark literature (\Cref{sec:related}). The
question requires frontier agents, because system-scale
builds are precisely the regime where lightweight open models
have not demonstrated reliable end-to-end capability
(\Cref{sec:threats}). The corpus yields artefacts of
independent interest to the programming-language and
chess-programming communities, including first-of-kind
engines in several languages and a reusable cross-PL
evaluation platform. Finally, inference prices have fallen
steadily since data collection, so replications of this
protocol should cost less than the original. We did not
instrument energy consumption, and token volumes plus
canonical USD are the resource proxies we report.

\section{Self-assessment against the LLM-study reporting
guidelines}
\label{app:guidelines}

We self-assessed this paper against the reporting checklist
(version 2026.06) that accompanies the community guidelines
for empirical studies in software engineering involving
LLMs~\citep{baltes2025llmguidelines}. \Cref{tab:guidelines}
maps each checklist area to the places in the paper that
address it and names the residual gaps plainly rather than
claiming blanket compliance. The recurring gaps trace back to
two structural facts about the study: it observes commercial
agents whose internals and decoding parameters are not
user-controllable, and raw \claudecode{} transcripts are
stored client-side by the product, so the released record of
\claudecode{} interaction is the committed derived snapshot.

\begin{table}[!htbp]
\centering\small
\caption{Checklist areas of the LLM-study reporting
guidelines~\citep{baltes2025llmguidelines}, where this paper
addresses them, and residual gaps.}
\label{tab:guidelines}
\begin{tabularx}{\linewidth}{>{\raggedright\arraybackslash}p{0.23\linewidth} >{\raggedright\arraybackslash}p{0.28\linewidth} >{\raggedright\arraybackslash}X}
\toprule
Checklist area & Where addressed & Status and residual gap \\
\midrule
Declare LLM usage (study, analysis, writing) &
\Cref{sec:background:agents,sec:corpus:evidence},
\Cref{app:llm-usage}, LLM-use declaration &
Addressed. \\
\addlinespace
Model and tool versions, configuration, dates &
\Cref{sec:background:agents,sec:corpus:acquisition},
timestamps in the data snapshot &
Partial. \claudecode{} CLI versions are not recorded in the
released snapshot. Decoding parameters are vendor defaults,
not exposed to users. \\
\addlinespace
Prompts and interaction design &
\Cref{sec:corpus:protocol}, \Cref{fig:protocol},
per-prompt audit in the artefact &
Addressed. Initial template published verbatim, all 374
follow-up prompts audited. \\
\addlinespace
Session traces &
Data and code availability statement, provenance notes in
the artefact &
Partial. \codex{} rollouts released with PII redaction.
\claudecode{} interaction released as the derived snapshot
of 2026-04-20, raw JSONLs are stored client-side and not
released. \\
\addlinespace
Benchmarks and metrics &
\Cref{sec:background:oracles,sec:corpus:canonical-eval,sec:rq4},
\Cref{app:anchor-eval} &
Addressed. The Elo construct, its calibration, and its limits
are discussed in \Cref{sec:threats}. \\
\addlinespace
Non-determinism and repetitions &
\Cref{sec:threats} (conclusion validity),
\Cref{app:anchor-eval:retry} &
Partial. One to three sessions per language cell, justified
by cost and by the existential form of the headline claims.
Elo rests on repeated games with confidence intervals. \\
\addlinespace
Human validation and coding reliability &
\Cref{sec:corpus:feature-analysis,sec:corpus:supervision},
\Cref{sec:threats} &
Partial. Single human coder, no human-versus-human agreement
statistic. Deterministic re-coding is possible from released
records. \\
\addlinespace
Open-weight baseline &
\Cref{sec:threats} &
Not done. Named as a measurable follow-up for which our
artefacts provide ground truth. \\
\addlinespace
Contamination and memorisation &
\Cref{sec:rq2:audit,sec:rq3:evasion}, \Cref{sec:threats} &
Addressed on the output side via the novelty audit.
Training-data fluency confound acknowledged as open. \\
\addlinespace
Replication package &
Data and code availability statement, \Cref{app:github-repos} &
Addressed. The known gap is the raw \claudecode{}
transcripts, which are stored client-side and not part of
the artefact. \\
\addlinespace
Resources and justification &
\Cref{sec:corpus:evidence,sec:rq5}, \Cref{app:llm-usage} &
Addressed. Tokens and canonical USD reported. Energy not
instrumented. \\
\bottomrule
\end{tabularx}
\end{table}

\end{document}